\documentclass[11pt,leqno]{article}
\usepackage{amssymb,amsfonts}%amsLatex
\usepackage{amsmath,amsthm,amsxtra}
\usepackage[all]{xy}

\newcommand{\CC}{\mathbb{C}}
\newcommand{\NN}{\mathbb{N}}
\newcommand{\QQ}{\mathbb{Q}}
\newcommand{\RR}{\mathbb{R}}
\newcommand{\ZZ}{\mathbb{Z}}

\newcommand{\D}{\mathcal{D}}

\newcommand{\fa}{\mathfrak{a}}
\newcommand{\fg}{\mathfrak{g}}
\newcommand{\fh}{\mathfrak{h}}
\newcommand{\fn}{\mathfrak{n}}

\newcommand{\ad}{\mathop{\rm ad \, }}
\newcommand{\bose}{\mathop{\rm bose \, }}
\newcommand{\charge}{\mathop{\rm charge \, }}
\newcommand{\diag}{\mathop{\rm diag \, }}
\newcommand{\energy}{\mathop{\rm energy \, }}
\newcommand{\fermi}{\mathop{\rm fermi \, }}
\renewcommand{\Im}{\mathop{\rm Im  \, }}

\newcommand{\low}{\mathop{\rm  low \, }}
\renewcommand{\Re}{\mathop{\rm Re  \, }}
\newcommand{\sdim}{\mathop{\rm sdim \, }}
\newcommand{\tr}{\mathop{\rm tr \, }}
\newcommand{\weight}{\mathop{\rm weight \, }}
\newcommand{\mac}{|m\rangle}
\newcommand{\sac}{|s\rangle}
\newcommand{\vac}{|0\rangle}

\newcommand{\str}{\mathop{\rm str}}

\newcommand{\st}[1]{\ensuremath{^{\scriptstyle \textrm{#1}}}}

\renewcommand{\hat}{\widehat}

\makeatletter
\renewcommand\section{\@startsection {section}{1}{\z@}%
                                   {-3.5ex \@plus -1ex \@minus -.2ex}%
                                   {2.3ex \@plus.2ex}%
                                   {\normalfont\large\bfseries}}
\renewcommand\subsection{\@startsection{subsection}{2}{\z@}%
                                     {-3.25ex\@plus -1ex \@minus -.2ex}%
                                     {0ex \@plus .0ex}%
                                     {\normalfont\normalsize\bfseries}}  

\@addtoreset{equation}{section}
\makeatother

\newtheorem{theorem}{Theorem}[section]
\newtheorem{definition}{Definition}[section]
\newtheorem{lemma}{Lemma}[section]
\newtheorem{proposition}{Proposition}[section]

\theoremstyle{remark}
\newtheorem{remark}{Remark}[section]
\newcommand{\alphaparenlist}{% changes enumerate 1st level to (a)...(z)
  \renewcommand{\theenumi}{\alph{enumi}}%
  \renewcommand{\labelenumi}{(\theenumi)}%
}
\newcommand{\arabicparenlist}{% changes enumerate 1st level to (1)...(9)
  \renewcommand{\theenumi}{\arabic{enumi}}%
  \renewcommand{\labelenumi}{(\theenumi)}%
}
\newcommand{\romanparenlist}{% changes enumerate 1st level to (i)...(ix)
  \renewcommand{\theenumi}{\roman{enumi}}%
  \renewcommand{\labelenumi}{(\theenumi)}%
}
\newcommand{\starlist}{% changes enumerate 1st level to (*)
  \renewcommand{\theenumi}{\rm\textup{(*)}}%
  \renewcommand{\labelenumi}{\theenumi}%
}
\newcommand{\romanlistii}{% changes enumerate 2nd level to i. ... x.
  \renewcommand{\theenumii}{\roman{enumii}}%
  \renewcommand{\labelenumii}{(\theenumii)}%
}
\makeatletter
\def\@maketitle{\newpage
 \null
 \vskip 2em
 \begin{center}%
  {\@date}%
  \vskip 3em
  {\Large\bf \@title \par}%
  \vskip 1.5em
  {\normalsize
   \lineskip .5em
   \begin{tabular}[t]{c}\@author
   \end{tabular}\par}%
  \vskip 2em

 \end{center}%
 \par
 \vskip 2.5em}
\makeatother
\begin{document}

\title{Integrable highest weight modules over affine
  superalgebras and Appell's function}

\author{Victor G. Kac\thanks{Department of Mathematics, M.I.T., 
    Cambridge, MA 02139, USA.~~kac@math.mit.edu}~~\thanks{Supported in part by NSF grant
    DMS-9970007.}~~and Minoru Wakimoto\thanks{Graduate School
    of Mathematics, Kyushu University, Fukuoka 812-81,
    Japan~~wakimoto@math.kyushu-u.ac.jp}}
%\thanks{
%    Department of Mathematics, M.I.T., 
%    Cambridge, MA 02139, 
%    $<$kac@math.mit.edu$>$ 
%%    Supported in part by NSF grant DMS-9970007.
%\date{}
%}

\maketitle

\begin{abstract}
  We classify integrable irreducible highest weight
  representations of non-twisted affine Lie superalgebras. We
  give a free field construction in the level~1 case.  The analysis
  of this construction shows, in particular, that in the
  simplest case of the $s\ell (2|1)$ level~1 affine superalgebra
  the characters are expressed in terms of the Appell elliptic
  function.  Our results demonstrate that the representation
  theory of affine Lie superalgebras is quite different from that 
  of affine Lie algebras.
\end{abstract}

\section{Introduction}
\label{intro}

In this paper we continue the study of integrable irreducible
highest weight modules over affine superalgebras that we began in 
\cite{KW}. 

First, let us recall the definition of an integrable module over
an ordinary affine Kac-Moody algebra $\hat{\fg}$ \cite{K3}.  Let
$\fg$ be a finite-dimensional simple or abelian Lie algebra over
$\CC$ with a symmetric invariant bilinear form $(.|.)$.  Recall
that the associated affine algebra is
\begin{equation}
\label{eq:0.1}
  \hat{\fg}= (\CC [t,t^{-1}] \otimes_{\CC}  \fg)
  \oplus \CC K \oplus \CC d
\end{equation}
with the following commutation relations ($a,b \in \fg ; m,n \in \ZZ$ and $ a(m)$
  stands for $ t^m \otimes a$) :
\begin{eqnarray}
  \label{eq:0.2}
\hspace*{8ex}  [a(m), b(n)] = [a,b](m+n) + m \delta_{m,-n} (a|b) K  , \quad
  [d,a(m)] = -ma (m) , \quad [K, \hat{\fg}] =0 \, .
\end{eqnarray}
We identify $\fg$ with the subalgebra $1 \otimes \fg$.  The
bilinear form $(.|.)$ extends from $\fg$ to a symmetric invariant 
bilinear form on $\hat{\fg}$ by:
\begin{eqnarray}
\nonumber
  (a(m)| b(n)) &=& \delta_{m,-n} (a|b) , (\CC [t,t^{-1}]
  \otimes \fg | \CC K + \CC d) =0 \, , \\
  \label{eq:0.3}
  (K|K) &=& (d|d) =0 , \quad (K|d) = -1 \, . 
\end{eqnarray}

Choose a Cartan subalgebra $\fh$ of $\fg$ and let $\fg = \fh
\oplus (\oplus_{\alpha \in \Delta} \fg_{\alpha})$ be the root space
decomposition, where $\fg_{\alpha}$ denotes the root space
attached to a root $\alpha \in \Delta \subset \fh^*$.  Let
\begin{equation}
  \label{eq:0.4}
  \hat{\fh} = \fh + \CC K + \CC d
\end{equation}
be the Cartan subalgebra of $\hat{\fg}$, and, as before, let
$\fg_{\alpha}(m) = t^m \otimes \fg_{\alpha}$.

A $\hat{\fg}$-module $V$ is called \emph{integrable} if the
following two properties hold \cite{K3}:
\begin{eqnarray}
  \label{eq:0.5}
 && \hat{\fh} \hbox{ is diagonalizable on } V \, ,\\
\label{eq:0.6}
  && \hbox{all } \fg_{\alpha} (m)  (\alpha \hbox{ a root of } \fg 
  \, ,  m \in \ZZ)
     \hbox{ are locally finite on } V \, . 
\end{eqnarray}
(Property (\ref{eq:0.6}) means that $\dim U (\fg_{\alpha}(m)) v < \infty$
for any $v \in V$.)

It is easy to show that these two properties imply
\begin{eqnarray}
  \label{eq:0.7}
  \fg \hbox{ is locally finite on } V (\hbox{ i.e., }
  \dim U (\fg) v < \infty \hbox{ for any } v \in V) \, .
\end{eqnarray}
Here and further $U (\fa)$ denotes the universal enveloping
algebra of a Lie (super)algebra $\fa$.  Note also that condition
(\ref{eq:0.6}) is vacuous if $\fg$ is abelian.

Let now $\fg = \fg_{\bar{0}} + \fg_{\bar{1}}$ be a
finite-dimensional Lie superalgebra over $\CC$ with an even
symmetric invariant bilinear for $(.|.)$ (for a background on Lie 
superalgebras see \cite{K1}).  Recall that ``even'' means that
$(\fg_{\bar{0}}| \fg_{\bar{1}})=0$, ``symmetric'' means that
$(.|.)$ is symmetric on $\fg_{\bar{0}}$ and skewsymmetric on
$\fg_{\bar{1}}$, and ``invariant'' means that $([a,b]|c) =
(a|[b,c]), a,b,c \in \fg$.  We shall assume, in addition, that
$\fg_{\bar{0}}$ is reductive:
\begin{equation}
  \label{eq:0.8}
  \fg_{\bar{0}} = \oplus^N_{j=0} \fg_{\bar{0}j} \, ,
\end{equation}
where $\fg_{\bar{0}0}$ is abelian and $\fg_{\bar{0}j}$ with $j
\geq 1$ are simple Lie algebras.

The affine superalgebra $\hat{\fg}$ associated to the Lie
superalgebra $\fg$ and the bilinear form $(.|.)$ is defined in
exactly the same way as in the Lie algebra case by relations
(\ref{eq:0.2}).  Likewise, the invariant even symmetric bilinear
form $(.|.)$ on $\hat{\fg}$ is defined by (\ref{eq:0.3}), and
the Cartan subalgebra $\hat{\fh}$ is defined by (\ref{eq:0.4})
after a choice of a Cartan subalgebra $\fh$ of $\fg_{\bar{0}}$.
Note that for each\break $j \in \{ 0,1, \ldots ,N \}$, the superalgebra 
$\hat{\fg}$ contains an affine Kac-Moody algebra
$\hat{\fg}_{\bar{0}j}$ associated to $\fg_{\bar{0}j}$.

We shall see that condition (\ref{eq:0.6}) of integrability is
too strong in the superalgebra case, as for most of the affine
superalgebras it allows only trivial highest weight modules.
This forces us to consider weaker conditions (cf.~\cite{KW}):

\begin{definition}
  \label{def:0.1}
  Given a subset $J \subset \{ 1, \ldots ,N \}$, a
  $\hat{\fg}$-module $V$ is called $J$-\emph{integrable} if it
  satisfies conditions (\ref{eq:0.5}) and (\ref{eq:0.7}) and if it is integrable as
  $\hat{\fg}_{\bar{0}j}$-module for all $j \in J$.
\end{definition}

Let $\displaystyle{\fg = \fh \oplus (\oplus_{\alpha \in \Delta}
  \fg_{\alpha})}$ be a root space decomposition of the Lie
superalgebra $\fg$ with respect to a Cartan subalgebra $\fh$ of $
\fg_{\bar{0}}$.  Choose a set of positive toots $\Delta_+$ in $
\Delta$ and let $\displaystyle{\fn_+=\oplus_{\alpha \in \Delta_+}
  \fg_{\alpha}}$.  For each $\Lambda \in \hat{\fh}^*$ one defines 
an \emph{irreducible highest weight module} $L (\Lambda)$
over $\hat{\fg}$ as the (unique) irreducible $\hat{\fg}$-module
for which there exists a non-zero vector $v_{\Lambda}$ such that 
\begin{equation}
  \label{eq:0.9}
  h v_{\Lambda}= \Lambda (h)v_{\Lambda} \hbox{ for }
  h \in \hat{\fh}, \,\, \fn_+ v_{\Lambda} =0, \,\,
  \fg (m) v_{\Lambda}=0 \hbox{ for } m>0 \, ,
\end{equation}
where, as before $\fg (m) =t^m \otimes \fg$.  The number
$k=\Lambda(K)$ is called the \emph{level} of $L(\Lambda)$ and of
$\Lambda$.
Note that $K=kI$ on $L(\Lambda)$ and that
$\bar{L}(\Lambda):=U(\fg)v_{\Lambda}$ is an irreducible highest weight
module over $\fg$.

%moved to p.34  rem 5.3
%It is easy to see that, when restricted to the derived subalgebra 
%$[\hat{\fg}, \hat{\fg}]$ of $\hat{\fg}$ the module $L(\Lambda)$
%remains irreducible.  Two $\hat{\fg}$-modules are called
%\emph{essentially equivalent} if they are equivalent as
%$[\hat{\fg}, \hat{\fg}]$-modules.  For example, the modules
%$L(\Lambda)$ and $L(\lambda + a \delta)$ are essentially equivalent for any $
%a \in \CC$.

In \S~\ref{sec:1} we describe a general approach to the
classification of irreducible integrable highest weight modules
over arbitrary Kac-Moody superalgebras, and in \S~\ref{sec:2} and 
\S~\ref{sec:6} give their complete classification in the affine
(non-twisted) case, using Serganova's odd reflections.  

In \S~\ref{sec:3} we give a free field
realization of all level~$1$ integrable highest weight modules
over $g \ell (m|n)^{\hat{}}$, which leads to a ``quasiparticle''
character formula for these modules and to a ``theta function'' type 
character formula.  This construction may be viewed as a
generalization of the classical boson-fermion correspondence
based on the oscillator algebra $g\ell (1)^{\hat{}}$ and of the
super boson-fermion correspondence based on $g\ell (1|1)^{\hat{}}$
 \cite{KL}.  The former produces the classical vertex operators
 and relates representation theory of $g\ell (1)^{\hat{}}$ to the 
 denominator identity for $s\ell(2)^{\hat{}}$, while the latter
 produces vertex operators for the symplectic bosons and relates
 representation theory of $g\ell (1|1)^{\hat{}}$ to the
 denominator identity for $s\ell (2|1)^{\hat{}}$
 (see~\cite{K4}).  

 In \S~\ref{sec:4} we show that the ``theta function'' type
 character formula for $s \ell (m|1)^{\hat{}} (m \geq 2)$ is a
 product of a theta function, a power of the eta function, and a
 more ``exotic'' function, called a multivariable Appell
 function. The classical Appell function appeared in the 1880's
 in the papers by Appell \cite{A} and by Hermite in their study
 of elliptic functions. Most recently this function has been
 discussed in \cite{P}.  The study of asymptotics of Appell's
 functions gives the high temperature asymptotics of integrable
 level~1 $s \ell (m|1)^{\hat{}}$-characters.  We also derive here 
 formulas for branching functions for integrable level~$1$ $s
 \ell (m|1)^{\hat{}}$-modules restricted to the even subalgebra.
 They turn out to be certain ``half'' modular functions.

In \S~\ref{sec:5} we relate integrable level~$ 
1$ modules over $g\ell (m|n)^{\hat{}}$ to the denominator
identity for $s\ell (m+1|n)^{\hat{}}$, and as a result, we derive for these 
modules 
yet another, a Weyl type, character formula.  

In \S~\ref{sec:7}
we give a free field realization of the two level~$1$ integrable
highest weight modules over $osp (m|n)^{\hat{}}$, which
generalizes the constructions for $so(m)^{\hat{}}$ and
$sp(n)^{\hat{}}$ from \cite{KP1}, \cite{F} and \cite{FF}.  These
lead to character formulas and high temperature asymptotics of
the characters.

In \S~\ref{sec:8} we show that integrability is a necessary
condition for an irreducible highest weight $\hat{\fg}$-module to
be a module over the associated vertex algebra, and that in the
level~$1$ case this condition is sufficient.  We thus get
examples of rational vertex algebras for which the $\CC$-span of
normalized (super)characters is not $SL (2,\ZZ)$-invariant.  The latter
property was proved in \cite{Z} under certain additional
assumptions, and it was generally believed that these assumptions 
were superfluous.

In \S~\ref{sec:9} we discuss some open problems.

It is interesting to note that in the ``super'' case a number of
new interesting phenomena occur.  The level gets quantized by the
integrability condition, but in almost all cases the number of
integrable modules is infinite.  This is the case for the lowest, 
level~$1$, integrable $s\ell (m|n)^{\hat{}}$-modules which
apparently causes the specialized characters and branching
functions to lose their customary modularity properties, which
are so ubiquitous in the affine Lie algebra case \cite{KP2},\cite{K3}.
However, in the cases when the number of characters of given
level is finite, like, for example, $k=1 \quad osp
(m|n)^{\hat{}}$ case, the specialized normalized characters are
still modular, though their $\CC$-span is no longer $SL
(2,\ZZ)$-invariant as in the affine Lie algebra case.

It is also interesting to note that while the characters of
affine Lie algebras are global sections of line bundles on
abelian varieties, the characters of affine Lie superalgebras are 
related to global sections of rank~$2$ vector bundles on abelian
varieties, as the work of Polishchuk \cite{P} on Appell's
function apparently indicates.
%but only for the lowest positive level
%($=1 $) the number of modules is finite (actually, always one!).
%The characters of $s\ell (m|n)^{\hat{}}$-integrable modules are
%not modular even at level~$1$ (though for $osp (m|n)^{\hat{}}$
%they are).  The calculation of characters of level~$k>1$ seems to
%be far away.  But even in the $k=1$ case the transformation
%properties of $s\ell (m|n)^{\hat{}}$ characters
%remain an open problem.

We are grateful to A.~Polishchuk for giving us the idea of
Lemma~\ref{lem:4.1} and other very useful remarks.  The first
named author wishes to thank ENS, IHES and MSRI for their hospitality.

\section{Integrability of highest weight modules over Kac-Moody superalgebras}
\label{sec:1}

Consider the following data:
\begin{displaymath}
  \D = \{ \fh , I , I_1, \Pi^{\vee}, \Pi \} \, ,
\end{displaymath}
where $\fh$ is a vector space, $I$ is an index set, $I_1$ is a
subset of $I$, $\Pi^{\vee} = \{ \alpha^{\vee}_i \}_{i \in I}$ and 
$\Pi = \{ \alpha_i \}_{i \in I}$ are linearly independent sets of 
vectors in $\fh$ and $\fh^*$ respectively indexed by $I$.  One
associates to these data a Lie superalgebra $\fg (\D)$ defined as 
the quotient of the Lie algebra on generators $e_i, f_i (i \in
I)$ and $\fh$, the generators $e_i$ and $f_i$ for $i \in I_1$
being odd and all other generators being even, and the
\emph{standard relations} $(i,j \in I, h \in \fh)$:
\begin{displaymath}
  [\fh, \fh] =0 , \quad [e_i , 
   f_j] = \delta_{ij} \alpha^{\vee}_i,
  \quad   [h,e_i]= \langle \alpha_i , 
 h \rangle e_i,   \quad [h, f_i] = - \langle \alpha_i,h \rangle f_i \, ,
\end{displaymath}
by the maximal graded with respect to the root space
decomposition intersecting $\fh$ trivially (cf.~\cite{K1}, \cite{K3}
).

The commutative $\ad$-diagonizable subalgebra $\fh$ of $\fg (\D)
$ is called the Cartan subalgebra, $\Pi$ and $\Pi^{\vee}$ are
called the sets of simple roots and coroots respectively,
elements $e_i$ and $f_i$ $(i \in I)$ are called Chevalley
generators, etc.  One defines the notions of roots and root
spaces in the usual way (cf.~\cite{K1}, \cite{K3}).  Let $\fn_+$
(resp.~$\fn_-$) denote the subalgebra of $\fg$ generated by the $
e_i$'s (resp.~$f_i$'s).  Then, as usual, one has the triangular
decomposition:
\begin{displaymath}
  \fg = \fn_- + \fh + \fn_+ \, .
\end{displaymath}
Let $a_{ij} = \langle \alpha_j , \alpha^{\vee}_i \rangle$.  The
matrix $A=(a_{ij})_{i,j \in I}$ is called the Cartan matrix of
the data $\D$ (and of $\fg (\D)$).

%The data $\D^{\vee} = \{ \fh^* , I,I_1, \Pi , \Pi^{\vee}\}$ are
%called the dual data; their Cartan matrix is the transpose of $A$.
%The roots of $\fg (\D^{\vee})$ are called coroots of $\fg (\D)$.

A root of $\fg (\D)$ is called \emph{even} (resp.~\emph{odd}) if the attached
root space is even  (resp.~odd).  For example a simple root
$\alpha_s$ is called odd iff $s \in I_1$.  An odd simple root
$\alpha_s$ (and the coroot $\alpha^{\vee}_s$) is called
\emph{isotropic} if $a_{ss} =0$.  In what follows we let
\begin{eqnarray*}
  p_{ij} = \left\{
    \begin{array}{rl}
      -1 & \hbox{if both $\alpha_i$ and $\alpha_j$ are odd,}\\
      1 & \hbox{othewise.}
    \end{array}\right.
\end{eqnarray*}

Note that $\fg (\D)$ has an anti-involution $\omega$ defined by $
\omega (e_i)=f_i, \omega (f_i)=e_i, \omega |_{\fh} =I$.  For that
reason properties of the $e_i$'s automatically hold for the
$f_i$'s.

\alphaparenlist
\begin{lemma}
  \label{lem:1.1}
  \begin{enumerate}
  \item %%a
    An odd simple root $\alpha_i$ is isotropic iff $[e_i,
    e_i]=0$.

\item %%b
  If $i \neq j$, then $[e_i,e_j]=0$ iff $a_{ij} =a_{ji}=0$.
  \end{enumerate}
\end{lemma}

\begin{proof}
  It is clear that $[f_j, [e_i,e_i]]=0$ if $j \neq i$, and one
  has: $[f_i, [e_i,e_i]]= 2a_{ii} e_i$, which proves (a).  The
  proof of (b) is similar.

\end{proof}

It is straightforward to check the following relation $(i,j \in I , 
 i \neq j)$:
\begin{equation}
  \label{eq:1.1}
  [[e_i,e_j], [f_i,f_j]] = p_{ij} 
     (a_{ij} \alpha^{\vee}_j - p_{ii}a_{ji}\alpha^{\vee}_i) \, 
     .
\end{equation}
Further on we shall always assume the following property of the
Cartan matrix $A$:
\begin{equation}
  \label{eq:1.2}
  a_{ij} =0 \hbox{ iff } a_{ji}=0 \, .
\end{equation}

Given $s \in I_1$ such that $a_{ss} =0$ (i.e.,~$\alpha_s$ is an odd 
isotropic simple root), define a new data
\begin{displaymath}
  r_s (\D) = \{ \fh ,I, r_s (I_1), r_s (\Pi^{\vee}),
  r_s (\Pi) \}
\end{displaymath}
and new Chevalley generators $r_s (e_i), r_s (f_i)$ of $\fg (\D)$ 
as follows (cf.~\cite{S}, \cite{PS}, \cite{KW}):
\begin{eqnarray*}
  i \in r_s (I_1) \hbox{ iff } i &\not\in & I_1 \hbox{ in case }
  a_{si} \neq 0, \quad i \in r_s (I_1) \hbox{ iff }
  i \in I_1 \hbox{ otherwise;}\\
  r_s (\alpha^{\vee}_s) &=& - \alpha^{\vee}_s , \quad r_s
  (\alpha_s) =-\alpha_s \, , \\[1ex]
  r_s (\alpha^{\vee}_i) &=& \alpha^{\vee}_i +
  \frac{a_{is}}{a_{si}} \alpha^{\vee}_s \hbox{ and }
  r_s (\alpha_i) = \alpha_i + \alpha_s \hbox{ if }
  a_{si} \neq 0 \, , \\[1ex]
  r_s (\alpha^{\vee}_i) &=& \alpha^{\vee}_i \hbox{ and }
  r_s (\alpha_i) = \alpha_i
  \hbox{ in all other cases;}\\
  r_s (e_s) &=& f_s , \quad r_s (f_s) =-e_s \, , \\
  r_s (e_i) &=& [e_s, e_i] \hbox{ and }
  r_s (f_i) = \frac{1}{p_{si}a_{si}}[f_s,f_i]
  \hbox{ if } a_{si} \neq 0 \, , \\[1ex]
  r_s (e_i) &=& e_i, \quad r_s (f_i)=f_i \hbox{ in all other cases.}
\end{eqnarray*}
Denote by $r_s (\fn_+)$ (resp.~$r_s (\fn_-)$) the subalgebra of
$\fg (\D)$ generated by the $r_s (e_i)$'s (resp.~$r_s (f_i)$'s).
The transformation $r_s$ is called an \emph{odd reflection} (with 
respect to $\alpha_s$).

\begin{lemma}
  \label{lem:1.2}
  \begin{enumerate}
  \item %%a
    The data $r_s (\D)$ satisfy (\ref{eq:1.2}).

\item %%b
  The new Chevalley generators satisfy the standard relations and 
  together with $\fh$ generate $\fg (\D)$, so that $\fg (r_s
  (\D)) \simeq \fg (\D)$.

\item  %%c
  One has the new triangular decomposition:
  \begin{displaymath}
    \fg (\D) =r_s (\fn_-) + \fh + r_s (\fn_+) \, .
  \end{displaymath}

\item %%d
  The data $r_s (r_s (\D))$ coincide with $\D$, and the Chevalley 
  generators $r_s (r_s (e_i))$ (resp.~$r_s (r_s (f_i))$)
  coincide, up to a non-zero factor, with $e_i$ (resp.~$f_i$).

  \end{enumerate}
\end{lemma}

\begin{proof}
  is straightforward using (\ref{eq:1.1}) and the relation
  \begin{displaymath}
    [a, [a,b]] = \frac{1}{2} [[a,a],b] \hbox{ if $a$ ia an odd element.}
  \end{displaymath}

\end{proof}

An element $\rho \in \fh^*$ such that
\begin{displaymath}
  \langle \rho , \alpha^{\vee}_i \rangle = \frac{1}{2} a_{ii}
  \hbox{ for all } i \in I
\end{displaymath}
is called a \emph{Weyl vector} for $\Pi^{\vee}$.

\begin{lemma}
  \label{lem:1.3}
If $\rho$ is a Weyl vector for $\Pi^{\vee}$, then $\rho +
\alpha_s$ is a Weyl vector for $r_s (\Pi^{\vee})$.
\end{lemma}

\begin{proof}
  It suffices to check that in the case
    $a_{is} \neq  0$ one has: $ \langle \rho + \alpha_s,
    a_{si} \alpha^{\vee}_i + a_{is} \alpha^{\vee}_s \rangle
    = $\break $\frac{1}{2} \langle \alpha_i + \alpha_s , 
         a_{si} \alpha^{\vee}_i + a_{is} \alpha^{\vee}_s \rangle$,
  which is immediate.

\end{proof}

Recall that for each $\Lambda \in \fh^*$ one defines an
irreducible highest weight module $L(\Lambda)$ over $\fg (\D)$
as the (unique) irreducible $\fg (\D)$-module for which there
exists a non-zero vector $v_{\Lambda}$ such that
\begin{equation}
  \label{eq:1.3}
  h v_{\Lambda} = \Lambda (h) v_{\Lambda} \hbox{ for }
  h \in \fh , \quad \fn_+ v_{\Lambda} =0 \, .
\end{equation}
The vector $v_{\Lambda}$, called a highest weight vector (with
respect to $\fn_+$) is determined uniquely up to a (non-zero)
constant factor by the condition $\fn_+ v_{\Lambda} =0$
(cf.~\cite{K3}).  The linear function $\Lambda$ is called the
\emph{highest weight} (with respect to $\fn_+$) of $L (\Lambda)$.

\begin{lemma}
  \label{lem:1.4}
Let $\alpha_s$ be an odd isotropic simple root and let $\fn'_+ =
r_s (\fn_+)$.

\alphaparenlist
\begin{enumerate}
\item %%a
  If $\langle \Lambda , \alpha^{\vee}_s \rangle =0$, then
  $v'_{\Lambda} = v_{\Lambda}$ is a highest weight vector with
  respect to $\fn'_+$, so that the highest weight remains the
  same:  $\Lambda'=\Lambda$.

\item %%b
  If $\langle \Lambda , \alpha^{\vee}_s \rangle \neq 0$, then
  $v'_{\Lambda} = f_s v_{\Lambda}$ is a highest weight vector
  with respect to $\fn'_+$ so that the highest weight vector
  becomes $\Lambda' = \Lambda - \alpha_s$.
\end{enumerate}
\end{lemma}

\begin{proof}
  is straightforward using the facts that $f^2_s v_{\Lambda} =
  \frac{1}{2} [f_s,f_s]v_{\Lambda}=0$, and $f_s v_{\Lambda}=0$
  iff\break$\langle \Lambda , \alpha^{\vee}_s \rangle =0$.
\end{proof}

As an immediate corollary of Lemmas~\ref{lem:1.3} and
\ref{lem:1.4} we obtain the following very useful formulas
(cf.~\cite{KW}):
\begin{eqnarray}
  \label{eq:1.4}
  \nonumber
  \Lambda' + \rho' &=& \Lambda + \rho \hbox{ if }
     \langle \Lambda , \alpha^{\vee}_s \rangle 
     ( = \langle \Lambda + \rho , \alpha^{\vee}_s \rangle )
     \neq 0 \, , \\
   \Lambda' + \rho' &=& \Lambda + \rho + \alpha_s \hbox{ if }
      \langle \Lambda , \alpha^{\vee}_s \rangle 
      (=  \langle \Lambda + \rho , \alpha^{\vee}_s \rangle ) =0 \, .
\end{eqnarray}

Let $\alpha \in \fh^*$ be a positive even root of $\fg (\D)$ such that 
there exist root vectors $e$ attached to $\alpha$ and $f$
attached to $-\alpha$ satisfying the following conditions:

\romanparenlist

\begin{enumerate}
\item %%i
  $\ad f$ is locally nilpotent on $\fg (\D)$,

\item %%ii
  $[e,f] = \alpha^{\vee} \in \fh, \quad [\alpha^{\vee},e]=2e,
  \quad [\alpha^{\vee},f]=-2f$.
\end{enumerate}

Then we call $f$ an \emph{integrable element} of $\fg (\D)$.  The 
following lemma is well-known (cf.~\cite{K3}).

\begin{lemma}
  \label{lem:1.5}
Let $f$ be an integrable element attached to a negative root
$\alpha$.

\alphaparenlist

\begin{enumerate}
\item %%a
  If $f$ is locally nilpotent on $L (\Lambda)$ then $\langle \Lambda 
  , \alpha^{\vee} \rangle \in \ZZ_+$.

\item %%b
  Provided that $\alpha$ is a simple root, $f$ is locally
  nilpotent on $L (\Lambda)$ iff $\langle \Lambda , \alpha^{\vee} 
  \rangle \in \ZZ_+$.
\end{enumerate}
\end{lemma}

Let $\beta = \alpha_s$ be an odd isotropic simple root.  It will
be convenient to use notation $r_{\beta}$  in place of $r_s$.
Consider a sequence of roots $\beta_0, \beta_1, \ldots , \beta_k$ 
such that $\beta_0$ is an odd isotropic simple root from
$\Pi^{(0)}:= \Pi$, $\beta_1$ is an odd isotropic simple root from 
$\Pi^{(1)}= r_{\beta_0} (\Pi^{(0)}), \ldots$ , $\beta_k$ is an
odd isotropic simple root from $\Pi^{(k)}= r_{\beta_{k-1}}$
$(\Pi^{(k-1)})$.  Given $\Lambda \in \fh^*$, let $\Lambda^{(0)}=\Lambda$
be the highest weight of $L (\Lambda)$ with respect to
$\fn^{(0)}_+ := \fn_+$, $\Lambda^{(1)}$ be the highest weight of
$L (\Lambda)$ with respect to $\fn^{(1)}_+ := r_{\beta_0}
(\fn_+), \ldots , \Lambda^{(k)}$ be the highest weight of $L
(\Lambda)$ with respect to $\fn^{(k)}_+ = r_{\beta_{k-1}}
(\fn^{(k-1)}_+)$.  Let $\rho^{(k)}$ be a Weyl vector for
$\Pi^{(k)}$.

\begin{proposition}
  \label{prop:1.1}
Let $\alpha$ be a positive root of $\fg (\D)$ and let $f$ be an
integrable root element attached to $- \alpha$.  Given $\Lambda \in \fh^*$, let
\begin{displaymath}
  S=\{ i \in [0,1, \ldots ,k-1] | \langle \Lambda^{(i)},
  \beta^{\vee}_i \rangle =0 \} \, .
\end{displaymath}
Suppose that $\alpha \in \Pi^{(k)}$.  Then the element $f$ is
locally nilpotent on $L (\Lambda)$ if and only if
\begin{displaymath}
  \langle \Lambda + \rho + \sum_{i \in S} \beta_i ,
  \alpha^{\vee} \rangle \in \NN = \{ 1,2,  \ldots \} \, .
\end{displaymath}

\end{proposition}

\begin{proof}
  It follows from (\ref{eq:1.4}) that
  \begin{displaymath}
    \Lambda^{(k)}+ \rho^{(k)} = \Lambda + \rho
    + \sum_{i \in S} \beta_i \, .
  \end{displaymath}
Since $\langle \Lambda^{(k)} + \rho^{(k)}, \alpha^{\vee}
\rangle = \langle \Lambda^{(k)}, \alpha^{\vee} \rangle +1$, the
proposition follows from Lemma~\ref{lem:1.5}b.

\end{proof}

\begin{proposition}
  \label{prop:1.2}
If, under the assumptions of Proposition~\ref{prop:1.1}, one has:
\begin{displaymath}
  \langle \Lambda + \rho , \alpha^{\vee} \rangle \in \NN \, ,
\end{displaymath}
then $f$ is integrable on $L (\Lambda)$.

\end{proposition}

\begin{proof}
  Due to Proposition~\ref{prop:1.1}, Proposition~\ref{prop:1.2}
  holds if $S= \emptyset$.  Let $N = \langle \Lambda ,
  \alpha^{\vee} \rangle$.  It is well-known (cf.~\cite{K3}) that
  $f$ is integrable on $L (\Lambda)$ iff
  \begin{equation}
    \label{eq:1.5}
    f^{N+1} v_{\Lambda} \hbox{ lies in a maximal submodule
       of the Verma module } M(\Lambda) \, .
  \end{equation}
But we have just shown that (\ref{eq:1.5}) holds for a Zariski
open set of $\lambda$ on the hyperplane $\langle \lambda ,
\alpha^{\vee} \rangle =N$.  Since (\ref{eq:1.5}) is a polynomial
condition, we conclude that it holds for all $\lambda$ on this hyperplane.
\end{proof}

\begin{proposition}
  \label{prop:1.3}
If, under the assumptions of Proposition~\ref{prop:1.1}, $f$ is
integrable on $L (\Lambda)$ and 
\begin{displaymath}
  \langle \Lambda + \rho , \beta^{\vee}_i \rangle \neq 0
  \hbox{ for } i=0,1, \ldots , s (\leq k) \, ,
\end{displaymath}
then $\langle \Lambda - \sum^{s}_{i=0} \beta_i, \alpha^{\vee}
\rangle \in \ZZ_+$.

\end{proposition}

\begin{proof}
  We have:  $\langle \Lambda , \beta^{\vee}_0 \rangle =
  \langle \Lambda + \rho , \beta^{\vee}_0 \rangle \neq 0$, hence, 
  by (\ref{eq:1.4}) we have:  $\Lambda + \rho = \Lambda^{(1)} +
  \rho^{(1)}$, etc.  Thus, $\Lambda^{(i)} + \rho^{(i)} = \Lambda
  + \rho$ for $i=1, \ldots ,s$.  Therefore, by Lemma~\ref{lem:1.4}b
, we have:
\begin{displaymath}
  \Lambda^{(s)}= \Lambda - \sum^s_{i=0} \beta_i \, .
\end{displaymath}
Now the proposition follows from Lemma~\ref{lem:1.5}a.

\end{proof}

The calculation of coroots is facilitated by the following simple 
fact.

\begin{proposition}
  \label{prop:1.4}
\alphaparenlist
\begin{enumerate}
\item %%a
  There exists a non-degenerate symmetric bilinear form $(.|.)$
  on $\fh$ such that, identifying $\fh$ and $\fh^*$ via this
  form, we have:
  \begin{equation}
    \label{eq:1.6}
    \alpha^{\vee}_i = \nu_i \alpha_i, \hbox{ where } \nu_i \in
    \CC^{\times} \, ,
  \end{equation}
if and only if 
  \begin{equation}
    \label{eq:1.7}
A = \diag (\nu_i)_{i \in I} B \, , \hbox{ where }
B=(b_{ij}) \hbox{ is a symmetric matrix.}  
 \end{equation}

One then has:  $(\alpha_i |
\alpha_j) = b_{ij}$.

\item %%b
  Let $\Pi^{\vee \prime} = \{ \alpha^{\vee \prime}_i \} =r_s (\Pi^{\vee})$ and 
  $\Pi = \{ \alpha'_i \} =r_s (\Pi)$ where $r_s$ is an odd reflection, and
  suppose that (\ref{eq:1.6}) holds.  Then
  \begin{displaymath}
    \alpha^{\vee \prime}_i = \nu_i \alpha'_i \, .
  \end{displaymath}

\item %%c
  Provided that (\ref{eq:1.7}) holds and $a_{ii}=2$ or $0$ for
  all $i \in I$, one has for any non-isotropic root $\alpha$
  which is obtained from a simple root by a sequence of odd
  reflections:  $\alpha^{\vee} =2 \alpha/(\alpha | \alpha)$.

\end{enumerate}
\end{proposition}

\begin{proof}
  (a) is proved in \cite{K3}, (b) and (c) are easily checked.
\end{proof}

\begin{remark}
  \label{rem:1.1}
A natural question is which of the Lie superalgebras $\fg (\D)$
are of ``Kac-Moody'' type?  The most natural answer, in our
opinion, is that they should satisfy the following conditions:

\romanparenlist
\begin{enumerate}
\item %%i
  $\fg (\D)_{\bar{0}}$ is a (generalized) Kac-Moody algebra,

\item %%ii
  the $\fg (\D)_{\bar{0}}$-module  $\fg (\D)_{\bar{1}}$ is integrable.
\end{enumerate}
This definition covers the basic classical finite-dimensional Lie 
superalgebras and the associated affine superalgebras (including
the twisted ones).  Unfortunately, a well developed theory of
generalized Kac-Moody superalgebras (see \cite{B}, \cite{R} and
references there) does not cover most of the latter superalgebras 
(because of the crucial assumption on the Cartan matrix that its
off diagonal entries are non-positive).

\end{remark}

\section{Classification of integrable irreducible highest weight
  modules over $g \ell (m|n)^{\widehat{}}$}
\label{sec:2}

Consider the Lie superalgebra $g \ell (m|n)$, where $m,n \geq 1$ 
(see \cite{K1}).  Let $e_{ij}$ $(1 \leq i , j \leq m+n)$ denote
its standard basis.  Denote by $\fh$ the Cartan subalgebra of $g
\ell (m|n)$ consisting of all diagonal matrices.  Let
$\epsilon_i$ $(1 \leq i \leq m+n)$ be the basis of $\fh^*$ dual to
the basis $u_i:=e_{ii}$ of $\fh$.  Then $g \ell (m|n)=\fg (\D)$
for the following data $\D = \{ \fh , I,I_1, \Pi^{\vee}, \Pi
\}$ (cf.~\cite{K1}).  We let $I= \{ 1,2, \ldots ,m+n-1 \}$, $I_1 
= \{ m \}$; $\alpha^{\vee}_i =u_i - u_{i+1}$ for $i \in I
\backslash I_1$, $\alpha^{\vee}_m=u_m + u_{m+1}$, $\alpha_i =
\epsilon_i - \epsilon_{i+1}$ for all $i \in I$.  Its Cartan
matrix is the following $(m+n-1) \times (m+n-1)$ matrix:
\begin{eqnarray*}
  A= \left(
    \begin{array}{rrrrrrrrr}
      2   &  -1   &  0 \\
     -1   &   2   & -1 \\
     &&  \ddots \\
     & & -1 & 2 & -1\\
     && & -1 & 0 & 1 \\
     &&& & -1 & 2 & -1 \\
     &&&&&  \ddots \\
     &&&&&&  -1 & 2
    \end{array} \right) \raisebox{-1.5ex}{m\hbox{-th row}.}
\end{eqnarray*}
%
%\begin{eqnarray*}
%  A= \left(
%    \begin{array}{rrrrrrrrrr}
%      2   &  -1   &  0 \\
%     -1   &   2   & -1 \\
%     &&  \ddots \\
%     & -1 & 2 & -1\\
%     && -1 & 0 & 1 \\
%     &&& -1 & 2 & -1 \\
%     &&&&& \ddots \\
%     &&&&& -1 & 2
%    \end{array} \right) m\hbox{-th row}.
%\end{eqnarray*}
%
The Chevalley generators are as follows:
\begin{displaymath}
  e_i = e_{i,i+1}, f_i = e_{i+1,i} \quad (i=1 , \ldots , m+n-1)
  \, .
\end{displaymath}
Note that $\alpha_m$ is the only odd simple root, and it is
isotropic.

Consider the supertrace form on $g \ell (m|n)$:
\begin{displaymath}
  (a|b) = \str ab \, .
\end{displaymath}
This is a non-degenerate invariant supersymmetric bilinear form on $g \ell 
(m|n)$ whose restriction to $\fh$ is non-degenerate and
symmetric.  Identifying $\fh$ and $\fh^*$ via this bilinear form, 
we have:
\begin{displaymath}
  \epsilon_i =u_i  \hbox{ for } i=1, \ldots ,m ;
  \epsilon_i =-u_i \hbox{ for } i=m+1 , \ldots , m+n \, .
\end{displaymath}
Hence we have:
\begin{equation}
  \label{eq:2.1}
  \alpha^{\vee}_i = \alpha_i \hbox{ for }  i=1, \ldots ,m , \quad
  \alpha^{\vee}_i = -\alpha_i \hbox{ for } i=m+1,\ldots , m+n-1 \, ,
\end{equation}
and we may use Proposition~\ref{prop:1.4}.  In particular,
\begin{displaymath}
  ((\alpha_i | \alpha_i))_{i,j \in I} = \diag (\underbrace{1,
    \ldots ,1}_m, -1 , \ldots , -1)A \, .
\end{displaymath}

Likewise, the affine superalgebra $g \ell (m|n)^{\widehat{}}$ is
isomorphic to $\fg (\widehat{\D})$, where the data $\widehat{\D}
= \{ \hat{\fh}, \hat{I}, \hat{I}_1, \hat{\Pi}^{\vee}, \hat{\Pi}
\}$ is an extension of the data $\D$ for $g \ell (m|n)$ defined
as follows (cf.~\cite{K3}).  The space $\hat{\fh}$ is defined by
(\ref{eq:0.4}), $\hat{I} = I \cup \{ 0 \}$, $\hat{I}_1 = \{ m,0
\}$, $\hat{\Pi}^{\vee} = \Pi^{\vee} \cup \{ \alpha^{\vee}_0 \}$,
$\hat{\Pi} = \Pi \cup \{ \alpha_0 \}$.  Here the $\alpha_i$ for
$i \in \Pi$ are extended from $\fh$ to $\hat{\fh}$ by letting
$\alpha_i (K) = \alpha_i (d)=0$, $ \alpha_0 = \delta - \theta ,
\alpha^{\vee}_0 =K- \theta^{\vee}$, where: $ \delta |_{\fh + \CC
  K} = 0 , \langle \delta ,d \rangle =1 , \theta = \epsilon_1 -
\epsilon_{m+n}$ is the highest root of $g \ell (m|n),
\theta^{\vee} =u_1 + u_{m+n}$.  We extend the bilinear form
$(.|.)$ from $g \ell (m|n)$ to $g \ell (m|n)^{\hat{}}$ by
(\ref{eq:0.3}).  Identifying $\hat{\fh}$ with $\hat{\fh}^*$ via
this symmetric bilinear form, we get:
\begin{equation}
  \label{eq:2.2}
  K= \delta , \quad \theta =\theta^{\vee} , 
      \quad \alpha_0 = \alpha^{\vee}_0 \, .
\end{equation}
We have the following expression of $\delta =K$ in terms of
simple roots and coroots:
\begin{equation}
  \label{eq:2.3}
  \delta =K=\sum^{m+n-1}_{i=0} 
  \alpha_i = \sum^m_{i=0} \alpha^{\vee}_i
  - \sum^{m+n-1}_{j=m+1} \alpha^{\vee}_j \, .
\end{equation}
The Cartan matrix for $\hat{\D}$ is
\begin{eqnarray*}
  \hat{A} = \left(
    \begin{array}{rrrcrrrrrrrrr}
0 & -1 & 0 & \cdots & 0 &1 \\
-1\\
0\\
\cdots &&& A\\
0\\
-1
    \end{array} \right) \, .
\end{eqnarray*}

As above, we have:
\begin{displaymath}
  ((\alpha_i | \alpha_j))_{i,j \in \hat{I}}
  = \diag (\underbrace{1, \ldots ,1}_{m+1}, -1 , \ldots ,
  -1)\hat{A} \, .
\end{displaymath}

The even part of $g \ell (m|n)$ is $g \ell (m) \oplus g \ell
(n)$, hence the even part of $g \ell (m|n)^{\hat{}}$ is the sum $
g \ell (m)^{\hat{}} + g \ell (n)^{\hat{}}$ with a common central
element $K$ and a common scaling element $d$.  Note that the
restriction of the supertrace form to $g \ell (m)$ (resp. $g \ell 
(n)$) is the normalized (resp. negative of the normalized)
invariant form, i.e.,~$(\alpha | \alpha)=2$ (resp. $(\alpha |
\alpha) =-2$) for any root $\alpha$.

The set of simple roots for $g \ell (m)^{\hat{}}$ (resp. $g \ell
(n)^{\hat{}}$) is empty if $m=1$ (resp. $n=1$), and for $m \geq
2$ (resp. $n \geq 2$) it is as follows:
\begin{displaymath}
\hat{\Pi}' =  \{ \alpha'_0 = \delta - \theta' ,
\alpha_1 , \ldots , \alpha_{m-1}\}
\end{displaymath}
(resp. $\hat{\Pi}'' = \{ \alpha''_0 = \delta - \theta'',
\alpha_{m+1}, \ldots , \alpha_{m+n-1}$), where $\displaystyle{\theta'=
\sum^{m-1}_{i=1} \alpha_i , 
\theta'' = \sum^{m+n-1}_{i=m+1} \alpha_i}$.  Assuming that $m \geq 
2$, we have: $(\theta' | \theta')=2$, hence $\theta' =
\theta^{'\vee}$, and we have:
\begin{equation}
  \label{eq:2.4}
  \alpha'_0 = \alpha^{'\vee}_0 = 
  \alpha_0 + \sum^{m+n-1}_{i=m} \alpha_i =
  \alpha^{\vee}_0 + \alpha^{\vee}_m
  - \sum^{m+n-1}_{i=m+n} \alpha^{\vee}_i \, .
\end{equation}

A $g \ell (m|n)^{\hat{}}$-module $L (\Lambda)$ is called
\emph{integrable}, if its restriction to $g \ell (m)^{\hat{}}$ is 
integrable and its restriction to $g \ell (m|n)$ is locally
finite.  In this section we shall classify all such modules.

As usual, define fundamental weights $ \Lambda_i \in
\hat{\fh}\quad (i=0,1, \ldots , m+n-1)$  by 
\begin{displaymath}
\langle \Lambda_i , \alpha^{\vee}_j \rangle =
\delta_{ij}, j=0 , \ldots , m+n-1 ,
\langle \Lambda_i ,d \rangle =0 \, ,
\end{displaymath}
and \emph{labels} of a weight $\Lambda$ by:
\begin{displaymath}
  k_i = \langle \Lambda , \alpha^{\vee}_i \rangle \, .
\end{displaymath}
The following necessary conditions of integrability of $L
(\Lambda)$ follow from Lemma~\ref{lem:1.5}a:
\begin{eqnarray}
  \label{eq:2.5}
  k_i \in \ZZ_+ \hbox{ for } i=1, \ldots , m-1, m+1, \ldots , m+n-1 
\, ,\\
\label{eq:2.6}
k':=k_0 + k_m- \sum^{m+n-1}_{i=m+1} k_i \in \ZZ_+ \, .
\end{eqnarray}
We assume in (\ref{eq:2.6}) that $m \geq 2$ and use
(\ref{eq:2.4}).

We call $k'$ the \emph{partial level} of $\Lambda$ since, using
(\ref{eq:2.3}), we see that the \emph{level} $k:= \langle \Lambda
,K \rangle$ is given by
\begin{equation}
  \label{eq:2.7}
  k= \sum^{m-1}_{i=1} k_i + k' \, .
\end{equation}
Hence, provided that $m \geq 2$, the level of an integrable $g
\ell (m|n)^{\hat{}}$-module is a non-negative integer.

\begin{lemma}
  \label{lem:2.1}
  Assume that $m \geq 2$.  Then conditions (\ref{eq:2.5}) and
  (\ref{eq:2.6}) along with the condition
  \begin{equation}
    \label{eq:2.8}
    k' \geq n
  \end{equation}
are sufficient for integrability of the $g \ell
(m|n)^{\hat{}}$-module $L (\Lambda)$.

\end{lemma}

\begin{proof}
  The lemma follows from Lemma~\ref{lem:1.5} applied to the
  simple roots $\alpha_i , i=1, \ldots ,m-1$, and
  Proposition~\ref{prop:1.2} applied to $\alpha^{\vee} =
  \alpha^{'\vee}_0$, since, due to (\ref{eq:2.4}) we have:
  \begin{equation}
    \label{eq:2.9}
    \langle \rho , \alpha^{'\vee}_0 \rangle =-n+1 \, .
  \end{equation}

\end{proof}

\begin{lemma}
\label{lem:2.2}
Let $L (\Lambda)$ be an integrable $g \ell (m|n)^{\hat{}}$-module
such that $k' <n$, and let $m \geq 2$.  Then the following
complementary condition holds:

\starlist
\begin{enumerate}
\item\hspace{2in} there exist $r,s \in \ZZ_+ $ such that 
\end{enumerate}

\romanparenlist
\begin{enumerate}
\item %%i
$k'=r+s $,

\item %%2
$k_0-k_{m+n-1}-k_{m+n-2}- \cdots - k_{m+n-r}-r=0$,

\item %%3
  $k_m-k_{m+1}-k_{m+2}- \cdots - k_{m+s}-s=0$.
\end{enumerate}

\end{lemma}

\begin{proof}
  Consider the following two sequences of roots of $g \ell
  (m|n)^{\hat{}}$:
  \begin{eqnarray*}
    \beta_0\hspace{-2ex} &=&\hspace{-2ex}  \alpha_0, \beta_1 \!= \alpha_0 + \alpha_{m+n-1},
       \beta_2 \!= \alpha_0 + \alpha_{m+n-1} +\alpha_{m+n-2},
       \ldots , \beta_{n-1}\!=\alpha_0 + \alpha_{m+n-1} +\ldots + 
       \alpha_{m+1};\\
  \beta'_0\hspace{-2ex} &=&\hspace{-2ex} \alpha_m , \beta'_1 = \alpha_m + \alpha_{m+1},
      \beta'_2 = \alpha_m + \alpha_{m+1} + \alpha_{m+2},
      \ldots , \beta'_{n-1}=\alpha_m + \cdots + \alpha_{m+n-1} \, .
  \end{eqnarray*}
It is clear by Proposition~\ref{prop:1.4} that $\beta^{\vee}_i =
\beta_i$ and $\beta^{'\vee} = \beta'_i$.  
Note that $\langle \Lambda + \rho , \beta^{\vee}_r \rangle$
(resp. $\langle \Lambda + \rho , \beta^{'\vee}_s \rangle$) is
equal to the left-hand side of (ii) (resp. (iii)).  Note that 
\begin{equation}
  \label{eq:2.10}
  \langle \beta_i , \alpha^{' \vee}_0 \rangle =1
  = \langle \beta'_i , \alpha^{'\vee}_0 \rangle ,
  i=0 , \ldots , n-1 \, .
\end{equation}
If  $\langle \Lambda + \rho , \beta^{\vee}_i \rangle \neq 0$ for all
$i$,
using (\ref{eq:2.6}) and (\ref{eq:2.10}) we would conclude, by
Proposition~\ref{prop:1.3}, that $k'-n \geq 0$, in contradiction
with the assumption of the lemma.  Hence (ii) holds for some
non-negative integer $r$ $(<n)$.  Similarly, (iii) holds for some 
non-negative integer $s$ $(<n)$.  Similarly, applying
Proposition~\ref{prop:1.3} to the union of sequences $\beta_i$
and $\beta'_i$ , we conclude that
\begin{equation}
  \label{eq:2.11}
  r+s \leq k' \, .
\end{equation}
Hence, adding up (ii) and (iii) we get
\begin{equation}
  \label{eq:2.12}
  k'+ \sum^{m+n-r-1}_{i=m+s+1} k_i =r+s \, .
\end{equation}
Now (i) follows from (\ref{eq:2.5}), (\ref{eq:2.11}) and (\ref{eq:2.12}).

\end{proof}

\begin{remark}
  \label{rem:2.1}
Condition~(*) on $\Lambda$ is equivalent to the following condition: 
 there exists a non-negative integer $s \leq k'$ $(\leq n-1)$ such
 that:
 \begin{displaymath}
   k_m = k_{m+1} + \cdots + k_{m+s} +s \hbox{ and }
   k_{m+s+1} = \cdots = k_{m+s+n-k'-1} =0 \, .
 \end{displaymath}
This condition implies that $\Lambda$ lies in a union of $k'+1$
hyperplanes of dimension $k'+m-1$.  Equivalently, there exists a
non-negative integer $r \leq k'$ $(\leq n-1)$ such that
\begin{eqnarray*}
  k_0 = k_{m+n-1} + k_{m+n-2}+ \cdots + k_{m+n-r}+r \hbox{ and }
  k_{m+n-r-1} = \cdots = k_{m+k'-r+1}=0 \, .
\end{eqnarray*}

\end{remark}

\begin{theorem}
  \label{th:2.1}
\alphaparenlist
\begin{enumerate}
\item %%a
  A $g \ell (1|n)^{\hat{}}$-module $L (\Lambda)$ is integrable
  iff $k_2, \ldots , k_n \in \ZZ_+$.

\item %%b
  Provided that $m \geq 2$, a $g \ell (m|n)^{\hat{}}$-module
  $L(\Lambda)$ is integrable iff conditions (\ref{eq:2.5}), (\ref{eq:2.6}) hold and, in
  the case $k'<n$, the complementary condition (*) holds.
\end{enumerate}
\end{theorem}

\begin{proof}
  In the case $m=1$, the only condition of integrability is local 
  finiteness of $g \ell (1,n)$ on $L (\Lambda)$ which is equivalent
  to $k_2, \ldots , k_n \in \ZZ_+$ due to Lemma~\ref{lem:1.5}b.
    It follows from Lemma~\ref{lem:2.2} that in the case $m \geq 2$,
    the conditions listed by Theorem~\ref{th:2.1}b are
    necessary.  In view of Lemma~\ref{lem:1.5}b, it remains to
    show that these conditions are sufficient for local
    nilpotency of $e_{-\alpha'_0}$.  Due to Lemma~\ref{lem:2.1}, we 
    may assume that
    \begin{equation}
      \label{eq:2.13}
      k' \leq n-1 \, .
    \end{equation}
Consider the sequence of odd roots
$\beta_0 , \ldots , \beta_{n-1}$ introduced in the proof of
Lemma~\ref{lem:2.2} and let $\Pi^{(0)} = \Pi$, $\Pi^{(1)}=
r_{\beta_0} (\Pi^{(0)}), \ldots , \Pi^{(n)} =r_{\beta_{n-1}}
(\Pi^{(n-1)})$, and notice that
\begin{displaymath}
  \alpha'_0 \in \Pi^{(n)} \, .
\end{displaymath}
Let $\Lambda^{(n)}$ be the highest weight vector of $L
(\Lambda)$ with respect to $\fn^{(n)}_+ = r_{\beta_{n-1}} \ldots
r_{\beta_0} (\fn)$.  Due to Lemma~\ref{lem:1.5}b, it remains to 
show that conditions listed by Theorem~\ref{th:2.1}b imply that
\begin{equation}
  \label{eq:2.14}
  \langle \Lambda^{(n)}, \alpha^{'\vee}_0 \rangle \in \ZZ_+ \, .
\end{equation}

Recall that by (\ref{eq:1.4}) we have:
\begin{equation}
  \label{eq:2.15}
  \Lambda^{(n)} + \rho^{(n)}= \Lambda + \rho + \sum_{i \in S}
  \beta_i \, ,
\end{equation}
where $S= \{ i \in [0,\ldots ,n-1]|\langle \Lambda^{(i)},
\beta^{\vee}_i \rangle =0 \}$.  Let $t_i =\langle \Lambda^{(i)},
\beta^{\vee}_i \rangle $ for short.  Then condition (*) gives
for some $r \in \ZZ_+$, $r <n$, that $t_r=0$.  In view of
Remark~\ref{rem:2.1}, we have:
\begin{equation}
  \label{eq:2.16}
  t_r=t_{r+1}= \cdots = t_{n-s-1}=0 \, .
\end{equation}
Hence, due to (\ref{eq:2.15}), (\ref{eq:2.9}), (\ref{eq:2.10}) and (\ref{eq:2.16}) we get:
\begin{displaymath}
  \langle \Lambda^{(n)} + \rho^{(n)} , \alpha^{'\vee}_0 \rangle
  = k'+(1-n)+ |S| \geq k' + (1-n) + (n-s-r) \geq 1 \, ,
\end{displaymath}
proving (\ref{eq:2.14}), since $\langle \rho^{(n)}, \alpha^{'\vee}_0 \rangle =1$.
\end{proof}

\begin{remark}
  \label{rem:2.2}
It follows from Theorem~\ref{th:2.1} that when $m \geq 2$, the
only integrable $g \ell (m|n)^{\hat{}}$-modules $L(\Lambda)$ of
level $k=0$ are those for which all labels are $0$, in which case $\dim L 
(\Lambda) =1$.
\end{remark}

\begin{remark}
\label{rem:2.3}
  If $m \geq 2$ and $n \geq 2$, then the only $L(\Lambda)$ which
  are integrable with respect to the 
whole even subalgebra are $1$-dimensional.  (It is because the $g 
\ell (m)^{\hat{}}$-integrability implies $k \geq 0$ and $g \ell
(n)^{\hat{}}$-integrability implies $k \leq 0$.)
\end{remark}

\begin{remark}
\label{rem:2.4}
%It follows from Theorem~\ref{th:2.1} that when $m \geq 2$, the
%complete list of highest weights of integrable $g \ell
%(m|n)^{\hat{}}$-modules of level $k=1$ is as follows (up to
%adding a multiple of $\delta$):
Define $\epsilon \in \hat{\fh}^*$ by letting $\epsilon |_{\fh}=$
supertrace, $\epsilon (K) = \epsilon (d) =0$.  It follows from Theorem~\ref{th:2.1} that when $m \geq 2$, the
complete list of highest weights of integrable $g \ell
(m|n)^{\hat{}}$-modules of level $1$, up to adding an arbitrary
linear combination of $\epsilon$ and $\delta$, is as follows:
\begin{eqnarray*}
  \Lambda_s (1 \leq s \leq m-1), \quad(a+1) \Lambda_m 
    + a \Lambda_{m+1} (a \in \ZZ_+) \, \quad
    (a+1) \Lambda_0 + a \Lambda_{m+n-1} (a \in \ZZ_+) \, .
\end{eqnarray*}

\end{remark}

\begin{remark}
\label{rem:2.5}
Consider the sequence of the sets of simple roots $\Pi^{(0)} =
\Pi , \ldots , \Pi^{(n)} =$\break $\{ \alpha'_0 , \ldots ,
\alpha'_{m+n-1} \}$, introduced in the proof of
Theorem~\ref{th:2.1}.  One has:
\begin{eqnarray*}
  \alpha'_0 &=& \alpha_0 + \alpha_1 , \alpha'_1 = \alpha_2, 
     \ldots , \alpha'_{m-2} = \alpha_{m-1}, \alpha'_{m-1}
     = \alpha_m + \alpha_{m+1} + \cdots + \alpha_{m+n-1}
     + \alpha_0 \, ,\\
   \alpha'_m &=& - (\alpha_{m+n} + \cdots + \alpha_{m+n-1}
      + \alpha_0), \alpha'_j = \alpha_j \hbox{ for }
      m+1 \leq j \leq m+n-1 \, .
\end{eqnarray*}
  Let $\Lambda'_j$ be the fundamental weights with respect to
  $\Pi^{(n)}$.  Given a weight $\Lambda$, denote by
  $\Lambda^{(n)}$ the highest weight of $L (\Lambda)$ with
  respect to $\Pi^{(n)}$ (or rather $\fn^{(n)}_+$).  Using
  Lemma~\ref{lem:1.4}, it is easy to see that the weights listed
  in Remark~\ref{rem:2.4} get changed under the map $\Lambda
  \mapsto \Lambda^{(n)}$ as follows:
  \begin{eqnarray*}
    \Lambda^{(n)}_j &=& \Lambda_j \quad (1 \leq j \leq m),
    ((a+1) \Lambda_0 + a \Lambda_{m+n-1})^{(n)}
    = (a+1) \Lambda_0 + a \Lambda_{m+n-1}- \alpha_0 \, , \\
    && ((a+1) \Lambda_m + a \Lambda_{m+1})^{(n)}
    = (a+1) \Lambda_m + a \Lambda_{m+1}+ \alpha'_m
    (a>0) \, .
  \end{eqnarray*}
In terms of the fundamental weights $\Lambda'_j$ the map $\Lambda 
\to \Lambda^{(n)}$ looks as follows:
\begin{eqnarray*}
  && \Lambda_j \mapsto \Lambda'_{j-1} \quad (1 \leq j \leq m) \,
  ,\\
  && (a+1) \Lambda_0 + \Lambda_{m+n-1} \mapsto (a+2)
  \Lambda'_0 + (a+1) \Lambda'_{m+n-1} (a \in \ZZ_+) \, ,\\
  && (a+1) \Lambda_m + a \Lambda_{m+1} \mapsto a \Lambda'_m
   + (a-1) \Lambda'_{m+1} \quad (a \in \NN) \, .
\end{eqnarray*}
It follows that all weights of level~1 listed by
Remark~\ref{rem:2.4} are conjugate to each other by odd
reflections.  
\end{remark}

\section{Free field realization of level $1$ integrable modules
  over $g \ell (m|n)^{\widehat{}}$.}
\label{sec:3}

Fix non-negative integers $m$ and $n$ such that $m+n \geq 1$ and
denote by $F$ the vertex algebra generated by $m$ pairs of odd
fields $\psi^i (z), \,  \psi^{i*} (z)$, $(i=1, \ldots ,m)$ and $n$
pairs of even fields $\varphi^j (z), \varphi^{j*}(z)$ $(j=1,
\ldots ,n)$, all pairwise local, subject to the following
operator product expansions (as usual, we list only the
non-trivial OPE):
\begin{eqnarray*}
\begin{array}{ll}
  \psi^i (z) \psi^{j*} (w) \sim \frac{\delta_{ij}}{z-w} , &  
   \psi^{i*} (z) \psi^j (w) 
  \sim \frac{\delta_{ij}}{z-w} \, , \\[1ex]
  \varphi^i (z) \varphi^{j*}(w) \sim
  -\frac{\delta_{ij}}{z-w} , & 
  \varphi^{i*}(z) \varphi^j(w) \sim \frac{\delta_{ij}}{z-w} \, .
\end{array}
\end{eqnarray*}
This is called a free fermionic vertex algebra in the book
\cite{K4} to which we refer for foundations of the vertex algebra
theory.

This vertex algebra has a family of Virasoro fields \cite{K4},
from which it is convenient to choose the following one:
\begin{eqnarray}
\label{eq:3.1}
  L(z) \equiv \sum_{k \in \ZZ} L_k z^{-k-2} &=&\tfrac{1}{2}
  \sum^m_{i=1} (: \partial \psi^i (z) \psi^{i*} (z) :
  + : \partial \psi^{i*}(z) \psi^i(z) :)\\[1ex]
\nonumber
  &+& \tfrac{1}{2}
  \sum^n_{j=1} (: \partial \varphi^j (z) \varphi^{j*} (z):
  - : \partial \varphi^{j*}(z) \varphi^j(z):) \, .
\end{eqnarray}
With respect to $L(z)$ the fields $\psi^i (z)$, $\psi^{i*}(z)$,
$\varphi^j(z)$ and $\varphi^{j*}(z)$ are primary of conformal
weight $1/2$.  We therefore write all these fields in the form $
x^i(z) =\sum_{k \in \frac{1}{2}+\ZZ} x^{(i)}_k z^{-k-1/2}$ where
$x=\psi$ , $\psi^*$, $\varphi$ or $\varphi^*$, and we have the
following conditions on the vacuum $\vac$:
\begin{displaymath}
  \psi^{(i)}_k \vac =0 , \quad \psi^{(i)*}_k \vac =0 , \quad
  \varphi^{(i)}_k \vac =0 , \quad \varphi^{(i)*}_k \vac =0
  \hbox{ for }k>0 \, .
\end{displaymath}
The operator $L_0$ is called the \emph{energy operator} or
(Hamiltonian) and its eigenvalues are called the \emph{energies}
of the corresponding eigenvectors.  The energy can be calculated
from the following relations:
\begin{equation}
  \label{eq:3.2}
  \energy \vac =0 , \quad \energy (\psi^{(i)}_k, \psi^{(i)*}_k,
  \varphi^{(j)}_k, \varphi^{(j)*}_k)=-k \, .
\end{equation}
The second relation means that $\psi^{(i)}_k$, etc., changes the
energy by $-k$, i.e.,
\begin{displaymath}
\energy \,\, (\psi^{(i)}_k v)= \energy (v)-k \, , \hbox{ etc.}
\end{displaymath}

Next, for each pair $i,j$ that may occur 
introduce the following fields of conformal weight $1$:
\begin{eqnarray*}
\begin{array}{ll}
  a^{ij+}(z) = : \psi^i(z) \psi^{j*}(z) :,& 
   a^{ij-}(z) = : \varphi^i(z) \varphi^{j*}(z) : \\
   E^{ij+}(z) = : \psi^i(z) \varphi^{j*}(z) :,&
   E^{ij-}(z) = : \varphi^i(z) \psi^{j*}(z) : \, .
\end{array}
\end{eqnarray*}

\begin{proposition}
  \label{prop:3.1}

\alphaparenlist

  \begin{enumerate}
  \item %%(a)
Consider the affine superalgebra $g \ell (m|n)^{\widehat{}}$ and let
$A(z)=$\break $\sum_{k \in \ZZ} (t^k \otimes A) z^{-k-1}$ for $A \in g
\ell (m|n)$.  Then the linear map $\sigma$ given by 
\begin{eqnarray*}
  \begin{array}{ll}
 e_{ij}(z) \mapsto a^{ij+}(z) \, ,
    & e_{i+m,j+m} (z) \mapsto a^{ij-}(z) \, , \\
 e_{i,j+m}(z) \mapsto E^{ij+}(z) \, , 
    &  e_{i+m,j}(z) \mapsto E^{ij-}(z), \quad K \mapsto 1, 
       \quad  d \mapsto L_0
  \end{array}
\end{eqnarray*}
defines a representation of $g \ell (m|n)^{\widehat{}}$ (of level~$1$) in 
the space $F$.

\item %%(b)
  Consider the standard $g \ell (m|n)$-module $\CC^{m|n}$ and its 
  contragredient module $\CC^{m|n*}$.  Consider the
  corresponding $\CC [t,t^{-1}] \otimes g \ell (m|n)$-modules
  $\CC [t,t^{-1}] \otimes \CC^{m|n}$ and $\CC [t,t^{-1}] \otimes
  \CC^{m|n*}$, and let $v(z) = \sum_{k \in \ZZ} (t^k \otimes
  v)z^{-k-1}$ for $v \in \CC^{m|n}$ or $\CC^{m|n*}$.  Then the
  linear maps $\nu$ and $\nu^*$ given by $(i=1, \ldots ,m ; j=1, \ldots ,n)$:
\begin{eqnarray*}
  \begin{array}{ll}
  v_i(z) \mapsto \psi^i (z) , 
      & v_{j+m}(z) \mapsto \varphi^j(z) \hbox{ and}\\
  v^*_i (z) \mapsto \psi^{i*}(z) , 
      & v^*_{j+m}(z) \mapsto  \varphi^{j*}(z)
  \end{array}
\end{eqnarray*}
are equivariant, i.e.,~they have the following property:
\begin{eqnarray*}
  \nu (A(z)v(w)) &=& [\sigma (A(z)), \nu (v(w))] , \quad
  v \in \CC^{m|n} \, , \\
  \nu^* (A(z)v^*(w)) &=& [\sigma (A(z)), \nu^* (v^* (w))],
  \quad v^* \in \CC^{m|n} \, .
\end{eqnarray*}
  \end{enumerate}

\end{proposition}

\begin{proof}

Both statements follow from the corresponding OPE, which are
easily derived from Wick's formula.  Below we give the less
trivial OPE needed for the proof of~(a).
\begin{eqnarray*}
  E^{ij+} (z) E^{k \ell -}(w) & \sim &
  \frac{\delta_{jk}a^{i \ell +}(w) + \delta_{i \ell}a^{kj-}(w)}
  {z-w }\, + \, \frac{\delta_{i \ell}  \delta_{jk}}{(z-w)^2}\\[2ex]
  a^{ij \pm}(z) E^{k \ell \pm}(w) & \sim &
  \frac{\delta_{jk}E^{i\ell \pm}(w)}{z-w} \, , \\[2ex]
  a^{ij \pm}(z) E^{k\ell \mp}(w) & \sim &
  \frac{-\delta_{\ell i}E^{kj \mp}(w)}{z-w} \, , \\[2ex]
  a^{ij\pm}(z) a^{k \ell \pm}(w) & \sim &
  \frac{\delta_{jk}a^{i \ell \pm}(w)
       -\delta_{\ell i}a^{kj \pm}(w)}{z-w} \, \pm\, 
     \frac{1}{(z-w)^2}\\[2ex]
     a^{ij \pm}(z)a^{k \ell \mp}(w) &\sim & 0 \, .
\end{eqnarray*}

\end{proof}

Introduce the \emph{total charge operator}
\begin{displaymath}
  a_0 = \sigma (I), \hbox{ where }
  I=\sum^{m+n}_{i=1} e_{ii} \in g \ell (m|n) \, .
\end{displaymath}
Its eigenvalues are called \emph{charges} of the corresponding
eigenvectors.  It is clear from Proposition~\ref{prop:3.1} that
the total charge can be calculated from the following relations:
\begin{equation}
  \label{eq:3.3}
  \charge \vac =0 , \quad 
  \charge (\psi^{(i)}_k,\varphi^{(j)}_k)=1, \quad
  \charge (\psi^{(i)*}_k, \varphi^{(j)*}_k) =-1 \, .
\end{equation}
Consider the charge decomposition of $F$, i.e.,~its decomposition 
in eigenspaces of $a_0$:
\begin{equation}
  \label{eq:3.4}
  F= \oplus_{s \in \ZZ} F_s \, .
\end{equation}
Since $a_0$ commutes with $\sigma (g \ell (m|n)^{\widehat{}})$, we
conclude that (\ref{eq:3.4}) is a decomposition in a direct sum
of $g\ell (m|n)^{\widehat{}}$-modules.

It is clear that $L_0$ commutes with $a_0$, hence each $F_s$ is $
L_0$-invariant, and since all eigenvalues of $L_0$ in $F$ lie in
$\tfrac{1}{2}\ZZ_+$, the same holds for eigenvalues of $L_0$ in $
F_s$, $s \in \ZZ$.  Note also that $L_0$ commutes with $\sigma (g 
\ell (m|n))$.  It is because all fields $a^{ij \pm}(z)$ and
$E^{ij \pm}(z)$ have conformal weight~$1$.  It follows that each
eigenspace of $L_0$ in $F_s$ is a $g \ell (m|n)$-module.  The
following proposition describes the lowest energy subspace $F^{\low}_s$
and the lowest weight vector $\sac$ in each $F_s$.

\begin{proposition}
  \label{prop:3.2}

\alphaparenlist

  \begin{enumerate}
  \item %%(a)
    Let $s \in \ZZ_+$.  Then, as a $g\ell (m|n)$-module,
    $F^{\low}_s$ is isomorphic to $\Lambda^s \CC^{m|n}$.
    Furthermore, any highest weight vector of $g\ell (m|n)^{\widehat{}}$ 
    in $F_s$ lies in $F^{\low}_s$ and is proportional to the
    vector
    \begin{displaymath}
      \sac =\psi^{(1)}_{-\frac{1}{2}} \ldots
            \psi^{(s)}_{-\frac{1}{2}} \vac \hbox{ with weight }
            \Lambda_0 + \epsilon_1 + \cdots + \epsilon_s - \tfrac{s}{2}
            \delta \, ,
    \end{displaymath}
provided that $s \leq m$, and to the vector
    \begin{displaymath}
      \sac = \left(
        \varphi^{(1)}_{-\frac{1}{2}} \right)^{s-m} \mac \hbox{ with
        weight } \Lambda_0 + \epsilon_1 + \cdots + \epsilon_m
      +  (s-m) \epsilon_{m+1} -\tfrac{s}{2} \delta
   \end{displaymath}
 provided that $s \geq m$.

  \item %%(b)
    Let $-s \in \ZZ_+$.  Then, as a $g \ell (m|n)$-module,
    $F^{\low}_s$ is isomorphic to $\Lambda^{-s}(\CC^{m|n})^*$.
    Furthermore, any highest weight vector of $g \ell
    (m|n)^{\widehat{}}$ in $F_s$  lies in $F^{\low}_s$ and is proportional to
    the vector
    \begin{displaymath}
      \sac =
    \left( \varphi^{(n)*}_{-\frac{1}{2}}\right)^{-s} \vac 
    \hbox{ with weight }
    \Lambda_0 +s \epsilon_{m+n}+\tfrac{s}{2} \delta \, .
   \end{displaymath}

  \end{enumerate}

\end{proposition}

\begin{proof}
  It is clear that, if $s$ (resp. $-s$) $\in \ZZ_+$, then
  $F^{\low}_s$ consists of homogeneous polynomials of degree
  $|s|$ in anticommuting operators $\psi^{(i)}_{-\frac{1}{2}}$
  (resp. $\psi^{(i)*}_{-\frac{1}{2}}$) and commuting operators
  $\varphi^{(j)}_{-\frac{1}{2}}$
  (resp. $\varphi^{(j)*}_{-\frac{1}{2}}$), applied to $\vac$.
  This proves (a) (resp. (b)), due to Proposition~\ref{prop:3.1}.

\end{proof}

\begin{remark}
  \label{rem:3.1}
The lowest energy in $F_s$ is $\tfrac{1}{2}|s|$ and the spectrum
of $L_0$ in $F_s$ is $\tfrac{1}{2}|s| + \ZZ_+$.

\end{remark}

\begin{remark}
  \label{rem:3.2}  Denote by $\Lambda_{(s)}$ the weight of $\sac$.
When restricted to $s \ell (m|n)^{\widehat{}}$, $\Lambda_{(s)}$ is
given by the following formulas:
\begin{eqnarray*}
\begin{array}{ll}
  \Lambda_s - \tfrac{s}{2} \delta& \hbox{ if }0 \leq s \leq m \, , 
  \\[1ex]
  (1+s-m) \Lambda_m + (s-m) \Lambda_{m+1} - \tfrac{s}{2} 
  \delta & \hbox{ if } s \geq m \, , \\[1ex]
  (1-s )\Lambda_0 -s \Lambda_{m+n-1}+\tfrac{s}{2} \delta
  &   \hbox{ if } s \leq 0 \, .
\end{array}
\end{eqnarray*}
We identify here $\Lambda_{m+1}$ with $-\Lambda_{0}$ in the case $n=1$.
\end{remark}

The following theorem is the central result of this section.

\begin{theorem}
 \label{th:3.1}
Suppose that $m \geq 1$.  Then each $g \ell
(m|n)^{\widehat{}}$-module $F_s$, $s \in \ZZ$, is an irreducible integrable
highest weight module of level~$1$.

\end{theorem}

\begin{remark}
  \label{rem:3.3}
The $g \ell (0|n)^{\widehat{}}$-modules $F_s$ are not
irreducible.  For example, one can show that in the case
$(m,n)=(0,2)$, one has the following decomposition as $g\ell
(2)^{\hat{}}$-modules (in the standard notation of \cite{K3}):
\begin{displaymath}
ch F_s=\sum^{\infty}_{j=0} ch L
(-(1+2j+|s|)\Lambda_0 + (2j+|s|) \Lambda_1)
q^{j^2+(|s|+1)j+|s|/2} \, .
\end{displaymath}

\end{remark}

\noindent 
E. Frenkel informed one of us that he had found this
decomposition too.

The proof of Theorem~\ref{th:3.1} is based on the (super)
boson-fermion correspondence, which we shall now recall
(cf.~\cite{K4}).

For each $i=1, \ldots ,m$ there exists a unique invertible odd
operator $e^{\epsilon_i}$ with inverse $e^{-\epsilon_i}$
satisfying the following three properties:
\begin{eqnarray}
  \label{eq:3.5}
  [e^{\epsilon_i}, \psi^j(z)] =0 \hbox{ if } i \neq j , \qquad
    [e^{\epsilon_i}, \varphi^j (z)]=0 \hbox{ for all }j \, ,\\[1ex]
    \label{eq:3.6}
  e^{\epsilon_i}\psi^{(i)}_k e^{-\epsilon_i}
    = \psi^{(i)}_{k-1},  \qquad 
    e^{\epsilon_i} \psi^{(i)*}_k
    e^{-\epsilon_i} = \psi^{(i)*}_{k+1} \, , \\[1ex]
    \label{eq:3.7}
  e^{\epsilon_i} \vac = \psi^{(i)}_{-\frac{1}{2}} \vac , \qquad
    e^{-\epsilon_i} \vac = \psi^{(i)*}_{-\frac{1}{2}} \vac \, .
\end{eqnarray}
It is easy to see that $e^{\epsilon_i} e^{\epsilon_j} =
-e^{\epsilon_j}  e^{\epsilon_i}$ if $i \neq j$.
 
We let for short $(i=1, \ldots ,m; j=1, \ldots ,n)$:
\begin{displaymath}
  \epsilon^i (z) =a^{ii+}(z) 
  = \sum_{k \in \ZZ} \epsilon^{(i)}_k z^{-k-1}, \qquad
  \epsilon^{j+m}(z) =a^{jj-}(z)
  = \sum_{k \in \ZZ} \epsilon^{(j+m)}_k z^{-k-1} \, .
\end{displaymath}
Then we have:
\begin{equation}
  \label{eq:3.8}
  [\epsilon^{(i)}_k, e^{\epsilon_j}] = \delta_{ij}\delta_{k0}
  e^{\epsilon_j}, \quad i=1,\ldots,m+n; \quad
  j=1,\ldots ,m \, .
\end{equation}

For each $i=1,\ldots,m+n$ introduce the following fields:
\begin{displaymath}
  \Gamma^+_{\epsilon_i} (z) = e^{\sum^{\infty}_{k=1}
  \frac{z^k}{k} \epsilon^{(i)}_{-k}} , \quad
  \Gamma^-_{\epsilon_i}(z) = e^{- \sum^{\infty}_{k=1}
  \frac{z^{-k}}{k} \epsilon^{(i)}_{k}} \, ,
\end{displaymath}
and for a linear combination with integer coefficients $\alpha =
\sum^m_{i=1} s_i \epsilon_i$ we let $\Gamma^{\pm}_{\alpha} (z) =
\Pi_i(\Gamma^{\pm}_{\epsilon_i})^{s_i}$ (recall that all
$\epsilon^{(i)}_k$ commute and all
$\epsilon^{(i)}_{-k}$ commute for $k \geq 1$, see
Proposition~\ref{prop:3.1}a).  The central fact of the classical
boson-fermion correspondence is the following formula, see e.g. \cite{K4}
$(i=1, \ldots ,m)$
\begin{equation}
  \label{eq:3.9}
  \psi^i(z) =e^{\epsilon_i}z^{\epsilon^{(i)}_0}
  \Gamma^+_{\epsilon_i}(z) \Gamma^-_{\epsilon_i}(z) , \quad
  \psi^{i*} (z)=e^{-\epsilon_i} z^{-\epsilon^{(i)}_0}
  \Gamma^+_{-\epsilon_i}(z) \Gamma^-_{-\epsilon_i}(z) \, .
\end{equation}
The key formulas of the super boson-fermion correspondence are
the following \cite{KL}, \cite{K4}\break($j=1,\ldots ,n$):
\begin{eqnarray}
  \label{eq:3.10}
  \varphi^j(z) &=& z^{\epsilon^{(i)}_0} e^{\epsilon_i}
    \Gamma^+_{\epsilon_i}(z) E^{ji-}(z)
    \Gamma^-_{\epsilon_i}(z) \\
\nonumber
   \varphi^{j*}(z) &=& z^{-\epsilon^{(i)}_0}e^{-\epsilon_i}
     \Gamma^+_{-\epsilon_i}(z) E^{ij+}(z)
     \Gamma^-_{-\epsilon_i}(z) \, ,
\end{eqnarray}
for each $i=1, \ldots, m$ (we assume here that $m \geq 1$).

\begin{proof}[Proof of Theorem~\ref{th:3.1}]
  Since the eigenspaces of $L_0$ in $F_s$ are finite-dimensional
  and $L_0$ commutes with $g\ell (m|n)$, it follows that $F_s$ is 
  a direct sum of finite-dimensional $g\ell (m|n)$-modules, hence 
  $g \ell (m|n)$ acts locally finitely on $F_s$.  Furthermore, we 
  have:
  \begin{displaymath}
    F=F^{\fermi} \otimes F^{\bose} \, ,
  \end{displaymath}
where $F^{\fermi}$ (resp. $F^{\bose}$) is the vertex algebra
generated by the $\psi^i (z), \psi^{i*} (z)$
(resp. $\varphi^j(z),$\break $\varphi^{j*} (z) $), and the
subalgebra $g \ell (m)^{\widehat{}}$ of $g\ell(m|n)^{\widehat{}}$ acts on $F$
via $\pi \otimes 1$, where the representation $\pi$ of $g
\ell (m)^{\widehat{}}$ on $F^{\fermi}$ is known to be integrable of level~$1$
(see \cite{KP1}).  Thus, the representation of $g \ell
(m|n)^{\widehat{}}$ in each $F_s $ is integrable.

The irreducibility of $F_s$, provided that $m \geq 1$, is proved
using (\ref{eq:3.9}) and (\ref{eq:3.10}) in exactly the same
fashion as the proof of Theorem~5.8a from \cite{K4}.

\end{proof}

\begin{remark}
  \label{rem:3.4}
We have got along the way the following vertex operator
construction of $g\ell (m|n)^{\widehat{}}$.  For each $\alpha =
\sum^m_{i=1} s_i \epsilon_i, \, s_i \in \ZZ$, introduce the usual 
vertex operator
\begin{displaymath}
  \Gamma_{\alpha} = e^{\alpha}z^{\alpha_0}
  \Gamma^+_{\alpha} \Gamma^-_{\alpha} \, .
\end{displaymath}
Then the following map defines an irreducible integrable highest
weight module of level~$1$ in each $F_s$:
\begin{eqnarray*}
\begin{array}{ll}
  e_{ii} (z)  \mapsto  \epsilon_i (z) 
     &  (i=1,\ldots ,m), \quad K \mapsto 1 \, , \\
  e_{ij} (z)  \mapsto  \Gamma_{\epsilon_i-\epsilon_j}
     & (i,j=1, \ldots ,m) \, , \\
  e_{i+m,j+m}(z)  \mapsto : \varphi^i (z) \varphi^{j*}(z):
     & (i,j=1, \ldots ,n) \, ,\\
  e_{j+m,i} (z)  \mapsto \Gamma_{-\epsilon_i}(z) \varphi^j (z)
     & (i=1,\ldots ,m; j=1, \ldots ,n) \, , \\
  e_{i,j+m}(z)  \mapsto  \Gamma_{\epsilon_i} (z) \varphi^{j*}(z) 
     & (i=1,\ldots ,m; j=1 ,\ldots ,n)\, .
\end{array}
\end{eqnarray*}

\end{remark}

Next, we give a standard derivation of a 
``quasiparticle'' character formula for the $g\ell
(m|n)^{\widehat{}}$-modules $F_s$, $s \in \ZZ$.

Given $a=(a_1,\ldots ,a_m)$, $b=(b_1,\ldots ,b_m) \in \ZZ^m_+$
and $c=(c_1 ,\ldots , c_n)$, $d=(d_1 , \ldots , d_n) \in \ZZ^n_+$, 
denote by $F(a,b,c,d)$ the linear span of vectors in $F$
 obtained from the vacuum vector $\vac$ by applying all
monomials in the $\psi^{(i)}_k$, $\psi^{(i)*}_k$,
$\varphi^{(i)}_k$, $\varphi^{(i)*}_k$ which contain $a_1$ factors
of the form $\psi^{(1)}_k$, $k \in \tfrac{1}{2} +\ZZ, \ldots ,
a_m$  factors of the form $\psi^{(m)}_k, b_1$ factors of the form 
$ \psi^{(1)*}_k, \ldots , b_m$ 
factors of the form $\psi^{(m)*}_k$, $c_1$  factors of the
form $\varphi^{(1)}_k, \ldots , d_n$  factors of the form
$\varphi^{(n)*}_k$.  These states lie in $F_s$ iff the following
condition holds:
\begin{equation}
  \label{eq:3.11}
  |a| - |b| + |c| - |d| =s \, ,
\end{equation}
where $|a|=\sum a_i$, etc.

It is clear that the state of minimal energy in $F(a,b,c,d)$ is
(up to a constant factor) the following vector:
\begin{eqnarray*}
  v(a,b,c,d) &=& (\psi^{(1)}_{-(a_1-\frac{1}{2})} \ldots 
       \psi^{(1)}_{-\frac{3}{2}} \psi^{(1)}_{-\frac{1}{2}})   \ldots
       (\psi^{(m)}_{-(a_m -\frac{1}{2})} \ldots 
       \psi^{(m)}_{-\frac{1}{2}})\\
  &&  \times (\psi^{(1)*}_{-(b_1-\frac{1}{2})} \ldots
       \psi^{(1)*}_{-\frac{1}{2}}) \ldots
       (\psi^{(m)*}_{ -(b_m-\frac{1}{2})} \ldots
       \psi^{(m)*}_{-\frac{1}{2}})\\
  &&
       \times (\varphi^{(1)}_{-\frac{1}{2}})^{c_1} \ldots
       (\varphi^{(n)}_{-\frac{1}{2}})^{c_n}
       (\varphi^{(1)*}_{-\frac{1}{2}})^{d_1} \ldots
       (\varphi^{(n)*}_{-\frac{1}{2}})^{d_n} \vac \, .
\end{eqnarray*}
All other basis elements from $F(a,b,c,d)$ are obtained from
$v(a,b,c,d)$ by adding to the lower indices of the factors
arbitrary non-negative integers.  Hence we have (since weight
$\vac = \Lambda_0$):
\begin{subequations} \label{eq:3.12}
  \begin{eqnarray}
    \label{eq:3.12a}
    ch F (a,b,c,d) &=& e^{\weight (v(a,b,c,d))}/\Pi (q) \, , \hbox{ where}\\
    \label{eq:3.12b}
\Pi (q) &=& (q)_{a_1} \ldots (q)_{a_m} 
          (q)_{b_1} \ldots (q)_{b_m} 
          (q)_{c_1} \ldots (q)_{c_n} 
          (q)_{d_1} \ldots (q)_{d_n} \, .
  \end{eqnarray}
\end{subequations}
Here and further we use the usual notation and assumptions:  
\begin{displaymath}
(q)_a = (1-q) \ldots (1-q^a),
 q=e^{-\delta}\hbox{  and  }  |q|<1 \, .  
\end{displaymath}

Noticing that 
\begin{eqnarray}
      \label{eq:3.13}
  \weight (\psi^{(i)}_k) &=& \epsilon_i + k \delta, \,\,
  \weight (\psi^{(i)*}_k) = -\epsilon_i + k \delta \, , \\
\nonumber
  \weight (\varphi^{(i)}_k) &=& \epsilon_{m+i} + k \delta , \,\,
  \weight (\varphi^{(i)*}_k) = -\epsilon_{m+i} +k \delta \, ,
\end{eqnarray}
we obtain from (\ref{eq:3.11}) and (\ref{eq:3.12}) the
``quasiparticle'' character formula for $F_s$:
\begin{equation}
      \label{eq:3.14}
      ch F_s = e^{\Lambda_0} \sum_{\substack{a,b \in
          \ZZ^{m+n}_+\\|a|-|b|=s}} 
        \frac{e^{\sum^{m+n}_{i=1} (a_i-b_i) \epsilon_i} 
          q^{\frac{1}{2} \sum^m_{i=1} (a^2_i + b^2_i) +
            \frac{1}{2} \sum^{m+n}_{i=m+1} (a_i+b_i)}}
        {\Pi^{m+n}_{i=1} \quad (q)_{a_i} (q)_{b_i}} \, .
\end{equation}

Another formula, which we call a theta function type character
formula, is derived as follows.  Let
\begin{displaymath}
  ch F=\sum_{s \in \ZZ} z^s ch F_s \, .
\end{displaymath}
Using (\ref{eq:3.3}) and (\ref{eq:3.13}), we obtain:
\begin{equation}
  \label{eq:3.15}
  ch F = e^{\Lambda_0} \Pi^{\infty}_{k=1}
  \frac{\Pi^m_{i=1} (1+ze^{\epsilon_i} q^{k-1/2})
        (1+z^{-1} e^{-\epsilon_i}q^{k-1/2})}
       {\Pi^n_{j=1} (1-ze^{\epsilon_{m+j}}q^{k-1/2})
         (1-z^{-1}e^{-\epsilon_{m+j}} q^{k-1/2})} \, .
\end{equation}
In order to compute the coefficient of $z^s$, we use the Jacobi
triple product identity
\begin{equation}
  \label{eq:3.16}
  \Pi^{\infty}_{k=1} (1+zq^{k-\frac{1}{2}})
  (1+z^{-1}q^{k-\frac{1}{2}}) =\frac{1}{\varphi (q)}
  \sum_{m \in \ZZ} 
  z^mq^{\frac{1}{2}m^2} \, ,
\end{equation}
and also the following well-known identity which can be derived
from the super boson-fermion correspondence \cite{K4}:
\begin{eqnarray}
  \label{eq:3.17}
  \lefteqn{\hspace{-6ex}\Pi^{\infty}_{k=1} (1+zq^{k-\frac{1}{2}})^{-1}
  (1+z^{-1}q^{k-\frac{1}{2}})^{-1}=\frac{1}{\varphi (q)^2}
  \sum_{m\in\ZZ}(-1)^m
  \frac{q^{\frac{1}{2}m(m+1)}}{1+zq^{m+\frac{1}{2}}}}\\
  \nonumber
  &=& \varphi (q)^{-2} \left( \sum_{m,k \geq 0} -\sum_{m,k<0} \right) 
  ((-1)^{m+k} z^k q^{\frac{1}{2}m(m+1)+(m+\frac{1}{2})k}) \, .\\
\noalign{\nonumber
\hbox{Here and further }}\\
\nonumber
\varphi (q) &=& \Pi^{\infty}_{j=1} (1-q^j) \, .
\end{eqnarray}
Substituting (\ref{eq:3.16}) and (\ref{eq:3.17}) in
(\ref{eq:3.15}), we get:
\begin{eqnarray*}
  ch F &=& \frac{e^{\Lambda_0}}{\varphi (q)^{m+2n}}
  \sum_{k \in \ZZ^m} \left( \sum_{p_1,a_1\geq 0}-
    \sum_{p_1,a_1<0} \right) \ldots
  \left( \sum_{p_n,a_n\geq 0}-
    \sum_{p_n,a_n<0} \right) \\[1ex]
  && (-1)^{|r|} z^{|k|+|p|}
    e^{\sum_ik_i\epsilon_i+\sum_jp_j\epsilon_{m+j}}
    q^{\frac{1}{2} \sum_ik^2_i+\frac{1}{2}\sum_j(a_j(a_j+1)
       +p_j(a_j+1/2))}\, ,
\end{eqnarray*}
where $ k=(k_1, \ldots ,k_m) \in \ZZ^m,
  p=(p_1, \ldots ,p_n), a=(a_1,\ldots,a_n) \in \ZZ^n$,
   and $ |k| =\sum_i k_i$.

The coefficient of $z^s$ is a rather complicated expression for $
ch F_s$, which, after letting $r=p+a \in \ZZ^n$, can be written
as follows:
\begin{eqnarray}
  \label{eq:3.18}
  ch F_s = \frac{e^{\Lambda_0 +s\epsilon_1}q^{s^2/2}}
      {\varphi  (q)^{m+2n}}
      \sum_{k \in \ZZ^{m-1}} \left(\sum_{r_1\geq p_1\geq 0} -
      \sum_{r_1<p_1<0}\right) \ldots \left(\sum_{r_n\geq p_n\geq 0} -
      \sum_{r_n < p_n<0}\right)\\[1ex]       
\nonumber
       (-1)^{|r+p|} e^{\sum_ik_i(\epsilon_i-\epsilon_1)
        + \sum_jp_j(\epsilon_{m+j}-\epsilon_1)}
  q^{\frac{1}{2}|k|^2 + \frac{1}{2}\sum_ik^2_i+
     \frac{1}{2}\sum_jr_j(r_j+1) +
     \sum_{i<j} p_ip_j + |k| |p|-s(|k| + |p|)} \, ,
\end{eqnarray}
where $ k = (k_2,\ldots ,k_m) \in \ZZ^{m-1}, p, r \in
 \ZZ^n$.

We rewrite (\ref{eq:3.18}) using translation operators
$t_{\alpha}$, $\alpha \in \fh^*$, defined by $(\lambda \in
\hat{\fh}^*)$:
\begin{equation}
  \label{eq:3.19}
  t_{\alpha}(\lambda)=\lambda + (\lambda | \delta)\alpha -
  (\tfrac{1}{2} (\alpha | \alpha) (\lambda | \delta) +
  (\lambda | \alpha)) \delta \, .
\end{equation}
Let $M^{\#} =\sum^{m-1}_{i=1} \ZZ \alpha_i$; recall that $M^{\#}$ 
acts on $\hat{\fh}^*$ via $\alpha \mapsto t_{\alpha}$ and the
image of this action is the translation subgroup of the Weyl
group of $s\ell (m)^{\hat{}}$.  It is straightforward to show
that (\ref{eq:3.18}) can be rewritten as follows:
\begin{eqnarray}
  \label{eq:3.20}
  ch F_s = \frac{1}{\varphi (q)^{m+2n}}
  \left( \sum_{r_1 \geq p_1 \geq 0} -
    \sum_{r_1<p_1<0} \right) \ldots 
  \left( \sum_{r_n \geq p_n \geq 0} -
    \sum_{r_n<p_n<0} \right)\\[2ex]
\nonumber
  (-1)^{|r|+|p|} q^{\frac{1}{2} \sum_j r_j (r_j+1) +
    \sum_{i<j} p_ip_j -|p|+a_s} 
    \sum_{\alpha \in M^{\#}} e^{t_{\alpha}  (\Lambda_{(s)} - 
      \sum^n_{j=1} p_j (\epsilon_1 - \epsilon_{m+j}))} \, ,
\end{eqnarray}
where $\Lambda_{(s)}$ is the weight of $\sac$ and $a_s
=s(r_n-p_n+1)+|p|$ if $s \leq 0, \, =0$ if $0<s \leq m$, and
$=(s-m)(r_1-p_1)$ if $s \geq m$.

In the case $n=1$ formula (\ref{eq:3.18}) can be simplified by
making use of the following lemma.

\begin{lemma}
  \label{lem:3.1}
Let $a,b \in \ZZ$.  Then $(j,k,n \in \ZZ)$:
\begin{eqnarray*}
\begin{array}{ll}
 \hbox{(a)} &(\sum_{k \geq j \geq a} - \sum_{k<j<a}) (-1)^{j+k}
        x^j q^{bk}q^{k(k+1)/2} 
        = \frac{x}{1+x} \varphi (q) \Pi^{\infty}_{n=1}
          (1+x^{-1}q^{n-b-1}) (1+xq^{n+b}) \, .\\[2ex]
\hbox{(b)} &
 \Pi^{\infty}_{n=1} (1+x^{-1}q^{n-b-1}) (1+xq^{n+b})
     = x^{-b}q^{-b(b+1)/2} \Pi^{\infty}_{n=1}
       (1+x^{-1}q^{n-1}) (1+xq^n)\\[1ex]
  &= x^{-b-1}q^{-b(b+1)/2} \Pi^{\infty}_{n=1}
  (1+x^{-1}q^n) (1+xq^{n-1}) \, .
\end{array}
\end{eqnarray*}

\end{lemma}

\begin{proof}
  If $k \geq a$ (resp. $k<a$), we have:
  \begin{displaymath}
          \sum^k_{j=a}(-1)^{j+k}x^j (\hbox{resp. }
          -\sum^{a-1}_{j=k+1}(-1)^{j+k}x^j)
          = \frac{x^{k+1}-(-1)^k x^a}{1+x} \, .
  \end{displaymath}
Hence the LHS of (a) is equal to
\begin{displaymath}
  \frac{1}{1+x}\sum_{k\in \ZZ} x^{k+1}q^{bk+k(k+1)/2}
     - \frac{x^a}{1+x} \sum_{k \in \ZZ} (-1)^k
     q^{bk+k(k+1)/2} \, .
\end{displaymath}
Noticing that the second summand is zero and applying to the
first summand the Jacobi triple product identity (\ref{eq:3.16}), 
we obtain~(a).  In the proof of~(b) we assume that $b>0$, the
case $b<0$ being similar:
\begin{displaymath}
  \Pi^{\infty}_{n=1} (1+x^{-1}q^{n-b-1})=
  \Pi_{m \geq 1-b} (1+x^{-1}q^{m-1})
  = \Pi^{\infty}_{m=1} (1+x^{-1}q^{m-1})
  \Pi^0_{m=1-b} (1+x^{-1}q^{m-1}) \, .
\end{displaymath}
The second product on the RHS is equal to
\begin{displaymath}
  \Pi^b_{n=1}(1+x^{-1}q^{-n}) = \Pi^b_{n=1} x^{-1} q^{-n}
   (1+xq^n) = x^{-b} q^{-b(b+1)/2} \Pi^b_{n=1} (1+xq^n) \, .
\end{displaymath}
Next, we have:
\begin{displaymath}
  \Pi^{\infty}_{n=1} (1+xq^{n+b})=\Pi^{\infty}_{m=1+b}
  (1+xq^m) =\Pi^{\infty}_{m=1} (1+xq^m)/
  \Pi^b_{m=1} (1+xq^m) \, .
\end{displaymath}
These equalities prove (b).
\end{proof}

Let now $n=1$.  Then (\ref{eq:3.18}) reads:
\begin{eqnarray}
  \label{eq:3.21}
  ch F_s &=& \frac{e^{\Lambda_0 +s \epsilon_1}q^{s^2/2}}
     {\varphi (q)^{m+2}} \sum_{k \in \ZZ^{m-1}}
     \psi (k)e^{\sum_i k_i (\epsilon_i-\epsilon_1)}
       q^{\frac{1}{2}|k| (|k|-2s)+\frac{1}{2}\sum_ik^2_i} \, ,\\
\noalign{\nonumber\hbox{where }}\\
\nonumber
\psi (k) &=& \left( \sum_{t \geq j \geq 0} -\sum_{t<j<0} \right)
(-1)^{j+t} \left( q^{|k|-s}e^{\epsilon_{m+1}-\epsilon_1}\right)^j
q^{t(t+1)/2} \, .
\end{eqnarray}
By Lemma~\ref{lem:3.1}a for $x=q^{|k|-s}
e^{\epsilon_{m+1}-\epsilon_1}, b=0$, we have:
\begin{displaymath}
  \psi(k)=\frac{q^{|k|-s} e^{\epsilon_{m+1}-\epsilon_1}
       \varphi (q)}
     {1+q^{|k|-s} e^{\epsilon_{m+1}-\epsilon_1}}
     \Pi^{\infty}_{n=1} \left(1+e^{\epsilon_1-\epsilon_{m+1}}
         q^{n-|k|+s-1}\right) \left(1+e^{\epsilon_{m+1}-\epsilon_1}
         q^{n+|k|-s}\right) \, .
\end{displaymath}
By Lemma~\ref{lem:3.1}b for $x=e^{\epsilon_{m+1}-\epsilon_1},
b=|k|-s$, we rewrite this as follows:
\begin{displaymath}
  \psi(k)= \frac{x^{-b}q^{-b(b-1)/2}\varphi (q)}
     {1+q^bx} \Pi^{\infty}_{n=1} (1+x^{-1}q^n)
     (1+xq^{n-1}) \, .
\end{displaymath}
Substituting this in (\ref{eq:3.21}), we obtain:
\begin{eqnarray}
  \label{eq:3.22}
  ch F_s &=& \frac{e^{\Lambda_0+s\epsilon_{m+1}}q^{-\frac{s}{2}}}
    {\varphi (q)^{m+1}} \Pi^{\infty}_{n=1}
    (1+e^{\epsilon_1-\epsilon_{m+1}}q^n)
    (1+e^{\epsilon_{m+1}-\epsilon_1} q^{n-1})\\[1ex]
\nonumber
&\times & \sum_{k \in \ZZ^{m-1}}
\frac{e^{\sum^m_{i=2} k_i(\epsilon_i-\epsilon_{m+1})}}
    {1+q^{|k|-s} e^{\epsilon_{m+1}-\epsilon_1}}
    q^{\frac{1}{2} \sum^m_{i=2} k_i (k_i+1)} \, .
\end{eqnarray}
This formula agrees with the one obtained in \cite{KL} (see also \cite{K4}) for
$g\ell(1|1)^{\hat{}}$.  

In the next section we use this formula
in the case $m \geq 2$ in order to derive character formulas for
all integrable level~$1$ $s \ell (m|1)^{\hat{}}$-modules in terms 
of the theta function and the (multivariable) Appell's functions,
and to obtain their high temperature 
asymptotics.

%In the case $m \geq 2$ and $s=0$ we
%obtain from (\ref{eq:3.20}) the following theta function type
%character formula for the $s \ell (m|1)^{\hat{}}$-module
%$L(\Lambda_0)$:
%%
%\begin{eqnarray}
%  \label{eq:3.21}
%  e^{-\Lambda_0} ch L (\Lambda_0)
%      &=& \frac{\Pi^{\infty}_{n=1} (1+e^{\epsilon_1
%          -\epsilon_{m+1}}q^n) (1+e^{\epsilon_{m+1}-\epsilon_1}q^{n-1})}
%      {\varphi (q)^m}\\[1ex]
%%
%\nonumber
%   &\times & \sum_{k\in \ZZ^{m-1}}
%   \frac{e^{\sum^m_{i=2}k_i (\epsilon_i-\epsilon_{m+1})}}
%     {1+q^{|k|} e^{\epsilon_{m+1}-\epsilon_1}  }
%     q^{\frac{1}{2} \sum^m_{i=2}k_i (k_i+1)} \, ,
%\end{eqnarray}
%%
%where, as before, $k=(k_2,\ldots,k_m)$ and $|k|=\sum^m_{i=2}
%k_i$.  

\section{Theta function type character formula for integrable
  level~$1$ $s \ell (m|1)^{\hat{}}$-modules and Appell's function.}
\label{sec:4}

Recall that Appell's function is defined by the following series 
(cf.~\cite{A} and \cite{P}):
\begin{displaymath}
  A (x,z,q) =\sum_{k \in \ZZ}
  \frac{q^{\frac{1}{2}k^2} z^k}{1+xq^k} \, ,
\end{displaymath}
which converges to a meromorphic function in the domain $x,z,q
\in \CC$, $ |q|<1$.  The classical theta function in one variable 
is a special case of this function:  
\begin{displaymath}
  \Theta (z) \equiv \Theta (z;q) = A (0,z,q) \, .
\end{displaymath}
Note that by (\ref{eq:3.16}) we have a product expansion:
\begin{equation}
  \label{eq:4.1}
  \Theta (zq^{1/2};q) = \varphi (q) \Pi^{\infty}_{k=1}
  (1+zq^k) (1+z^{-1} q^{k-1}) \, .
\end{equation}

We shall need also the following multivariable generalization of
Appell's function.  Let $B$ be an $N \times N$ symmetric matrix
such that $\Re B$ is positive definite and let $\ell$ be a linear 
function of $\CC^N$.  We define the series
\begin{displaymath}
  A_{B,\ell} (x; z_1 , \ldots ,z_N;q)=\sum_{k \in \ZZ^N}
  \frac{q^{\frac{1}{2}k^TBk} z_1^{k_1} \ldots z_N^{k_N}}
     {1+xq^{\ell (k)}} \, ,
\end{displaymath}
which converges to a meromorphic function provided that $|q|<1$.
Again, letting $x=0$, we get the multivariable theta function.

Consider now the $g \ell (m|1)^{\hat{}}$-modules $F_s$,
$s\in\ZZ$, and assume in this section that $m \geq 2$.  We have 
(see Remark~\ref{rem:3.2}):
\begin{eqnarray*}
  \Lambda_{(s)}= \left\{
    \begin{array}{ll}
      \Lambda_s-\frac{s}{2} \delta 
         & \hbox{if } 0 \leq s \leq m \\
         -(s-m) \Lambda_0 + (1+s-m) \Lambda_m 
         -\frac{s}{2} \delta
         & \hbox{if } s \geq m \\
         (1-s)\Lambda_0 + s \Lambda_m +\frac{s}{2} \delta
         & \hbox{if }s \leq 0 
    \end{array}\right. \, .
\end{eqnarray*}
For $m \geq 2$, we have:  $g \ell (m|1)^{\hat{}}=s \ell
(m|1)^{\hat{}}+g \ell (1)^{\hat{}}$ (sum of ideals), hence:
\begin{displaymath}
ch F_s = ch L (\Lambda_{(s)}) \varphi (q)^{-1} \, ,  
\end{displaymath}
where $L(\Lambda_{(s)})$ denotes the irreducible $s\ell
(m|1)^{\hat{}}$-module with highest weight $\Lambda_{(s)}$.
Hence formula (\ref{eq:3.22}) gives us the following expression
for $ch L (\Lambda_{(s)})$ in terms of the theta function $\Theta 
(z;q)$ (we use (\ref{eq:4.1})) and the
(multivariable) Appell's function:
\begin{equation}
  \label{eq:4.2}
  ch L (\Lambda_{(s)}) =
  \frac{e^{\Lambda_0 +s\epsilon_{m+1}}q^{-\frac{s}{2}}}
  {\varphi (q)^{m+1}} \Theta (z_1q^{\frac{1}{2}};q)
  A_{I,\ell}(z^{-1}_1 q^{-s}; z_2 q^{\frac{1}{2}},\ldots ,
  z_mq^{\frac{1}{2}};q) \, ,
\end{equation}
where $z_i =e^{\epsilon_i-\epsilon_{m+1}}$ $(i=1,\ldots ,m)$,
$B=I$ is the $(m-1) \times (m-1)$ identity matrix and $\ell (k)
=\sum_i k_i$, $k \in \ZZ^{m-1}$.  (Note that in the simplest case 
$m=2$ we get in this expression the classical Appell's function.)

Next, we derive yet another character formula for the $s\ell
(m|1)^{\hat{}}$-module $L(\Lambda_0)$ in the case $m \geq 2$, in
terms of classical theta functions and certain ``half'' modular
forms.  We use for this (\ref{eq:3.20}) for $n=1$:
\begin{eqnarray}
  \label{eq:4.3}
  chL (\Lambda_0)= \frac{1}{\varphi (q)^{m+1}}
  \left( \sum_{r \geq p \geq 0} -
        \sum_{r< p<0} \right) 
      \sum_{\alpha \in M^{\#}} (-1)^{r+p}
    e^{t_{\alpha} (\Lambda_0-p
      (\epsilon_1-\epsilon_{m+1}))}
        q^{\frac{1}{2}r(r+1)} \, .
\end{eqnarray}
Introduce the following elements of $M^{\#}$:
\begin{displaymath}
  \beta_k =k\epsilon_1-\sum^k_{i=1} \epsilon_1 , \quad
  k=1, \ldots ,m \, ,
\end{displaymath}
and the element $\mu = \epsilon_{m+1}-\tfrac{1}{m}(\epsilon_1 +
\cdots + \epsilon_m) \in \fh^*$, which is orthogonal to
$M^{\#}$.  The even part of $s \ell (m|1)^{\hat{}}$ is a sum of
ideals $s \ell (m)^{\hat{}}$ and $(\CC \mu)^{\hat{}}$.

Write $p=jm-k$, where $j \in \ZZ$, $1 \leq k \leq m$ and note
that 
\begin{equation}
  \label{eq:4.4}
  \Lambda_0 -p(\epsilon_1 -\epsilon_{m+1})+j\beta_m-\beta_k
  = \dot{\Lambda}_k +p\mu \, .
\end{equation}
where $\dot{\Lambda}_k$ denote fundamental weights of $s \ell
(m)^{\hat{}}$ and we identify $\dot{\Lambda}_m$ with
$\dot{\Lambda}_0$.  Adding to $\alpha$ the element $j
\beta_m-\beta_k$ in (\ref{eq:4.3}), and using (\ref{eq:4.4}), we
rewrite (\ref{eq:4.3}) as follows:
\begin{eqnarray*}
  ch L (\Lambda_0) &=& \frac{1}{\varphi (q)^{m+1}}
  \sum^m_{k=1} \Biggl( \sum_{\substack{j,r \in \ZZ\\r+k \geq jm\\
j>0}} - \sum_{\substack{j,r \in Z\\ r+k<jm\\ j\leq 0}} \Biggr)\\[2ex]
  &&(-1)^{r+k+jm} \sum_{\alpha \in M^{\#}}
  e^{t_{\alpha}(\dot{\Lambda}_k) + (jm-k)\mu}
  q^{\frac{r(r+1)}{2} -\frac{(jm-k) (jm-k-j)}{2}
    -\frac{k(j-1)}{2}} \, .
\end{eqnarray*}
Denoting by $\dot{L}(\dot{\Lambda},a)$ the irreducible $s \ell
(m)\sphat+(\CC \mu)\sphat$-module with highest weight
$\dot{\Lambda}+a \mu$, and recalling that (\cite{K3}, Proposition~12.13):
\begin{equation}
  \label{eq:4.5}
  ch \dot{L} (\dot{\Lambda}_k,a)
     =\frac{1}{\varphi (q)^m} \sum_{\alpha \in M^{\#}}
     e^{t_{\alpha}(\dot{\Lambda}_k)+a \mu} \, ,
\end{equation}
we obtain:
%
%\begin{eqnarray}
%  \label{eq:4.6}
%  ch L (\Lambda_0) &=& \sum^m_{k=1}
%  \left( \sum_{\substack{j,r \in \ZZ\\ r+k \geq jm\\ j>0}}
%    - \sum_{\substack{j,r \in \ZZ\\r+k \leq jm\\ j\leq
%        0}}\right)\\[2ex]
%%
%\nonumber
%    && (-1)^{r+k+jm} ch \dot{L} (\dot{\Lambda}_k,jm-kj)
%     q^{\frac{r(r+1)}{2} 
%       - \frac{(jm-k) (jm-kj-j)}{2} 
%       -\frac{k(j-1)}{2}}/ \varphi (q) \, .
%\end{eqnarray}
%%
%
\begin{eqnarray}
  \label{eq:4.7}
  ch L(\Lambda_0)\!\!\! &=&\!\!\! \sum^m_{k=1} \sum_{\substack{p \in \ZZ\\p+k|m}}
  b_{k,p}(q) ch \dot{L}(\dot{\Lambda}_k,p),\\
\noalign{ \nonumber \hbox{where}}\\
\nonumber
 \hspace*{8ex}  b_{k,p} (q)\!\!\! &=&\!\!\! \frac{q^{(\frac{1}{2}-\frac{1}{2m})p^2
     + \frac{1}{2m} k^2+\frac{1}{2}k}}{\varphi (q)}
   \times \sum_{r \geq p} (-1)^{r-p} q^{\frac{1}{2}r(r+1)} 
(\hbox{resp. }
   \times \sum_{r<p} (-1)^{r-p-1}
   q^{\frac{1}{2}r(r+1)}),
\end{eqnarray}
if  $p \geq 0$ (resp.  $p<0$).

Thus, the branching functions $b_{k,p}(q)$ are ``half'' modular
functions, in a sharp contrast with the case of affine Lie
algebras \cite{K3}. Recalling that the series $\sum_{\alpha \in M^{\#}}
e^{t_{\alpha}(\Lambda)}$ converges to a classical theta function 
\cite{K3}, we see that the character of the ``basic'' $s\ell
(m|1)^{\hat{}}$-module is a finite linear combination of
classical theta functions with coefficients ``half'' modular functions.

The basic specialization of (\ref{eq:3.22}) gives 
the specialized character formulas for $
s \ell (m|1)^{\hat{}}$-modules $L(\Lambda_{(s)})$:%, where
%$\Lambda_{(s)}$, $s \in \ZZ$, are the weights of $\sac$ described
%by Remark~\ref{rem:3.2}:
%
\begin{equation}
\label{eq:4.8}
  tr_{L(\Lambda_{(s)})}q^{L_0} =
  2q^{-\frac{s}{2}} \frac{\varphi (q^2)^2}{\varphi (q)^{m+2}}
  \sum_{k \in \ZZ^{m-1}}
\frac{q^{\frac{1}{2} \sum_i k_i (k_i+1)}}{1+q^{|k|-s}} \, .
\end{equation}
%
%
%\begin{equation}
%\label{eq:3.23}
%  tr_{L(\Lambda_{(s)})} q^{L_0} =
%  2q^{-\frac{s}{2}} \left( \frac{\varphi (q^2)}
%    {\varphi (q)^2}\right)^2 \sum_{k \in \ZZ}
%  \frac{q^{\frac{1}{2}k(k+1)}}{1+q^{k-s}} \, .
%\end{equation}
%
%In order to compute asymptotics as $\tau \downarrow 0$ and
%transformation properties of these functions, one needs to know
%them for the functions
%%
%\begin{displaymath}
%  \chi_s(\tau)=\sum_{k \in \ZZ}
%  \frac{q^{\frac{1}{2}k(k+1)}}{1+q^{k-s}} \, , 
%  \hbox{ where } q=e^{2\pi i \tau} \, .
%\end{displaymath}
%%
%This is apparently unknown.
where, as before, $|k|=\sum_ik_i$.In the remainder of this section we discuss asymptotics of (\ref{eq:4.8}).

Given a positive definite quadratic form $B(x)$ on $\RR^N$, an
affine linear function $\ell (x)$ on $\RR^N$ and an element
$\alpha$ of $\RR^N$, consider the following series, where $q=e^{2 
  \pi i \tau}$:
\begin{displaymath}
  f_{B,\ell,\alpha} (\tau) =\sum_{\gamma \in \ZZ^N + \alpha}
  \frac{q^{\frac{1}{2}B(\gamma)}}{1+q^{\ell (\gamma)}} \, .
\end{displaymath}
This series converges on the upper half plane $\Im \tau >0$ to a
specialization of the multivariable Appell's function.
From the transformation properties of theta series one gets (see
e.g.~\cite{K3}):
\begin{equation}
  \label{eq:4.9}
  f_{B,0,\alpha} (i \beta) = \frac{1}{2} (\det B)^{-1/2}
  \beta^{-N/2} + o (\beta) \hbox{ as } \beta \to 0 \quad
  (\Re \beta >0) \, .
\end{equation}

In order to get the asymptotics of the functions
$f_{B,\ell,\alpha} (\tau)$, let $\displaystyle{f^{\pm}_{B,\ell,\alpha}
(\tau)=\frac{1}{2} \sum_{\substack{\gamma \in \ZZ^N+\alpha\\ \pm \ell
(\gamma )>0}} q^{\frac{1}{2} 
  B(\gamma)}}$.
It is easy to derive from (\ref{eq:4.9}) by induction on $n$ the 
following asymptotics:
\begin{equation}
  \label{eq:4.10}
  f^{\pm}_{B,\ell,\alpha} (i \beta) = \frac{1}{4}(\det B)^{-1/2}
    \beta^{-N/2} + p_{\pm} (\beta^{-1/2}) +o (\beta) \, ,
\end{equation}
as $\beta \to 0, \beta \in \RR_+$, where $p_{\pm} (x)$  is a
polynomial in $x$ of degree strictly less than $N$.

The idea of the following lemma is due to A.~Polishchuk.

\begin{lemma}
  \label{lem:4.1}
$|f_{B,\ell,\alpha} (i\beta)-f_{B,0,\alpha} (i
\beta)|<p(\beta^{-1/2})$ for $\beta \in \RR, 0<\beta <a$, where $
a$ is a positive number and $p(x)$ is a polynomial in $x$ of
degree strictly less than $N$.

\end{lemma}

\begin{proof}
  Let $g(\beta) =f_{B,\ell,\alpha}
  (i\beta)-f_{B,0,\alpha}(i\beta)$.  We have:
  $g(\beta)=g^+(\beta)-g^{-}(\beta)$, where
  \vspace*{-0.75ex}
  \begin{displaymath}
    g^{\pm} (\beta) =\sum_{\substack{\gamma \in \ZZ^N +\alpha \\
        \pm \ell (\gamma)>0}} e^{-\pi \beta B (\gamma)}
    \frac{1-e^{\mp 2 \pi \beta \ell (\gamma)}}
    {1+e^{\mp 2\pi \beta \ell (\gamma)}} \, .
  \end{displaymath}
  \vspace*{-0.75ex}
Furthermore, we have:
\vspace*{-0.75ex}
\begin{displaymath}
  0 \leq g^{\pm}(\beta) \leq \sum_{\substack{\gamma \in
      \ZZ^N+\alpha\\ \pm \ell (\gamma)>0}}
  e^{-\pi \beta B (\gamma)}-\sum_{\substack{\gamma \in
      \ZZ^N+\alpha\\ \pm \ell (\gamma)>0}}
  e^{-\pi \beta (B(\gamma) \pm 2 \ell (\gamma))} \, .
\vspace*{-0.75ex}
\end{displaymath}
The first sum on the right is just
$f^{\pm}_{B,\ell,\alpha}(i\beta)$, which has asymptotics
(\ref{eq:4.10}).  But the second sum on the right has asymptotics 
of this form too since it can be written as a product of a power
of $q$ and a function $f^{\pm}_{B,\ell',\alpha}(i\beta)$ for some 
other affine linear function $\ell'$, by ``completing the squares''.  The 
lemma is proved.
\end{proof}

We shall write $f(\tau) \sim g (\tau)$ if $\lim_{\substack{\beta
    \to 0\\\beta \in \RR_+}} f (i\beta)/g(i\beta) =1$.
Lemma~\ref{lem:4.1} and (\ref{eq:4.9}) imply:
\begin{equation}
  \label{eq:4.11}
  f_{B,\ell,\alpha} (\tau) \sim \frac{1}{2} (\det B)^{-1/2}
  \beta^{-N/2} \, .
\end{equation}
Since 
\begin{equation}
  \label{eq:4.12}
\eta (\tau) \sim \beta^{-1/2} e^{-\pi/12 \beta} \, ,
\end{equation}
we deduce from (\ref{eq:4.11}) and (\ref{eq:4.8}) the following
asymptotics along the imaginary axis $\tau =i\beta , \beta \in
\RR_+$:
\begin{equation}
  \label{eq:4.13}
  tr_{L(\Lambda_{(s)})} q^{L_0} \sim \frac{1}{2} 
  \beta^{\frac{1}{2}} e^{\frac{\pi}{12 \beta}(m+1)} \, .
\end{equation}

%\begin{remark}
%  \label{rem:4.1}
%A.~Polishchuk pointed out that (\ref{eq:4.11}) still holds if
%$\tau$ tends to $0$ along any ray (in the upper half plane), but
%the limit may even not exist if $\tau$ tends to $0$ along a
%parabola, and this apparently happens for the sum in the right
%hand side of (\ref{eq:4.8}) (which shows that this is  not a
%modular form).
%\end{remark}

\section{A Weyl type character formula for integrable level~$1$ $ 
 g\ell (m|n)^{\hat{}}$-modules.}
\label{sec:5}

In this section we derive a Weyl type character formula (\ref{eq:5.12}) for
principal integrable level~$1$-modules over $g \ell
(m|n)^{\hat{}}$ provided that $m \geq n$.  We use for that
formula (\ref{eq:3.15}) for $ch F$ and the denominator identity
for $s\ell (m+1|n)^{\hat{}}$.  In order to compare these two
formulas, we consider the  labelings of simple roots of $
g \ell (m|n)^{\hat{}}$ and $s\ell (m+1|n)^{\hat{}}$ given below:
\begin{table*}[hb]
\vspace*{-8ex}$
\begin{array}{c c}
\setlength{\unitlength}{0.16in}
\begin{picture}(20,8)
\put(0,1.5){\line(4,1){6.65}}
\put(15,1.5){\line(-4,1){6.65}}
\put(7.5,4.5){\makebox(0,0)[c]{0}}
\put(7.5,3.5){\makebox(0,0)[c]{$\bigotimes$}}
\put(0,1){\makebox(0,0)[c]{$\bigcirc$}}
\put(2,1){\makebox(0,0)[c]{$\bigcirc$}}
\put(6.85,1){\makebox(0,0)[c]{$\bigcirc$}}
\put(9.25,1){\makebox(0,0)[c]{$\bigotimes$}}
\put(15,1){\makebox(0,0)[c]{$\bigcirc$}}
\put(0.4,1){\line(1,0){1.1}}
\put(2.4,1){\line(1,0){1.1}}
\put(5,1){\line(1,0){1.2}}
\put(7.45,1){\line(1,0){1.2}}
\put(10.25,1){\line(1,0){1.2}}
\put(13.25,1){\line(1,0){1.2}}
\put(4.5,1){\makebox(0,0)[c]{$\cdots$}}
\put(12.5,1){\makebox(0,0)[c]{$\cdots$}}
\put(0,0){\makebox(0,0)[c]{$1$}}
\put(2,0){\makebox(0,0)[c]{$2$}}
\put(6.5,0){\makebox(0,0)[c]{$m-1$}}
\put(9.25,0){\makebox(0,0)[c]{$m$}}
\put(15.1,0){\makebox(0,0)[c]{$m+n-1$}}
%
%\put(1,1.35){\makebox(0,0)[c]{$-1$}}
%\put(3,1.35){\makebox(0,0)[c]{$-1$}}
%\put(8,1.35){\makebox(0,0)[c]{$-1$}}
%\put(10,1.35){\makebox(0,0)[c]{$-1$}}
%\put(12,1.35){\makebox(0,0)[c]{$1$}}
%\put(14,1.35){\makebox(0,0)[c]{$1$}}
%\put(19,1.35){\makebox(0,0)[c]{$1$}}
%
%\put(5,3.4){\makebox(0,0)[c]{$-1$}}
%\put(15,3.4){\makebox(0,0)[c]{$1$}}
\end{picture}
&   
\setlength{\unitlength}{0.16in}
\begin{picture}(20,8)
\put(0,1.5){\line(4,1){6.65}}
\put(15,1.5){\line(-4,1){6.65}}
\put(7.5,4.5){\makebox(0,0)[c]{0}}
\put(7.5,3.5){\makebox(0,0)[c]{$\bigotimes$}}
\put(0,1){\makebox(0,0)[c]{$\bigcirc$}}
\put(0.5,1){\line(1,0){1.1}}
\put(2,1){\makebox(0,0)[c]{$\bigcirc$}}
\put(2.5,1){\line(1,0){1.1}}
\put(4,1){\makebox(0,0)[c]{$\bigcirc$}}
\put(4.5,1){\line(1,0){1.1}}
\put(6,1){\makebox(0,0)[c]{$\cdots$}}
\put(6.5,1){\line(1,0){1.1}}
\put(8,1){\makebox(0,0)[c]{$\bigcirc$}}
\put(8.5,1){\line(1,0){1.1}}
\put(10,1){\makebox(0,0)[c]{$\bigotimes$}}
\put(10.5,1){\line(1,0){1.1}}
\put(12.5,1){\makebox(0,0)[c]{$\cdots$}}
\put(13.5,1){\line(1,0){1.1}}
\put(15,1){\makebox(0,0)[c]{$\bigcirc$}}
\put(0,0){\makebox(0,0)[c]{$*$}}
\put(2,0){\makebox(0,0)[c]{$1$}}
\put(4,0){\makebox(0,0)[c]{$2$}}
\put(8,0){\makebox(0,0)[c]{$m-1$}}
\put(10,0){\makebox(0,0)[c]{$m$}}
\put(15.1,0){\makebox(0,0)[c]{$m+n-1$}}
\end{picture}
\end{array}$
\end{table*}\\
Putting
\begin{equation}
    \label{eq:5.1}
z=-e^{-\epsilon_1}q^{-\frac{1}{2}}y
  \hbox{ and } y=e^{-\alpha_*} \, ,
\end{equation}
we can rewrite formula (\ref{eq:3.15}) as follows:
\begin{equation}
  \label{eq:5.2}
  e^{-\Lambda_0} chF =\Pi^{\infty}_{k=1}\quad
  \tfrac{\Pi^{m-1}_{i=0}
    (1-e^{\alpha_*+\alpha_1+\cdots + \alpha_i}q^k)
    (1-e^{-\alpha_*-\alpha_1-\cdots-\alpha_i} q^{k-1})}
  {\Pi^{n-1}_{j=0}
    (1+e^{\alpha_*+\alpha_1+\cdots +\alpha_{m+j}}q^k)
    (1+e^{-\alpha_*-\alpha_1-\cdots -\alpha_{m+j}} q^{k-1})}\, .
\end{equation}

Denote by $W^{\#}$ (resp. $\tilde{W}^{\#}$) the subgroup of the
Weyl group of $g\ell (m|n)$ (resp.~$s\ell (m+1|n)$) generated by
reflections $r_{\alpha}$ in roots $\alpha = \alpha_1,\ldots
,\alpha_{m-1}$ (resp.~$\alpha_*,\alpha_1,\ldots,\alpha_{m-1}$)
and by $M^{\#}$ (resp.~$\tilde{M}^{\#}$) the subgroups of these
Weyl groups generated by translations $t_{\alpha}$ in integral
linear combinations of these roots.  Let $\tilde{\rho}$ denote a
Weyl vector for $s\ell (m+1|n)^{\hat{}}$ and let $R$ denote the
denominator for $g\ell (m|n)^{\hat{}}$.  It is clear by
(\ref{eq:5.2}) that the denominator $\tilde{R}$ of $s\ell (m+1|n)^{\hat{}}$
is given by 
\begin{equation}
  \label{eq:5.3}
  \tilde{R} =e^{-\Lambda_0} R ch F \, .
\end{equation}

In order to write down the denominator identity for $s\ell
(m+1|n)^{\hat{}}$, introduce the roots $\beta_{ij}=\alpha_i
+\alpha_{i+1}+ \cdots + \alpha_j$, $1\leq i  \leq j \leq m+n-2$
(here $\beta_{ii}=\alpha_i$), and let
$\beta_i=\beta_{m-n+i,m+i-1}$, $i=1, \ldots ,n$.  Let
\begin{equation}
  \label{eq:5.4}
  \tilde{\rho}'=\tilde{\rho} +\sum_{\substack{m-n+2\leq i \leq m \leq j
    \leq m+n-2\\ j-i \leq n-2}}\quad \beta_{ij} \, .
\end{equation}
Then the denominator identity for $s\ell (m+1|n)^{\hat{}}$ looks
as follows:
\begin{equation}
  \label{eq:5.5}
  e^{-\Lambda_0}e^{\tilde{\rho}'} R ch F =
  \sum_{w \in \tilde{W}^{\#}\ltimes \tilde{M}^{\#}}
  \epsilon (w) w \frac{e^{\tilde{\rho}'}}
  {\Pi^n_{j=1} (1+e^{-\beta_j})} \, .
\end{equation}
This identity can be derived from the denominator identity given
in \cite{KW} by making use of odd reflections as follows.

The denominator identity for $s \ell (m+1|n)^{\hat{}}$ in \cite{KW} is given for the choice of the set of
simple roots with a maximal number of grey nodes:
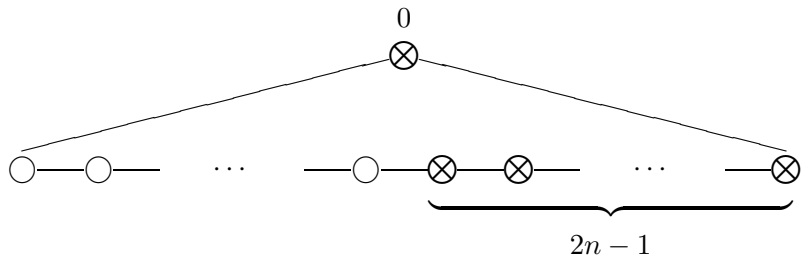
\begin{center}
\setlength{\unitlength}{0.2in}
\begin{picture}(20,8)(0,-1)
\put(0,1.5){\line(4,1){9.6}}
\put(20,1.5){\line(-4,1){9.6}}
\put(10,5){\makebox(0,0)[c]{0}}
\put(10,4){\makebox(0,0)[c]{$\bigotimes$}}
\put(0,1){\makebox(0,0)[c]{$\bigcirc$}}
\put(2,1){\makebox(0,0)[c]{$\bigcirc$}}
\put(9,1){\makebox(0,0)[c]{$\bigcirc$}}
\put(11,1){\makebox(0,0)[c]{$\bigotimes$}}
\put(13,1){\makebox(0,0)[c]{$\bigotimes$}}
\put(20,1){\makebox(0,0)[c]{$\bigotimes$}}
\put(0.4,1){\line(1,0){1.2}}
\put(2.4,1){\line(1,0){1.2}}
\put(7.4,1){\line(1,0){1.2}}
\put(9.4,1){\line(1,0){1.2}}
\put(11.4,1){\line(1,0){1.2}}
\put(13.4,1){\line(1,0){1.2}}
\put(18.4,1){\line(1,0){1.2}}
\put(5.5,1){\makebox(0,0)[c]{$\cdots$}}
\put(16.5,1){\makebox(0,0)[c]{$\cdots$}}
\put(15.4,-1){\makebox(0,0)[c]{$2n-1$}} 
\put(15.4,0){\makebox(0,0)[c]{$\underbrace{\hspace*{1.9in}}$}}
\end{picture}
\end{center}
Let $\gamma_1,\ldots \gamma_n$ be the (unique subset) of the set
of simple roots without $\alpha_0$ such that $(\gamma_i |
\gamma_j)=0$ for all $i,j$, and let $\tilde{\rho}''$ be its Weyl
vector.  Then the identity reads:
\begin{equation}
  \label{eq:5.6}
  e^{\tilde{\rho}''} \tilde{R} = 
      \sum_{w \in \tilde{W}^{\#}\ltimes \tilde{M}^{\#}}
        \epsilon (w) w \frac{e^{\tilde{\rho}''}}
        {\Pi^n_{j=1} (1+e^{-\gamma_j})} \, .
\end{equation}
In order to derive (\ref{eq:5.5}) from (\ref{eq:5.6}), we apply a
  sequence of odd reflections which transforms the initial
  diagram with two grey nodes to the above final diagram with
  $2n$ grey nodes.  In order to explain this sequence, denote by
  $\Pi^{(i,j)}$ a set of simple roots containing $\beta_{i,m+j}$
  and by $\Pi^{(i,j+1)}$ the set of simple roots obtained from it 
  by the odd reflection in $\beta_{i,m+j}$.  Denoting by
  $\tilde{\Pi}$ and $\tilde{\Pi}'$ the initial and the final sets
  of simple roots, we have the following sequence:
  \begin{eqnarray*}
    \tilde{\Pi} &=& \Pi^{(m,0)} \to \Pi^{(m,1)} \to
        \cdots \to \Pi^{(m,n-2)} \to \Pi^{(m,n-1)}\\
     &=& \Pi^{(m -1,0)} \to \Pi^{(m-1,1)} \to
         \cdots \to \Pi^{(m-1,n-2)}\\
     &=& \Pi^{(m-2,0)} \to \Pi^{(m-2,1)} \to
         \cdots \Pi^{(m-2,n-3)}\\
     &=& \Pi^{(m-3,0)} \to \Pi^{(m-3,1)} \to \cdots\\
 \cdots    &=& \Pi^{(m-n+2,0)} \to \Pi^{(m-n+2,1)}
          = \tilde{\Pi}' \, .
  \end{eqnarray*}
Using Lemma~\ref{lem:1.3}, one sees that $\tilde{\rho}'$ and
$\tilde{\rho}$ are related by formula~(\ref{eq:5.4}) and using
Lemma~\ref{lem:1.4}a, we see that
\begin{equation}
  \label{eq:5.7}
  \Lambda'_0 = \Lambda_0.
\end{equation}

Using the decompositions
\begin{displaymath}
  \tilde{W}^{\#}= W^{\#} \sqcup
  (\sqcup^{m-1}_{j=0}  W^{\#}r_{\alpha_*+\alpha_1 + \cdots + \alpha_j})
      \hbox{ and }
      \tilde{M}^{\#}=\ZZ \alpha_* +M^{\#} \, ,
\end{displaymath}
we obtain from (\ref{eq:5.5}):
\begin{eqnarray}
  \label{eq:5.8}
  e^{-\Lambda_0} R ch F &=& I_0 - \sum^{m-1}_{i=0} I\!I_i \, ,\\
\noalign{  \nonumber
\hbox{where}}\\
  \nonumber
  I_0 &=& e^{-\tilde{\rho}'} \sum_{k \in \ZZ}
  \sum_{\alpha \in M^{\#}} \sum_{w \in W^{\#}}
  \epsilon (w) t_{k\alpha_*} t_{\alpha} w
  \frac{e^{\tilde{\rho}'}}{\Pi^n_{j=1}(1+e^{-\beta_j})}\\
  \nonumber
  I\!I_i &=& e^{-\tilde{\rho}'} \sum_{k \in \ZZ}
     \sum_{\alpha \in M^{\#}} \sum_{w \in W^{\#}}
     \epsilon (w) t_{k \alpha_*} t_{\alpha}
     wr_{\alpha_* +\alpha_1+\cdots +\alpha_i}
     \frac{e^{\tilde{\rho}'}}{\Pi^n_{j=1} (1+e^{-\beta_j})} \, .
\end{eqnarray}

In order to compute $ch L (\Lambda_0)$ of the $g \ell
(m|n)^{\hat{}}$-module $L (\Lambda_0)$ we compare the constant terms
(i.e.,~$y^0$-terms) in the decomposition of both sides of
(\ref{eq:5.8}) is the series in powers of $y$.

We will show that
\begin{eqnarray}
  \label{eq:5.9}
  \hbox{constant term of } I\!I_i &=& 0 \hbox{ for all } i \, ,\\
  \label{eq:5.10}
  \hbox{constant term of } I_0 &=& \sum_{w \in W^{\#}\ltimes M^{\#}}
     \epsilon (w) w \frac{e^{\tilde{\rho}'}}
     {\Pi^n_{j=1} (1+e^{-\beta_j})}\, .
\end{eqnarray}
Using that $\tilde{\rho}'$ when restricted to the Cartan
subalgebra of $g\ell (m|n)^{\hat{}}$ coincides with $\Lambda_0 +
\rho'$, where $\rho'$ is related to the Weyl vector $\rho$ of
$g\ell (m|n)^{\hat{}}$ by (\ref{eq:5.4}) with $\sim$ removed, we 
obtain from (\ref{eq:5.5}), (\ref{eq:5.9}) and (\ref{eq:5.10}):
\begin{equation}
  \label{eq:5.11}
  e^{\rho'} R ch  L (\Lambda_0) =
  \sum_{w \in W^{\#} \ltimes M^{\#}}
  \epsilon (w) w 
  \frac{e^{\Lambda_0 + \rho'}}{\Pi^n_{j=1} (1+e^{-\beta_j})}\, .
\end{equation}
Applying to this formula odd reflections as above, we obtain an
equivalent character formula for the choice of the diagram with $
2n$ grey nodes as above (where $\rho$ is a Weyl vector for this
choice of the diagram):
\begin{equation}
  \label{eq:5.12}
  e^{\rho} R ch L (\Lambda_0)=
  \sum_{w \in W^{\#} \ltimes M^{\#}} \epsilon (w) w
  \frac{e^{\Lambda_0+\rho}}{\Pi^n_{j=1}(1+e^{-\alpha_{m-n+2j-1}})}\,.
\end{equation}

The proof of (\ref{eq:5.9}) and (\ref{eq:5.10}) is
straight-forward, and we explain it in the case $n=1$.  We~have:
\begin{displaymath}
  e^{-\tilde{\rho}}t_{-k\alpha_*}
  r_{\alpha_*+\alpha_1+\cdots + \alpha_{m-1}}
  \left( \frac{e^{\tilde{\rho}}}{1+e^{-\alpha_m}}\right)
  = \frac{e^{-mk\alpha_*} e^{-m(\alpha_*+\alpha_1+\cdots
    +\alpha_{m-1})} q^{k^2m-k+mk}}
    {1+e^{-(\alpha_*+\alpha_1+\cdots+\alpha_m)} q^k}\, .
\end{displaymath}
Hence:
\begin{displaymath}
  e^{-\tilde{\rho}} \sum_{k \in \ZZ}t_{-k\alpha_*}
  r_{\alpha_*+\alpha_1+\cdots+\alpha_{m-1}}
  \left( \frac{e^{\tilde{\rho}}}{1+e^{-\alpha_m}} \right) 
  = \left( \sum_{k,s \geq 0} - \sum_{k,s<0}\right) 
     (-1)^s y^{m(k+1)+s} (\ldots ) \, ,
\end{displaymath}
where $(\ldots)$ doesn't involve $y$.  But the constant
term of the last expression is $0$ since\break $m(k+1)+s>0$ if $k,s
\geq 0$ and $m(k+1)+s \leq s<0$ if $k,s <0$.  Thus, constant 
term of $I\!I_{m-1}=0$.  Furthermore, we have for $0 \leq i \leq
m-2$:
\begin{displaymath}
  e^{-\tilde{\rho}}t_{k\alpha_*} 
  r_{\alpha_*+\alpha_1+\cdots+\alpha_i}
  \frac{e^{\tilde{\rho}}}{1+e^{-\alpha_m}}=
  y^{-mk+i+1} (\ldots) \, .
\end{displaymath}
Since $-mk+i+1 \neq 0$ if $0 \leq i \leq m-2$, we see that constant term 
of $I\!I_i$ is $0$ for all $i$.  Finally:
\begin{displaymath}
  e^{-\tilde{\rho}}t_{k\alpha_*}
  \frac{e^{\tilde{\rho}}}{1+e^{-\alpha_m}} =
  y^{-mk} \frac{q^{k^2m+k}}{1+e^{-\alpha_m}}\, ,
\end{displaymath}
hence the constant term of this expression is equal
$\delta_{k,0} (1+e^{-\alpha_m})^{-1}$, which proves (\ref{eq:5.10}).

Using odd reflections one may derive from (\ref{eq:5.11}) or (\ref{eq:5.12}) the Weyl-type
character formulas for all other level~$1$ principal integrable
modules.  For example, in the case $\hat{\fg} = s \ell
(m|1)^{\hat{}}$, we let $k=mj+s \in \ZZ_+$, where $j \in \ZZ_+$
and $0 \leq s \leq m-1$; then
\begin{eqnarray}
  \label{eq:5.13}
  e^{\rho}Rch((k+1)\Lambda_0-k\Lambda_m) &=&
  \sum_{w \in \hat{W}} \epsilon (w)w
  \frac{e^{\Lambda +\rho}}
       {1+e^{-j\delta - \alpha_{m-s}-\ldots - \alpha_m}} \, , \\
\label{eq:5.14}
   e^{\rho}Rch (-k \Lambda_0 + (k+1) \Lambda_m) &=&
   \sum_{w \in \hat{W}} \epsilon (w) w
   \frac{e^{\Lambda + \rho}}
      {1+e^{-j\delta -\alpha_0 -\ldots -\alpha_s}} \, , 
\end{eqnarray}
and for $ 1 \leq j \leq m-1$ we have:
\begin{equation}
\label{eq:5.15}
e^{\rho} Rch (\Lambda_j) =
\sum_{w \in \hat{W}} \epsilon (w)w
\frac{e^{\Lambda_j+\rho}}{1+e^{-\alpha_m}} \, .
\end{equation}

\section{Classification of integrable highest weight modules over 
  affine superalgebras.}
\label{sec:6}

In this section we consider affine superalgebras of type
$A(m,n)^{\hat{}}$, $B(m,n)^{\hat{}}$, $C(n)^{\hat{}}$,
$D(m,n)^{\hat{}}$, $D(2,1;a)^{\hat{}}$, $F(4)^{\hat{}}$,
$G(3)^{\hat{}}$.  We shall exclude from consideration the well
understood case of $B(0,n)^{\hat{}}$ (see \cite{K2} and Section~\ref{sec:9.5}).  In all
cases except for $A(n,n)^{\hat{}}\,\,$ these are the affine superalgebras
$\hat{\fg}$ defined by (\ref{eq:0.1}), (\ref{eq:0.2}) for $\fg =
A(m,n)(m \neq n), B(m,n), C(n), D(m,n), D(2,1;a), F(4)$ and $
G(3)$ respectively (see \cite{K1} for a construction of the
simple finite-dimensional Lie superalgebras $\fg$).  In the
$A(n,n)^{\hat{}}$ case it is more convenient to take $\fg = s
\ell (n+1|n+1)$ in order not to lose the most interesting
modules.  The Lie superalgebra $\fg$ carries a unique, up to a
constant factor, non-zero invariant bilinear 
form $( .|.)$.  This form extends to $\hat{\fg}$ by formula~(\ref{eq:0.3})and it is normalized by the values of $(\alpha_i |
\alpha_i)$, given in Table~\ref{tab:1} (see below).

It is convenient to depict Cartan matrices of affine
(super)algebras by (generalized) Dynkin diagrams
(cf.~\cite{K1}).  We shall assume that the diagonal entries of a
Cartan matrix $\hat{A}$ are always $2$ or $0$ (one can achieve
this by rescaling simple coroots).  The Dynkin diagram of
$\hat{A}$ is a graph whose nodes label the index set $\hat{I}=\{ 0,1,2, 
\ldots \}$ and are of the form $0$, $\otimes$ or {\raisebox{-.75ex}{\Huge{\textbullet}}}
corresponding to cases $a_{ii} =2$, $i \not\in \hat{I}_1$;
$a_{ii}=0$ (then $i \in \hat{I}_1$); and $a_{ii}=2$, $i \in \hat{I}_1$,
respectively.  These nodes are called white, grey and black
respectively, so that $\hat{I}_1$ consists of non-white nodes.
We let $I=\hat{I} \backslash \{ 0 \}$, $I_1=\hat{I}_1 \backslash
\{ 0 \}$.  As usual, $I$ labels simple roots
$\alpha_1,\alpha_2,\ldots$ of $\fg$, $I_1$ labels odd simple
roots of $\fg$, and $\alpha_0=\delta-\theta$, where $\theta$ is
the highest root of $\fg$.  In the cases $a_{ij}=a_{ji}=0, i \neq j$, the
$i$\st{th} and $j$\st{th} nodes are not connected.  In the cases
$a_{ii}=a_{jj}=2 $, $i \neq j$, the nodes are, as usual, connected by $|
a_{ij}a_{ji}|$ edges with an arrow pointing to $j$\st{th} node if 
$|a_{ji}|>1$.  In the remaining cases the nodes are joined as
follows:
\begin{displaymath}
  \otimes \!\!-\!\!\bigcirc = \!\left( 
    \begin{array}{rr}
0 & a \\ -1 & 2
    \end{array} \right)\! ,\,
\otimes\!\! \Rightarrow\!\! \bigcirc = \!\left(
\begin{array}{rc}
0 & a\\ -2 & 2
\end{array} \right)\! = \otimes \!\!\Rightarrow \!\!
\raisebox{-.5ex}{\hbox{{\Huge{\textbullet}}}} , \,
\otimes\!\! - \!\!\otimes = \!\left(
  \begin{array}{cc}
0 & a\\ b & 0
  \end{array}\right) \!,\,
\bigcirc \!\!\Rightarrow
\!\!\raisebox{-.5ex}{\hbox{{\Huge{\textbullet}}}} =
\!\left(
\begin{array}{rr}
2 \!\!&\!\! -1\\ -2 \!\!& \!\!2
\end{array} \right).
\end{displaymath}
%
%We shall omit the label $a$ of an edge if $a=-1$ and often denote 
%the graph $\otimes \overset{2}{-}\bigcirc$ by $\otimes \Leftarrow 
%\bigcirc$.

In Table~\ref{tab:1} below we list the Dynkin diagrams of the symmetrizable
Cartan matrices of the affine Lie superalgebras $\hat{\fg}$ under
consideration.  The labels against the nodes $i$ are $(\alpha_i |
\alpha_i)$, and the labels against the edges connecting $i$ and
$j$ are $(\alpha_i | \alpha_j)$.  Recall that $\alpha^{\vee}_i =
2 \alpha_i/ (\alpha_i | \alpha_i)$ if $a_{ii} \neq 0$; we let
$\alpha^{\vee}_i = \alpha_i$ if $a_{ii}=0$.  We also give the
coefficients of the decomposition of the root $\delta$ in terms
of simple roots.  The nodes are numbered by $\hat{I}=\{ 0,1,
\ldots \}$ in increasing order from left to right, except when it
is impossible to do, in which case nodes are numbered by the
subscripts of their labels.

\begin{table}[htbp]
  \begin{center}
    \caption{}
    \label{tab:1} %%%{Table 1.}
\begin{tabular}{l c c}
\hspace*{3ex}$\hat{\fg}$ & Dynkin diagram & $\delta$\\
\hline\\[-8ex]
\raisebox{2ex}{$A(m,n)^{\hat{}}$} &
$
\setlength{\unitlength}{0.2in}
\begin{picture}(20,8)
\put(0,1.5){\line(4,1){9.6}}
\put(20,1.5){\line(-4,1){9.6}}
\put(10,5){\makebox(0,0)[c]{0}}
\put(10,4){\makebox(0,0)[c]{$\bigotimes$}}
\put(0,1){\makebox(0,0)[c]{$\bigcirc$}}
\put(2,1){\makebox(0,0)[c]{$\bigcirc$}}
\put(9,1){\makebox(0,0)[c]{$\bigcirc$}}
\put(11,1){\makebox(0,0)[c]{$\bigotimes$}}
\put(13,1){\makebox(0,0)[c]{$\bigcirc$}}
\put(20,1){\makebox(0,0)[c]{$\bigcirc$}}
\put(0.4,1){\line(1,0){1.2}}
\put(2.4,1){\line(1,0){1.2}}
\put(7.4,1){\line(1,0){1.2}}
\put(9.4,1){\line(1,0){1.2}}
\put(11.4,1){\line(1,0){1.2}}
\put(13.4,1){\line(1,0){1.2}}
\put(18.4,1){\line(1,0){1.2}}
\put(5.5,1){\makebox(0,0)[c]{$\cdots$}}
\put(16.5,1){\makebox(0,0)[c]{$\cdots$}}
\put(0,0){\makebox(0,0)[c]{\raisebox{-1.75ex}{$2_1$}}}
\put(2,0){\makebox(0,0)[c]{$2$}}
\put(9,0){\makebox(0,0)[c]{$2$}}
\put(11,0){\makebox(0,0)[c]{\raisebox{-1.5ex}{$0_{m+1}$}}}
\put(13,0){\makebox(0,0)[c]{$-2$}}
\put(20,0){\makebox(0,0)[c]{$-2_{m+n+1}$}}
\put(1,1.35){\makebox(0,0)[c]{$-1$}}
\put(3,1.35){\makebox(0,0)[c]{$-1$}}
\put(8,1.35){\makebox(0,0)[c]{$-1$}}
\put(10,1.35){\makebox(0,0)[c]{$-1$}}
\put(12,1.35){\makebox(0,0)[c]{$1$}}
\put(14,1.35){\makebox(0,0)[c]{$1$}}
\put(19,1.35){\makebox(0,0)[c]{$1$}}
\put(5,3.4){\makebox(0,0)[c]{$-1$}}
\put(15,3.4){\makebox(0,0)[c]{$1$}}
\end{picture}
$
& \raisebox{2ex}{$\displaystyle{\sum^{m+n+1}_{i=0}} \alpha_i$ }\\
\\[4ex]
$B(0,n)^{\hat{}}$ 
&  \arraycolsep=-1pt 
$\begin{array}[]{ccccccccccccccccccccccccccccccccccccccccc}
 & -2 &  & -1 &
     &  &  &     -1 & & -1 && &&\\
\bigcirc& \Rightarrow& \bigcirc &-&
     \bigcirc & - & \,\ldots\, & - & \bigcirc &
     \Rightarrow &   \raisebox{-.5ex}{\hbox{{\Huge{\textbullet}}}} \\
4 &&2&&2&&&&2&&1
 \end{array}$
& $\displaystyle{\alpha_0 +2  \sum^{n}_{i=1}} \alpha_i$ \\[5ex]
$B(m,n)^{\hat{}}$ 
 &  \arraycolsep=-1pt 
$\begin{array}[]{ccccccccccccccccccccccccccccccccccccccccc}
 & 2 &  & 1 &
     & 1 &  & 1 & & 1 &
      & -1 & & -1 &&-1 &&-1\\
\bigcirc& \Rightarrow& \bigcirc &-&
     \bigcirc & - & \ldots & - & \bigcirc &-&
     \bigotimes & - & \bigcirc &-&\ldots & - & \bigcirc &
     \Rightarrow & \bigcirc\\
-4 &&-2&&-2&&&&-2&&0_n&&2&&&&2&&1
 \end{array}$
& $\displaystyle{\alpha_0 +2  \sum^{m+n}_{i=1}} \alpha_i$ \\
$m>0$\\[2ex]
$C(n)^{\hat{}}$ 
&  \arraycolsep=-1pt 
$\begin{array}[]{rcccccccccccccccccccccc}
  \bigotimes\\[-.5ex]
  -2| & \diagdown1\\[-1ex]
 & 1 && 1 &&&&1 && 2\\[-1ex]
      \bigotimes & - & \bigcirc & - & \bigcirc & -
     & \ldots & - & \bigcirc & \Leftarrow & \bigcirc\\
    0_1 && -2 && -2 &&&& -2 && -4_n  \end{array}$
& $\displaystyle{\alpha_0 + \alpha_1 +2 \sum^{n-1}_{i=2} \alpha_i + \alpha_n}$\\[-8ex]
%%%%%%%%%%%%%%%%%%
%
\raisebox{4ex}{$D(m,n)^{\hat{}}$ }& 
$
\setlength{\unitlength}{0.15in}
\begin{picture}(20,8)(-1,0)
%%%%%%%%%%%%%%%
\put(.5,2){\makebox(0,0)[c]{$\bigcirc$}}
\put(1.25,2){\makebox(0,0)[c]{$\Rightarrow$}}
\put(2,2){\makebox(0,0)[c]{$\bigcirc$}}
\put(2.5,2){\line(1,0){1.1}}
\put(4.4,2){\makebox(0,0)[c]{$\cdots$}}
\put(5,2){\line(1,0){1.1}}
\put(6.5,2){\makebox(0,0)[c]{$\bigcirc$}}
\put(7,2){\line(1,0){1.1}}
\put(8.5,2){\makebox(0,0)[c]{$\bigotimes$}}
\put(9,2){\line(1,0){1.1}}
\put(1.25,2.75){\makebox(0,0)[c]{$2$}}
\put(2.75,2.75){\makebox(0,0)[c]{$1$}}
\put(5.5,2.75){\makebox(0,0)[c]{$1$}}
\put(7.5,2.75){\makebox(0,0)[c]{$1$}}
\put(9.5,2.75){\makebox(0,0)[c]{$-1$}}
\put(0.5,1.15){\makebox(0,0)[c]{$-4$}}
\put(2,1.15){\makebox(0,0)[c]{$-2$}}
\put(6.25,1.15){\makebox(0,0)[c]{$-2$}}
\put(8.75,1.15){\makebox(0,0)[c]{$0_n$}}
\put(10.5,1.15){\makebox(0,0)[c]{$2$}}
\put(15,1.15){\makebox(0,0)[c]{$2$}}
\put(17,1.15){\makebox(0,0)[c]{$2$}}
\put(10.5,2){\makebox(0,0)[c]{$\bigcirc$}}
\put(11,2){\line(1,0){1.2}}
\put(13,2){\makebox(0,0)[c]{$\ldots$}}
\put(13.5,2){\line(1,0){1.1}}
\put(15,2){\makebox(0,0)[c]{$\bigcirc$}}
\put(15.5,2){\line(1,0){1.1}}
\put(17,2){\makebox(0,0)[c]{$\bigcirc$}}
\put(14,2.75){\makebox(0,0)[c]{$-1$}}
\put(15,2.2){\line(0,1){1.1}}%%%vertical line
\put(15,3.75){\makebox(0,0)[c]{$\bigcirc$}}
\put(16.5,4){\makebox(0,0)[c]{$2_{m+n}$}}
\put(15.75,2.75){\makebox(0,0)[c]{$-1$}}
\end{picture}
$
& \raisebox{4ex}{\hspace{-2ex}$\displaystyle{\alpha_0 + 2 \sum^{m+n-2}_{i=1} \alpha_i +
\alpha_{m+n-1} + \alpha_{m+n}}$}\\[-10ex]
%%%%%%%%%%%%%%%%%%%%%%
\raisebox{2ex}{$D(2,1;a)^{\hat{}}$} &
\setlength{\unitlength}{0.15in}
\begin{picture}(20,8)(-8,1)
%%%%%%%%%%%%%%%
\put(1,1){\makebox(0,0)[c]{$2_2$}}
\put(3.2,1){\makebox(0,0)[c]{$0_1$}}
\put(5.2,1){\makebox(0,0)[c]{$2a_3$}}
\put(1,2){\makebox(0,0)[c]{$\bigcirc$}}
\put(1.5,2){\line(1,0){1.1}}
\put(3,2){\makebox(0,0)[c]{$\bigotimes$}}
\put(3.5,2){\line(1,0){1.1}}
\put(5,2){\makebox(0,0)[c]{$\bigcirc$}}
\put(1.75,2.5){\makebox(0,0)[c]{$-1$}}
\put(3.95,2.5){\makebox(0,0)[c]{$-a$}}
\put(3,4){\makebox(0,0)[c]{$\bigcirc$}}
\put(5.5,4){\makebox(0,0)[c]{$-2(a+1)$}}
\put(3,2.5){\line(0,1){1.2}}%%%vertical line
\put(4.35,3.3){\makebox(0,0)[c]{$a+1$}}
\end{picture}
&\raisebox{2ex}{ $\alpha_0 +2\alpha_1+\alpha_2+\alpha_3$}\\
%%%%%%%%%%%%%%%%%%%%%%
%%%%%%%%%%%%%%%%%%%%%%
\raisebox{4ex}{$F(4)^{\hat{}}$} &
\setlength{\unitlength}{0.2in}
\begin{picture}(5,6)(2,0)
\put(1,2){\makebox(0,0)[c]{$\bigcirc$}}
\put(1.5,2){\line(1,0){1.1}}
\put(3,2){\makebox(0,0)[c]{$\bigotimes$}}
\put(3.5,2){\line(1,0){1.1}}
\put(5,2){\makebox(0,0)[c]{$\bigcirc$}}
\put(6,2){\makebox(0,0)[c]{$\Longleftarrow$}}
\put(7,2){\makebox(0,0)[c]{$\bigcirc$}}
\put(7.5,2){\line(1,0){1.1}}
\put(9,2){\makebox(0,0)[c]{$\bigcirc$}}

\put(1.95,2.75){\makebox(0,0)[c]{$\tfrac{3}{2}$}}
\put(3.75,2.75){\makebox(0,0)[c]{$-\tfrac{1}{2}$}}
\put(6,2.75){\makebox(0,0)[c]{$-1$}}
\put(7.75,2.75){\makebox(0,0)[c]{$-1$}}
\put(1,1){\makebox(0,0)[c]{$-3$}}
\put(3,1){\makebox(0,0)[c]{$0$}}
\put(5,1){\makebox(0,0)[c]{$1$}}
\put(7,1){\makebox(0,0)[c]{$2$}}
\put(9,1){\makebox(0,0)[c]{$2$}}
\end{picture}
&\raisebox{4ex}{$\alpha_0 + 2 \alpha_1+ 3 \alpha_2 + 2 \alpha_3 + \alpha_4$}\\[4ex]
%%%%%%%%%%%%%%%%%%%%%%%%%%%%%%%%%%%%%%%%%%
$G(3)^{\hat{}}$ & 
   \arraycolsep=-1pt 
$\begin{array}[]{crcrcrcccccclccclcccccc}
  & \tfrac{4}{3} && -\tfrac{1}{3} && -1\\
\bigcirc &-\!- & \bigotimes & -\!- & \bigcirc & \Lleftarrow & \bigcirc
\\
-\tfrac{8}{3} && 0 && \tfrac{2}{3} && 2
 \end{array}$
& $\alpha_0 + 2 \alpha_1 + 4 \alpha_2 +2 \alpha_3$\\[8ex]
%%
%$Q(n)^{\hat{}}$ & \\
\end{tabular}
  \end{center}
\end{table}

\begin{remark}
  \label{rem:6.1}
Recall the definition of the orthosymplectic Lie subalgebra $ osp 
(M|N)$ \cite{K1}.  Let $V=V_{\bar{1}} \oplus V_{\bar{0}}$ be a
superspace, where $\dim V_{\bar{1}}=N, \dim V_{\bar{0}}=M$, and let
 $(.|.)$ be a non-degenerate bilinear  form on $V$ such
that $(V_{\bar{0}} | V_{\bar{1}})=0$, the restriction of $(.|.)$ 
to $V_{\bar{1}}$ is symmetric and to $V_{\bar{0}}$ is
skewsymmetric, so that $N=2n$ is even; let $m=[M/2]$.  Then
$(\alpha = 0,1)$:
\begin{eqnarray*}
  osp (M|N)_{\bar{\alpha}} = \{ a \in g \ell (M|N)_{\bar{\alpha}}
     | (a(x)|y)+ (-1)^{\alpha p(x)} (x|a(y))=0, \quad x,y \in V \} \, .
\end{eqnarray*}
For the definition in a matrix form, consider the following
$(M+N) \times (M+N)$ matrices:
\begin{displaymath}
  C= \left(
    \begin{array}{ll}
      C_1 & 0 \\
      0 & C_2
    \end{array} \right) , \quad 
  F= \left(
    \begin{array}{ll}
      I_M & 0 \\
      0 & -I_N
    \end{array} \right) \, ,
\end{displaymath}
where $C_1$ (resp. $C_2$) is a $M \times M$ (resp. $N \times N$) 
symmetric (resp. skewsymmetric) matrix.  Then 
\begin{displaymath}
  osp (M|N)_{\bar{\alpha}} = \{
    a \in g \ell (M|N)_{\bar{\alpha}} |
    F^{\alpha}a^{\intercal} C+Ca =0 \},\quad  \alpha =0,1 \, .
\end{displaymath}
Recall that $B(m,n) = osp (2m+1|2n)$, $C(n) =osp (2|2n)$ and
$D(m,n)=osp (2m|2n)$.  The invariant bilinear form on $osp
(M|N)$ that is used in Table~\ref{tab:1} and throughout the
paper is
\begin{displaymath}
  (a|b) = \tfrac{1}{2} \str ab \, .
\end{displaymath}

\end{remark}

\begin{table}[htbp]
  \begin{center}
    \caption{  } %% {Table 2.}
    \vspace{1ex}
    \label{tab:2}
\begin{tabular}{c c|c c}
  \hline\hline
     $\fg$ & $\fg_{\bar{0}}$ & $\fg$ &  $\fg_{\bar{0}}$\\[1ex]
\hline
$A(m,n)$ & $A_m + A_n + \CC$ & $D(2,1;a)$ & $D_2 + A_1$\\[2ex]
$C(n)$ & $\CC + C_n$ & $F(4)$ & $ B_3 + A_1$\\[2ex]
$B(m,n)$ & $B_m + C_n$ & $G(3)$ & $G_2 + A_1$\\[2ex]
$D(m,n)$ & $D_m + C_n$ & &\\
    \hline
\end{tabular}
  \end{center}
\end{table}
\noindent

The even parts $\fg_{\bar{0}}$ of the Lie superalgebras $\fg$ are listed
in Table~\ref{tab:2}.  In the case of $D(2,1;a)$, the subalgebra $D_2$ corresponds to
$\alpha_2$ and $\alpha_3$ (see Table~\ref{tab:1}).  We denote by
$\fg'_{\bar{0}}$ (resp. $\fg''_{\bar{0}}$) the first
(resp. second) non-zero summand of $\fg_{\bar{0}}$ in the
decomposition of Table~\ref{tab:2}.  Note that the invariant
bilinear form $(.|.)$ (which can be read off from
Table~\ref{tab:1}) is normalized in such a way that it is
positive definite on $\fg'_{\bar{0}}$ and negative definite on
$\fg''_{\bar{0}}$ (except that for $D(2,1;a)$ we should assume
that $a \in \RR_{>}$), and the maximal square length of a root is 
$2$, except for the cases $B(1,n)$ when it is $1$ and $D(2,1;a)$
when it is max $(2,2a)$.  If the Killing form on $
\fg$ is non-degenerate, then the form $(.|.)$ is a positive
(resp. negative) multiple
of the Killing form, in the cases $\fg \cong s \ell (m|n)$ with
$m>n$, $osp (m|n)$ with $m>n+2$, $F(4)$ and $G(3)$ (resp. $s \ell 
(m|n)$ with $m<n$ and $osp (m|n)$ with $m<n+2$).

An irreducible highest weight module over $\hat{\fg}$ is called
\emph{principal} (resp. \emph{subprincipal}) \emph{integrable}
module if it is integrable with respect to $\hat{\fg'_{\bar{0}}}$ 
(resp. $\hat{\fg''_{\bar{0}}}$) and locally finite with respect to 
$\fg$ (cf.~Definition~0.1).  As we shall see, the non-trivial
principal (resp. subprincipal) highest weight modules have
positive (resp. negative) level, except for the cases
$\hat{\fg}=A (0,n)^{\hat{}}$ and $C(n)^{\hat{}}$
(resp. $A(n,0)^{\hat{}}$). It is easy to see that in these cases
the only conditions of integrability are $k_i \in \ZZ_+$ if $i
\in I \backslash I_1$; we shall exclude these cases from further considerations. 

Let $\theta'$ be the highest root of $\fg'_{\bar{0}}$; in cases $
D(2,n)$ and $D(2,1;a)$, which are the only cases when
$\fg'_{\bar{0}}$ is not simple, we have:
$\fg'_{\bar{0}}=A_1+A_1$, and the highest roots are
$\theta'_+=\alpha_{n+1}$ and $\theta'_-=\alpha_{n+2}$ (where $n=1$ 
for $D(2,1;a)$).  The root $\theta'$ (resp. root $\theta'_{\pm}$)
gives rise to a simple root $\alpha'_0=\delta-\theta'$
(resp. simple roots $\alpha'_{0\pm}=\delta -\theta'_{\pm}$) of
$\fg'_{\bar{0}}$.  The corresponding coroot is
$\alpha^{\prime\vee}_0 = 2 \alpha'_0 / (\alpha'_0 | \alpha'_0)$.  

In all cases except for $A(m,n)^{\hat{}}$ and $C(n)^{\hat{}}$
there is a (unique) simple root $\alpha''$ of $\fg''_0$ (which
is a simple Lie algebra), which is not a simple root of $\fg$
\cite{K1}.  As we have seen in \S\ref{sec:2}, the principal
integrability in the case of $A(m,n)^{\hat{}}$ follows from local 
nilpotency of the root vector $f'$ attached to the root
$-\alpha'_0$.  In all
other cases one has to check in addition the local nilpotency of
the root vector $f''$ attached to the root $-\alpha''$ (in order
to ensure the local finiteness with respect to
$\fg''_{\bar{0}}$; that with respect to $\fg'_{\bar{0}}$ follows 
automatically from the integrability with respect to $\hat{\fg'_{\bar{0}}}$).  For that reason, as we have seen in
\S\ref{sec:1}, it is important to introduce the following
numbers, where $\rho$ is a Weyl vector for $\hat{\fg}$:
\begin{displaymath}
  b'=-\langle \rho, \alpha^{\prime\vee}_0 \rangle ,\quad
  b'_{\pm}=- \langle \rho,
  \alpha^{\prime\vee}_{0^{\pm}}\rangle ,\quad
  b''=- \langle \rho, \alpha^{\prime\prime\vee} \rangle \, .
\end{displaymath}
The values of the numbers $b'$ and $b'_{\pm}$ (resp. $b''$) are
given in Table~\ref{tab:3} (resp.~4), the first line for $D(2,
\ldots)^{\hat{}}$ in Table~\ref{tab:4} being for $b'_+$ and the second
for $b'_-$.  Table~\ref{tab:3} contains also the formula for the
\emph{partial levels}
\begin{displaymath}
  k'= \langle \Lambda , \alpha^{\prime\vee}_0 \rangle \, ,
  k'_{\pm} = \langle \Lambda , \alpha^{\prime\vee}_{0 \pm} \rangle \, ,
\end{displaymath}
and the level $k=\langle \Lambda ,K \rangle$ of a weight
$\Lambda$ in terms of its labels $k_i$ and $k'\, , k'_{\pm}$.  Table~\ref{tab:4}
contains also a formula for 
\begin{displaymath}
  k''=\langle \Lambda , \alpha^{\prime\prime\vee}\rangle
\end{displaymath}
and the level $k$ in terms of the $k_i$ and $k''$.

\begin{table}[htbp]
  \begin{center}
    \caption{ } %% {Table 3.}
    \vspace{1ex}
    \label{tab:3}
\begin{tabular}{c| c| c| c  }
  \hline\hline
$\hat{\fg}$ & $k'$ & $b'$ & $k$\\
\hline
$A(m,n)^{\hat{}}$ & $\displaystyle{k_0+k_{m+1}- \sum^{m+n+1}_{i=m+2} k_i}$
    & $n$ & $\displaystyle{k'+\sum^m_{i=1} k_i}$\\[1ex]
$(m \geq 1)$&&&\\[3ex]
%
%$B(0,n)^{\hat{}}$ & $\frac{1}{2} k_n$ & $-\frac{1}{2}$
%    & $k' + \displaystyle{\sum^{n-1}_{i=0} k_i}$\\[2ex]
%
$B(1,n)^{\hat{}}$ & $\displaystyle{4k_n +k_{n+1}-4 \sum^{n-1}_{i=0} k_i}$
    & $4n-1$ & $\frac{1}{2} (k'+k_{n+1})$\\[3ex]
$B(m,n)^{\hat{}}$ & $\displaystyle{2k_n + k_{n+1}-2 \sum^{n-1}_{i=0} k_i}$
    & $2n-1$ & $\displaystyle{k'+k_{n+1} + 2\sum^{n+m-1}_{i=n+2}
        k_i + k_{n+m}}$ \\
$(m \geq 2)$&&&\\[3ex]
$D(2,n)^{\hat{}}$ & $\displaystyle{2k_n+k_{n+1}-2 \sum^{n-1}_{i=0} k_i}$
    & $2n-1$ & $k'_+ + k_{n+2}$\\[1ex]
    & $\displaystyle{2k_n+k_{n+2}-2 \sum^{n-1}_{i=0} k_i}$ & $2n-1$
    & $k'_- + k_{n+1}$\\[3ex]
$D(m,n)^{\hat{}}$ & $\displaystyle{2k_n +k_{n+1} -2 \sum^{n-1}_{i=0} k_i}$
    & $2n-1$ & $\displaystyle{k'+k_{n+1}+2 \sum^{m+n-2}_{i=n+2}
        k_i+k_{n+m-1} + k_{n+m}}$\\[3ex]
$D(2,1;a)^{\hat{}}$ & $-(a+1)k_0+2k_1+ak_3$
    & $1$ & $k'_+ + k_2$ \\[2ex]
    & $a^{-1} (-(a+1) k_0 +2k_1 + k_2)$ & $1$
    & $a(k'_- + k_3)$\\[2ex]
$F(4)^{\hat{}}$ & $-\frac{3}{2} k_0 + 2k_1 + \frac{1}{2} k_2$
    & $1$ & $k'+k_2 +2k_3 + k_4 $\\[2ex]
$G(3)^{\hat{}}$ & $-\frac{4}{3} k_0+2k_1+\frac{1}{3}k_2$
    & $1$ & $k'+k_2+k_3$\\[1ex]
     \hline
\end{tabular}
  \end{center}
\end{table}

\begin{table}[htbp]
  \begin{center}
    \caption{ } %% {Table 4.}
    \vspace{1ex}
    \label{tab:4}
\begin{tabular}{c| c| c | c }
  \hline\hline
     $\hat{\fg}$ & $k''$ & $b''$ & k \\
\hline
%$B(0,n)^{\hat{}}$ & $\frac{1}{2} k_n$ & $-\frac{1}{2}$\\[2ex]
%
$B(m,n)^{\hat{}}, m \geq 1$ & $\displaystyle{-k_n-\sum^{n+m-1}_{i=n+1}
k_i - \tfrac{1}{2} k_{m+n}}$ & $m-\tfrac{1}{2}$ &
$\displaystyle{-2(k''+ \sum^{n-1}_{i=0}k_i )}$\\[3ex]
$C(n)^{\hat{}}$ & $-\tfrac{1}{2}(k_0 + k_1)$ & 0 
    & $\displaystyle{-2(k'' +  \sum^{n}_{i=2} k_i)}$\\[3ex]
$D(m,n)^{\hat{}}$ & $\displaystyle{-k_n-\sum^{n+m-2}_{i=n+1}
k_i-\tfrac{1}{2} (k_{m+n-1}+k_{m+n})}$ & $m-1$ &
$\displaystyle{-2(k'' + \sum^{n-1}_{i=0} k_i)}$\\[3ex]
$D(2,1;a)^{\hat{}}$ & $-(a+1)^{-1} (2k_1+k_2+ak_3)$
    & $1$ & $-(a+1)(k'' + k_0)$\\[2ex]
 $F(4)^{\hat{}}$ & $-\frac{4}{3} k_1-k_2-\frac{4}{3}
       k_3 - \frac{2}{3} k_4$ & $3$ & $-\frac{3}{2} (k'' +k_0)$\\[3ex]
$G(3)^{\hat{}}$ & $-\frac{3}{2} k_1-k_2-\frac{3}{2}k_3$
    & $\frac{5}{2}$ & $-\frac{4}{3} (k''+k_0)$\\[1ex]
     \hline
\end{tabular}
  \end{center}
\end{table}

\begin{theorem}
  \label{th:6.1}
  For an affine superalgebra from Table~\ref{tab:1} (recall that
  $A(0,n)\sphat$, $B(0,n)\sphat$ and $C(n)\sphat$ are excluded) the labels
  $\{k_i\}_{i=\hat{I}}$ of the highest weight of a principal integrable
  irreducible highest weight module $ L (\Lambda)$ are characterized by the
  following four series of conditions:

\arabicparenlist
\begin{enumerate}
\item %%1
$k_i \in \ZZ_+$ if $i \in I \backslash I_1$,

\item %%2
$k'$ (resp. $k'_{\pm}$ in case $D(2, \ldots)^{\hat{}}\, ) \in
\ZZ_+$, $k'' \in \ZZ_+$ (see Tables~\ref{tab:3} and~\ref{tab:4}),

\item %%3
if $k'$ (resp. one of the $k'_{\pm}$) $\leq b'$ (see Table~\ref{tab:4}),
there are the supplementary conditions:

\begin{list}{}{}
\item $A(m,n)^{\hat{}}, m \geq 1$:  there exists $s \in \ZZ_+$, $s \leq k'$, such
  that:
  \begin{displaymath}
    k_{m+1}=k_{m+2} + \cdots + k_{m+1+s}+s, \, \, 
    k_{m+s+2}= \cdots = k_{m+s+1+n-k'}=0 \, ,
  \end{displaymath}

%\item $C(n)^{\hat{}}$: $k_0=k_1=0$ if $k'=0$,\vspace{1ex} 

\item $B(m,n)^{\hat{}}$, $m \geq 2$, and $D(m,n)^{\hat{}}, m \geq 3$:  one of the
  four possibilities hold:

\romanlistii
\begin{enumerate}
\item %%i
  there exist $r,s \in \ZZ_+$, $r<s$, such that
  \begin{displaymath}
    k'=r+s , \, \, k_j=0 \hbox{ for } r+1 \leq j \leq m+n
      \hbox{ and } j \neq s, \,\, k_s=1 \, ,
  \end{displaymath}

\item %%ii
there exist $r,s \in \ZZ_+$, $r \leq s$, such that
\begin{displaymath}
  k'=r+s, \,\, k_j=0 \hbox{ for } r+1 \leq j \leq m+n ,\,  
           k_r  \neq 0 \, ,
\end{displaymath}

\item %%iii
there exist $r \in \ZZ_+$ such that
\begin{displaymath}
  k'=n+r, \,\, k_j=0 \hbox{ for } r+1\leq j \leq n-1, \,\,
  k_n \neq 0  , \,  k_n+k_{n+1}+1 =0\, ,
\end{displaymath}

\item %%iv
  there exist $r \in \ZZ_+$ such that
  \begin{displaymath}
    k' \geq n+r , \,\, k_j=0 \hbox{ for }
       j \geq r+1 , \, k_r \neq 0 \, .
  \end{displaymath}

\end{enumerate}
 \item $B(1,n)^{\hat{}}$:  the same as for $B(m,n)^{\hat{}}$ with $m>1$ with the
   following changes:  $k'$ is replaced by $\frac{1}{2}k'$
   everywhere and $k_{n+1}$ is replaced by $\frac{1}{2} k_{n+1}$
   in~(iii),
\vspace{1ex}

\item  $D(2,n)^{\hat{}}$:  the same as for $D(m,n)^{\hat{}}$ with $m>2$ with the
  following additions:
  \begin{displaymath}
    k'_+ =k'_- , \,\, k_n+k_{n+2} +1=0 \hbox{ in (iii)} \, ,
  \end{displaymath}

\item $D(2,1;a)^{\hat{}}$:  one of the four possibilities holds:

\romanparenlist
\begin{enumerate}
\item %%i
  $k'_+ k'_-=0$, then all $k_i=0$,

%\item %%ii
 % if $a \not\in \QQ_{>0}$, then $k'_+ \neq 1, k'_- \neq 1$,

\item %%ii
  $a \in \QQ_{>0}$, $a,a^{-1} \not\in \NN$ and one of
  $k'_{\pm}$ equals $1$, then $k'_+ =k'_-=1$

 and one has

 \begin{list}{(*)}{}
\item $k_0=-(a+1)^{-1} r-1, k_1=-r, k_2=r-1, k_3 = a^{-1}r-1$ for
  some $r \in \NN \cap a \NN$,

 \end{list}

\item %%iii
$a^{-1} \in \NN$ and $k'_+=1$, then either (*) holds or
$k_0=-(a+1)^{-1}, k_1=k_2=k_3=0$,

\item %%iv
  $a \in \NN$ and $k'_-=1$, then either (*) holds or $k_0
  =-a(a+1)^{-1}, k_1=k_2=k_3=0$,

\end{enumerate}

\vspace{1ex}

\item $F(4)^{\hat{}}$:  one of the two possibilities holds:
\romanlistii
\begin{enumerate}
\item %%i
  $k'=0$, then all $k_i=0$,

\item %%ii
  $k'=1$, then $k_0=-\tfrac{2}{3}, k_1=k_2=k_3=k_4=0$,
\end{enumerate}

\vspace{1ex}

\item $G(3)^{\hat{}}$:  one of the two possibilities holds:

\romanparenlist

\begin{enumerate}
\item %%i
  $k'=0$, then all $k_i=0$,

\item %%ii
  $k'=1$, then $k_0=-\tfrac{3}{4}, k_1=k_2=k_3=0$,
\end{enumerate}

\end{list}

\item %%(4)
 if $k'' \leq b''$ (see Table~\ref{tab:4}), there are the
  supplementary conditions:

  \begin{list}{}{}
  \item $B(m,n)^{\hat{}}$:  $k_j=0$ for all $j \geq n+k''+1$, 

\vspace{1ex}

\item  $C(n)^{\hat{}}$: $k_0=k_1=0$,

\vspace{1ex}

\item  $D(m,n)^{\hat{}}$:  one of the two possibilities holds:
\romanlistii
\begin{enumerate}
\item %%i
  $k'' \leq m-2$, then $k_j=0$ for all $j \geq n+k''+1$,

\item %%ii
  $k''=m-1$, then $k_{m+n-1}=k_{m+n}$,

\end{enumerate}

\vspace{1ex}

$D(2,1;a)^{\hat{}}$:  one of the two possibilities holds:

\romanlistii
\begin{enumerate}
\item %%i
  $k''=0$, then $k_1=k_2=k_3=0$,

\item %%ii
  $k'' = 1$, then  $a \in \QQ$ and $k_2+1=|a| (k_3+1)$,
\end{enumerate}

\vspace{1ex}
\item $F(4)^{\hat{}}$:  one of the three possibilities holds:

\romanlistii
\begin{enumerate}
\item %%i
$k''=0$, then $k_1=k_2=k_3=k_4=0$,

\item %%ii
$k''=2$, then $k_2=k_4=0$,

\item %%iii
$k''=3$, then $k_2=2k_4+1$,

\end{enumerate}

\vspace{1ex}

$G(3)^{\hat{}}$:  one of the two possibilities holds:

\romanlistii
\begin{enumerate}
\item %%i
$k''=0$, then $k_1=k_2=k_3=0$,

\item %%ii
$k''=2$, then $k_2=0$.
\end{enumerate}

  \end{list}

\end{enumerate}
\end{theorem}

\begin{proof}
  In the case  $\hat{\fg} =A (m,n)^{\hat{}}$, the theorem follows 
  from Theorem~\ref{th:2.1}.  In general, the proof is based on
  similar arguments.  Below we shall give details in the case
  $\hat{\fg}=B(1,n)^{\hat{}}$; in the rest of the cases arguments 
  are the same.

The even part of $B(1,n)$ is $A_1 +C_n$ and its simple roots are
$\alpha_{n+1}$ for $A_1$ and $\{ \alpha_1,\alpha_2, \ldots ,
\alpha_{n-1}$,\break$\alpha''= 2 (\alpha_n + \alpha_{n+1}) \}$ for
$C_n$.  The simple roots of $\hat{A}_1$ are $\{
\alpha'_0=\delta-\alpha_{n+1}, \alpha_{n+1} \}$.  Due to
Lemma~\ref{lem:1.5}a, the local finiteness (resp. integrability)
with respect to $C_n$ (resp. $\hat{A}_1$) implies that $k_1 ,
\ldots , k_{n-1}, k'' \in \ZZ_+$ (resp. $k_{n+1}, k' \in
\ZZ_+$).  Hence, conditions~(1) and~(2) are necessary.
Furthermore, it follows from Proposition~\ref{prop:1.2}, that in
the cases $k'>b'=4n-1$ (resp. $k''>b'' =\tfrac{1}{2}$) the element $f'$
(resp. $f''$) is locally nilpotent.  It remains to show that in
the case of inequality
\begin{equation}
  \label{eq:6.1}
  k' \leq 4n-1
\end{equation}
the element $f'$ is locally nilpotent iff condition~(3) holds,
and in the case $k''=0$, $f''$ is locally nilpotent iff~(4) holds.
We shall concentrate on the first claim, the second being easier
(cf. also \cite{K1}).

Introduce the following isotropic roots:
\begin{displaymath}
  \beta_j = \sum^n_{i=n-j} \alpha_i \,\,(j=0,\ldots ,n), \,\,
  \beta_{n+j}=\beta_n+ \sum^j_{i=1} \alpha_i\,\,
  (j=1,\ldots,n-1) \, .
\end{displaymath}
We have:  $\beta^{\vee}_j=\beta_j$ for all $j$, and
\begin{eqnarray}
  \label{eq:6.2}
(\beta_i | \alpha_{n+1}) = -1 \hbox{ for all } i,
(\beta_i | \beta_{i+1}) = \left\{
  \begin{array}{ccl}
    1 & \hbox{if} & 0 \leq i \leq n-2 , \\
    2 & \hbox{if} & i=n-1 \, .
  \end{array}\right.
\end{eqnarray}

Let $\Pi^{(0)} = \Pi$, then $\beta_0 \in \Pi^{(0)}$ and we let
$\Pi^{(1)} = r_{\beta_0} \Pi^{(0)}$.  Similarly $\beta_1 \in 
\Pi^{(1)}$ and we let $\Pi^{(2)} = r_{\beta_1} \Pi^{(1)},
\ldots , \Pi^{(2n)} = r_{\beta_{2n-1}} \Pi^{(2n-1)}$.  We
have:
\begin{equation}
  \label{eq:6.3}
  \alpha'_0 \in \Pi^{(2n)} , \,\, \alpha^{\prime\vee}_0 =
  2\alpha'_0 \, .
\end{equation}
Let $\Lambda^{(j)}$ denote the highest weight of $L(\Lambda)$
with respect to $\fn^{(j)}_+$.  It can be computed by making use
of Lemma~\ref{lem:1.4}.  Introduce the following numbers:
\begin{displaymath}
  u_j= \langle \Lambda^{(j)}, \alpha^{\prime\vee}_0 \rangle\,\,
(j=0,\ldots,2n), \quad t_j=\langle \Lambda^{(j)}, \beta_j \rangle 
\,\, (j=0,\ldots,2n-1) \, .
\end{displaymath}
Using (\ref{eq:6.2}), $\langle\Lambda,
\alpha^{\prime\vee}_0\rangle =k'$ 
and Lemma~\ref{lem:1.4}, we 
get the
following recurrent formula for the $u_j$'s:
\begin{equation}
  \label{eq:6.4}
  u_0=k', \,\, u_{j+1}=u_j-2 (\hbox{ resp. } =u_j) \hbox{ if }
  t_j \neq 0 (\hbox{ resp. } =0) \, .
\end{equation}
In view of Lemma~\ref{lem:1.5}, the local
nilpotency of $f'$ follows from $u_{2n} \in \ZZ_+$.  This,
clearly, holds if $k' \geq 4n$ (by (\ref{eq:6.4})), which again
shows that in this case there are no supplementary conditions.

>From now on we may assume that
$k' \leq 4n-1$.  We may also assume that conditions (1), (2) and
(4) hold.  We shall derive a recurrent formula for the $t_i$.
Using that, by Lemma~\ref{lem:1.4},
$\Lambda^{(i+1)}=\Lambda^{(i)}-\beta_i$ (resp. $=\Lambda^{(i)}$)
if $t_i \neq 0$ (resp. $=0$) and that $\alpha_{n-i-1}
\in \Pi^{(i)}$, we obtain:

\begin{subequations}\label{eq:6.5}
  \begin{equation}  
    \label{eq:6.5a}
    t_{i+1} = \left\{
      \begin{array}{lll}
        t_i-k_{n-i-1}-1 & \hbox{if} &  t_i \neq 0 \\
        t_i-k_{n-i-1} & \hbox{if} & t_i=0
      \end{array}\qquad (0 \leq i \leq n-2) \right.
  \end{equation}

  \begin{equation}  
    \label{eq:6.5b}
    t_n = t_{n-1}-2k_0-2 \,\, 
    (\hbox{resp. } =t_{n-1}-2k_0) \hbox{ if }
    t_{n-1} \neq 0 \,\, (\hbox{resp. }=0) \, .
  \end{equation}
\end{subequations}

For $t_j, \,\, j \geq n$, the recurrent formula involves numbers
\begin{displaymath}
  s_i= (\Lambda^{(n)}| \alpha_i) \hbox{ for }1 \leq i \leq n-1,\,\,
  s_n = (\Lambda^{(n)}| \alpha_n+\alpha_{n+1}) \, ,
\end{displaymath}
which, using the above arguments, can be expressed in terms of
the labels of $\Lambda$ as follows:
\begin{subequations}\label{eq:6.6} 
  \begin{equation}  
    \label{eq:6.6a}
    s_i = \left\{
      \begin{array}[]{lll}
        -k_i & \hbox{if} & t_{n-i-1}t_{n-i} \neq 0 \hbox{ or }
           t_{n-i-1}=t_{n-i}=0 \quad (1 \leq i \leq n-1)\, , \\
        -k_i -1 & \hbox{if} & t_{n-i-1} \neq 0\, , 
           t_{n-i} = 0 \, , \\
        -k_i +1 & \hbox{if} & t_{n-i-1}=0 , \,\, t_{n-i} \neq 0 \, ,
      \end{array}\right.
  \end{equation}
  \begin{equation}
    \label{eq:6.6b}
    s_n=-k'' +1 (\hbox{resp. }-k''), \hbox{ where }
    k''=-k_n-\tfrac{1}{2} k_{n+1}, \,\, \hbox{ if }k_n \neq 0 \,\,
    (\hbox{resp. }=0)\, .
  \end{equation}
\end{subequations}
Then we have $(0 \leq i \leq n-1)$:
\begin{subequations} \label{eq:6.7} 
\begin{equation} 
  \label{eq:6.7a}
  t_{n+i+1} = t_{n+i}+s_{i+1}-1 \,\,(\hbox{resp. } t_{n+i}+s_{i+1})
     \hbox{ if } t_{n+i} \neq 0 \,\,(\hbox{resp. }=0)\, ,
\end{equation}
where we let
\begin{equation}
  \label{eq:6.7b}
  t_{2n} = \tfrac{1}{2} u_{2n} \, .
\end{equation}
\end{subequations}

Note that $t_0=k_n \leq 0$ since
\begin{displaymath}
  k''=- k_n-\tfrac{1}{2} k_{n+1} \in \ZZ_+ \hbox{ and }
  k_{n+1} \in \ZZ_+ \, .
\end{displaymath}
Since $k_i \in \ZZ_+$ for $i \neq 0, n$, formulae~(\ref{eq:6.5}) imply
\begin{equation}
  \label{eq:6.8}
  0 \geq t_0 \geq t_1 \geq \cdots \geq t_{n-1} \, .
\end{equation}
Furthermore, we have
\begin{equation}
  \label{eq:6.9}
  t_n \geq t_{n+1} \geq \cdots \geq t_{2n}
     = \tfrac{1}{2} u_{2n} \,.
\end{equation}
In order to show this, it suffices to prove that $s_i \leq 0$, $1 
\leq i \leq n$.  But, due to (\ref{eq:6.6a}), $s_i >0$ can take
place for $1 \leq i \leq n-1$ only when $t_{n-i-1}=0, \,\, t_{n-i} 
  \neq 0, \, k_i=0$, which is impossible by (\ref{eq:6.5a}).
Also, $s_n =-k''+1 ($resp. $-k'')$ if $k_n \neq 0$ (resp. $
=0$) cannot be positive since in this case $k''=0$, which implies 
that $k_n=0$ (see supplementary conditions~(4)).  

Suppose now
that $f'$ is locally nilpotent.  Then $u_{2n} \geq 0$ (by
Lemma~\ref{lem:1.5}), hence we have from (\ref{eq:6.4}):
\begin{equation}
  \label{eq:6.10}
  t_j =0 \hbox{ for some } 0 \leq j \leq 2n-1 \, .
\end{equation}

Due to (\ref{eq:6.10}), (\ref{eq:6.8}) and (\ref{eq:6.9}), we
have the following three possibilities for some $0 \leq i_0 \leq
n-1$ and $n \leq j_0 \leq 2n-1$:
\begin{eqnarray*}
  \begin{array}{cl}
(\alpha) & t_0 = \cdots = t_{i_0} =0, \,\, 
           t_{i_{0}+1} \neq 0, \ldots , t_{2n-1} \neq 0 \, , \\
(\beta)  & t_0 \neq 0, \ldots , t_{j_0-1} \neq 0 , \,\,
           t_{j_0} = \cdots = t_{2n} =0  \, , \\
(\gamma) & t_0 = \cdots = t_{i_0}=0,\,\,
           t_{i_0+1} \neq 0 ,\ldots , t_{j_0-1} \neq 0,\,\,
           t_{j_0}=\ldots =      t_{2n}=0 \, .
  \end{array}
\end{eqnarray*}
The possibilities (i), (ii), (iii) and (iv) of the supplementary
conditions~(3) correspond respectively to the following cases:

\romanparenlist
\begin{enumerate}
\item %%i
  $(\gamma)$ when $i_0+j_0<2n-1$, where we put $i_0=n-s-1, j_0=n+r$,

\item %%ii
  $(\gamma)$ when $i_0+j_0\geq 2n-1$, where we put $i_0=n-r-1,
  j_0=n+s$,

\item %%iii
  $(\beta)$, where we put $j_0=n+r$,

\item %% iv
  $(\alpha)$, where we put $i_0 =n-r-1$.
\end{enumerate}

We consider in detail only case~(ii), the treatment of all other
cases being similar.  We have:
\begin{displaymath}
  t_0 = \cdots = t_{n-r-1}=0 ,\,\,
  t_{n-r} \neq 0, \ldots , t_{n+s-1} \neq 0,\,\,
  t_{n+s}=\cdots = t_{2n-1}=0
\end{displaymath}
for some integer $s$ such that $0 \leq r \leq s \leq n-1$.
Hence, by (\ref{eq:6.5a}) we have:
\begin{equation}
  \label{eq:6.11}
  k_r \neq 0, \,\, k_{r+1} = \cdots = k_n =0 \, , 
\end{equation}
and, by (\ref{eq:6.6a}), we have:
\begin{displaymath}
  s_i=-k_i \hbox{ if } i \neq r, \,\, 1 \leq i \leq n-1,\,\,
     s_r = -k_r+1,\,\, s_n=0 \, .
\end{displaymath}
The recurrent formulas (\ref{eq:6.5}) and (\ref{eq:6.7}) can be
now rewritten respectively as follows:
\begin{eqnarray*}
  \begin{array}{lccl}
   &  t_{n-r} &=& -k_r \, ,\\
   &  t_{n-r+1} &=& t_{n-r}-k_{r-1}-1 \, ,\\
   &  \cdots\\
   &  t_{n-1} &=& t_{n-2} -k_1-1 \, , \\
   &  t_n &=& t_{n-1}-2k_0-2 \, , \\
   &  t_{n+1} &=& t_n -k_1 -1 \, , \\
   &  \cdots \\
   &  t_{n+r-1} &=& t_{n+r-2}-k_{r-1}-1 \, , \\
   &  t_{n+r} &=& t_{n+r-1}-k_r \, , \\
   &  t_{n+r+1} &=& t_{n+r} - k_{r+1}-1 \, , \\
   &  \cdots\\
    0 = & t_{n+s} &=& t_{n+s-1} -k_{s}-1 +\delta_{r,s} \, .
  \end{array}
\end{eqnarray*}
Summing up these equalities, we get:  $\displaystyle{2
  \sum^r_{i=0} k_i=-(r+s)}$, which, in view of (\ref{eq:6.11}),
implies $k'=2(r+s)$.

Suppose now that conditions (1)--(4) hold.  We have to show that
$u_{2n} \geq 0$.  As before, we may assume that $k' \leq 4n-1$,
hence, due to (3), (\ref{eq:6.10}) holds.  Since (\ref{eq:6.8})
and (\ref{eq:6.9}) hold, we again have only the possibilities
$(\alpha)$, $(\beta)$ and $(\gamma)$.  In cases $(\beta)$ and
$(\gamma)$, $u_{2n} =2t_{2n}=0$, hence only case~$(\alpha)$
remains.  This case corresponds to (3)iv
 when we have:  $t_0=k_n=0$,\break$ t_1=t_0-k_{n-1}=0, \ldots ,
 t_{n-r-1} = t_{n-r-2}-k_{r+1}=0$.  Hence $v:= \#\{0 \leq j \leq
 2n-1 | t_j \neq 0 \} \leq n+r$ and $u_{2n}=u_0-2v \geq k'-2(n+r) 
 \geq 0$ (we have used here (\ref{eq:6.4}) and (3)iv).

\end{proof}

\begin{theorem}
  \label{th:6.2}
For an affine superalgebra $\hat{\fg}$ from Table~\ref{tab:4} the labels
$\{ k_i \}_{i \in \hat{I}}$ of the highest weight of a
subprincipal integrable irreducible highest weight module are
characterized by the following conditions:

\arabicparenlist
\begin{enumerate}
\item %%1
$k_i \in \ZZ_+$ if $i \in \hat{I} \backslash I_1$, $k'' \in
\ZZ_+$,

\item %%2
  if $k'' \leq b''$ (see Table~\ref{tab:4}), there are supplementary
  conditions described by (4) of Theorem~\ref{th:6.1}.%; in the
%  case $\hat{\fg} = C(n)^{\hat{}} $ the supplementary conditions
%  are $k_0=k_1=0$ if $k''=0$.
\end{enumerate}
\end{theorem}

\begin{proof}
  The only simple root of $\hat{\fg''_{\bar{0}}}$ which is not
  simple for $\hat{\fg}$ is $\alpha''$.  Hence the proof of
  Theorem~\ref{th:6.1} proves Theorem~\ref{th:6.2} as well.
\end{proof}

\begin{remark}
  \label{rem:6.2}
It follows from Theorem~\ref{th:6.1} that the level $k$ of a
principal integrable $\hat{\fg}$-module $L (\Lambda) $ is a
non-negative number which is an integer in all cases, except for
$B(1,n)^{\hat{}}$, when it is a half-integer; moreover, if $k=0$, 
then all labels of $\Lambda$ are $0$, hence $L(\Lambda)$ is
$1$-dimensional; also, if $k>0$, then $k \geq 1$. 
\end{remark}

\begin{remark}
  \label{rem:6.3}
It is easy to see that, when restricted to the derived subalgebra 
$[\hat{\fg}, \hat{\fg}]$ of $\hat{\fg}$ the module $L(\Lambda)$
remains irreducible.  Two $\hat{\fg}$-modules are called
\emph{essentially equivalent} if they are equivalent as
$[\hat{\fg}, \hat{\fg}]$-modules.  For example, the modules
$L(\Lambda)$ and $L(\Lambda + a \delta)$ are essentially equivalent for any $
a \in \CC$.  Theorem~\ref{th:6.1} gives the following complete list of
principal integrable modules of level~$1$ up to essential
equivalence:

\arabicparenlist
\begin{enumerate}
\item %%1
  $A(m,n)^{\hat{}}, m \geq 1$:  $\Lambda_s\,\, (1 \leq s \leq m)$, $(a+1)
  \Lambda_{m+1} + a \Lambda_{m+2}(a\in
 \ZZ_{+})$, and $(a+1)
  \Lambda_0 + a \Lambda_{m+n+1} (a \in \ZZ_{+})$,

%\item %
%  $A(1,n) $ and $C(n)$:  $a \Lambda_0 + (1-a) \Lambda_1\quad  (a \in
%  \CC)$,

\item %%2
$B(m,1)^{\hat{}}$ and $D(m,1)^{\hat{}}$:  $- \tfrac{1}{2} \Lambda_0$ and
$-\tfrac{3}{2} \Lambda_0 - \Lambda_1$,

\item %%3
  $B(m,n)^{\hat{}}$ and $D(m,n)^{\hat{}}, n \geq 2$:  $-\tfrac{1}{2} \Lambda_0$ and 
  $-\tfrac{3}{2} \Lambda_0 + \Lambda_1$,

\item %%4
  $D(2,1;a)^{\hat{}}, a^{-1} \in \NN$:  $-(a+1)^{-1} \Lambda_0$ 
  and  $- \tfrac{a+2}{a+1} \Lambda_0 -
  \Lambda_1 + \tfrac{1-a}{a} \Lambda_3$
  $a^{-1} $,

\item %%5
  $F(4)^{\hat{}}$:  $- \tfrac{2}{3} \Lambda_0$,

\item %%6
  $G(3)^{\hat{}}$:  $-\tfrac{3}{4} \Lambda_0$.
  
\end{enumerate}
One can show (cf.~Remark~\ref{rem:2.5}) that in all cases, all weights 
are conjugate to each other by odd
reflections.  Thus, for each 
of the affine superalgebras $A(m,n)^{\hat{}} (m \geq 1)$,
$B(m,n)^{\hat{}}$, $D(m,n)^{\hat{}}$, $D(2,1;a)^{\hat{}} (a \in
\QQ_>)$, $F(4)^{\hat{}}$ and $G(3)^{\hat{}}$ all, up to
essential equivalence, principal integrable modules of level~$1$,
can be obtained from one of them by making different choices of
the set of positive roots.  Note also that in all cases the
``basic'' module $L(u \Lambda_0)$, where $u$ is such that
$u\Lambda_0$ has level~$1$, is a principal integrable module.
\end{remark}

\begin{remark}
  \label{rem:6.4}
Using the symmetry of $A(m,n)^{\hat{}}$ which exchanges the
subalgebras $\hat{A}_m$ and $\hat{A}_n$, one gets the
classification of the subprincipal integrable modules $L(\Lambda)$ for
this affine subalgebra:
\begin{displaymath}
  k_i \in \ZZ_+ \hbox{ for } i \in \hat{I} \backslash
  \hat{I}_1,\quad
  k'':=- \sum^{m+1}_{i=0} k_i \in \ZZ_{+} \, ,
\end{displaymath}
and there exists $s \in \ZZ_+$, $s \leq k''$, such that
\begin{displaymath}
  k_0+k_1+ \cdots + k_s +s=0 \hbox{ and }
  k_{s+1}=\cdots = k_{s+m-k''+1} =0 \, .
\end{displaymath}
One has:  $\displaystyle{k=-(k''+ \sum^{m+n+1}_{i=n+2}k_i)}$.

\end{remark}

\begin{remark}
\label{rem:6.5}
  All principal integrable highest weights of level~$2$ (up to
  essential equivalence) for $B(1,2)^{\hat{}}$ are $-(1+a)
  \Lambda_0 + a \Lambda_1$, where $a \in \ZZ_+$.  Thus, in sharp
  contrast to the level~$1$ case, there are infinitely many
  essentially inequivalent principal integrable highest weight
  modules of level~$\geq 2$.

\end{remark}

\begin{remark}
  \label{rem:6.6}
It follows from Theorem~\ref{th:6.2} and Remark~\ref{rem:6.3}
that the level $k$ of a subprincipal integrable
$\hat{\fg}$-module $L(\Lambda)$ is a non-positive number,
provided that $a>-1$ for $D(2,1;a)$; moreover, $\dim L
(\Lambda)=1$ if $k=0$.  Thus, in view of Theorem~\ref{th:6.1},
the only $L(\Lambda)$ over  $\hat{\fg}\neq A(m,0)^{\hat{}}$,
$A(0,n)^{\hat{}}$ or $C(n)^{\hat{}}$, which are integrable over $
\hat{\fg_{\bar{0}}}$ are $1$-dimensional.
\end{remark}

\begin{remark}
  \label{rem:6.7}
Using the same arguments, one can show that the non-symmetrizable 
``twisted'' affine superalgebra of type~$Q$ (which is the
universal central extension of the Lie superalgebra
$\displaystyle{\sum_{n \in \ZZ} (Q(n)_{\bar{0}} t^{2n} +
  Q(n)_{\bar{1}} t^{2n+1})}$), with the Cartan matrix
\begin{displaymath}
  \left(
    \begin{array}{ccccc}
      0 & 1 & 0 &  \cdots & -1 \\
      -1 &&\\
      0 &&& A_n\\
      \cdots\\
      0\\
-1
    \end{array} \right)
\end{displaymath}
has no non-trivial integrable (with respect to its even part)
highest weight modules.

\end{remark}

\begin{remark}
  \label{rem:6.8}
Consider the $\ZZ/2\ZZ$-gradation of $F(4)$ of type $(0,0,0,1,0)$
and that of $G(3)$ of type $(0,0,0,1)$, cf.~Table~\ref{tab:1} and
\cite{K3}.  The $0$\st{th} piece in the first (resp. second) case 
is isomorphic to $D(2,1;1/2) \oplus A_1$ (resp. to $D(2,1;1/3)$),
and its representation on the $1$\st{st} piece is the module
$\CC^{10} \boxtimes \CC^2$ (resp.~$\CC^{14}$), where $\CC^{10}$
(resp. $\CC^{14}$) is the lowest-dimensional non-trivial module
over $D(2,1;1/2)$ (resp. $D(2,1;1/3)$).  This reduces to some extent
the construction of the principal integrable level~$1$ module
over $F(4)^{\hat{}}$ and $G(3)^{\hat{}}$ to that of
$D(2,1;a)^{\hat{}}$.  The free field construction of the
principal integrable level~$1$ modules over $osp (m,n)^{\hat{}}$
(covering the $B-C-D$ cases) will be given in Section~\ref{sec:7}.

\end{remark}

%%%%%%%%%%%%%%%%%%%%%%%%% section 6 %%%%%%%%%%%%%%%%%%%%%%%%%%%

\section{Free field realization of level~$1$ integrable modules
  over $osp (M|N)^{\hat{}}$.}
\label{sec:7}

Let $V$ be the superspace and let $(.|.)$ be the bilinear form
on $V$ considered in Remark~\ref{rem:6.1}.  Recall an equivalent
definition of $osp (M|N)$ via the Clifford superalgebra:
\begin{displaymath}
  C \ell V = T(V) / \langle [x,y]-(x|y)1|x,y \in V \rangle \, .
\end{displaymath}
The Lie superalgebra $osp (M|N)$ is identified with the
$\CC$-span of all quadratic elements of $C \ell V$ of the form:
\begin{displaymath}
  : \alpha \beta : \equiv \alpha \beta + (-1)^{p(\alpha)p(\beta)} \beta \alpha, 
  \quad \hbox{ where } \alpha, \beta \in V \, .
\end{displaymath}
Such an element is identified with an operator from $osp (M|N)$
by the formula:
\begin{equation}
  \label{eq:7.1}
  (: \alpha \beta :) v= [: \alpha \beta :,v], \quad v \in V \, .
\end{equation}
Denote by $\Phi_V$ the vertex algebra generated by pairwise local 
fields $\gamma (z)$, where $\gamma \in V_{\bar{0}} \cup
V_{\bar{1}}$ and $\gamma (z)$ is even (resp. odd) if $\gamma \in
V_{\bar{0}}$ (resp. $V_{\bar{1}}$), subject to the following OPE:
\begin{displaymath}
  \gamma (z) \gamma' (\omega) \sim 
     \frac{(\gamma | \gamma')}{z-w} \, .
\end{displaymath}
This is called the vertex algebra of free superfermions in
\cite{K4}.

\begin{remark}
  \label{rem:7.1}
The vertex algebra $F$ considered in \S~\ref{sec:3} is isomorphic 
to $\Phi_V$, where $\dim V_{\bar{0}}=2n, \dim V_{\bar{1}}=2m$ and 
the bilinear form is given by:
\begin{eqnarray*}
  (\varphi^{i*}| \varphi^{j}) &=& - (\varphi^i|\varphi^{j*}) 
  = \delta_{ij} \,\, (i,j=1, \ldots, n)\, , \\
  (\psi^{i*}|\psi^j)&=&(\psi^i | \psi^{j*}) = \delta_{ij}
  \,\, (i,j =1, \ldots ,m) \, , 
  \hbox{ all other inner products }=0 \, .
\end{eqnarray*}
Furthermore, in the case when $\dim V_{\bar{1}}=2m+1$ the vertex
algebra $\Phi_V$ is isomorphic to $F \otimes \Phi$, where $\Phi$
is a vertex algebra generated by one odd field $\psi (z)$ with
the OPE $
  \psi(z) \psi (w) \sim \frac{1}{z-w}$.  This corresponds to adding an odd vector $\psi$ with $(\psi |
\psi) =1$ orthogonal to all the above basis vectors.
\end{remark}

As in \S~\ref{sec:3}, we construct the Virasoro field $L(z)
\equiv \sum_{j \in \ZZ} L_j z^{-j-2}$ with respect to which all $
\gamma (z)$ are primary of conformal weight $1/2$.  Choose a
basis $\varphi^i, \varphi^{i*} \,\, (i=1, \ldots ,n)$ of
$V_{\bar{0}}$, and a basis $\psi^i, \psi^{i*} \,\, (i=1,\ldots
,m)$ and $\psi$ if $M$ is odd, with inner products described by
Remark~\ref{rem:7.1}.  Then $L(z)$ is given by
formula~(\ref{eq:3.1}) if $M$ is even.  In the case $M$ is odd, one 
should add to the expression (\ref{eq:3.1}) the term
$\tfrac{1}{2}: \partial \psi (z) \psi (z):$.  As in
\S~\ref{sec:3}, we shall write $\displaystyle{\gamma(z) = \sum_{k \in
  \frac{1}{2} +\ZZ} \gamma_k z^{-k-1/2}, \,\, \gamma \in
V_{\bar{0}} \cup V_{\bar{1}}}$.

We shall need also the following well-known fact (see
e.g.~\cite{K4}, formula (5.1.5)).

\begin{lemma}
  \label{lem:7.1}
Let $\psi^+, \psi^- \in V_{\bar{1}}$ be such that $(\psi^{\pm}|
\psi^{\pm}) =0, \,\, (\psi^+ |\psi^-)=1$.  Let $\displaystyle{\alpha (z)
\equiv \sum_{n \in \ZZ} \alpha_n z^{-n-1}}$\break $ \displaystyle{=: \psi^+ (z) \psi^-(z)}:$.
  Then one has:
\begin{displaymath}
  : \alpha (z) \alpha (z): = :\partial \psi^+ (z)
     \psi^-(z) : + :\partial \psi^-(z) \psi^+(z): \, .
\end{displaymath}
Consequently, the fields $\psi^{\pm}(z)$ are primary of conformal 
weight $1/2$ with respect to the Virasoro field $\displaystyle{\ell (z) \equiv
\sum_{n \in \ZZ} \ell_n z^{-n-2} = \tfrac{1}{2} : \alpha (z) 
\alpha (z):}$.  In particular, we have
\begin{equation}
  \label{eq:7.2}
  [\ell_0 , \psi^{\pm}_n] = -n \psi^{\pm}_n \, .
\end{equation}

\end{lemma}

Note that
\begin{displaymath}
  \gamma_j \vac =0 \hbox{ for } j>0, \,\, \gamma \in V \, .
\end{displaymath}
Hence $\Phi_V$ is obtained by applying polynomials in the
$\gamma_{-j}$, $\gamma \in V, \,\, j>0$, to the vacuum vector
$\vac$.  We have the decomposition
\begin{equation}
  \label{eq:7.3}
  \Phi_V = \Phi^+_V \oplus \Phi^-_V \, ,
\end{equation}
where $\Phi^+_V$ (resp. $\Phi^-_V$) is obtained by applying even
(resp. odd) degree polynomials in the $\gamma_{-j} $ to $\vac$.

\begin{theorem}
  \label{th:7.1}
  \alphaparenlist
  \begin{enumerate}
  \item %%a
    Consider the affine superalgebra $osp (M|N)^{\hat{}}$ and let 
    $:\alpha \beta : (z) =$\break $\displaystyle{\sum_{k \in \ZZ} (t^k \otimes : \alpha
    \beta:)z^{-k-1}}$ for $:\alpha \beta : \in osp (M|N)$.  Then
    the linear map $\sigma$ given by $(\alpha , \beta \in V)$:
    \begin{displaymath}
      :\alpha \beta : (z) \mapsto :\alpha (z) \beta (z) :\, ,
      \,\, K \mapsto 1 , \,\, d \mapsto L_0
    \end{displaymath}
defines a principal integrable representation of $osp
(M|N)^{\hat{}}$ of level~$1$ in the space $\Phi_V$ for which
$\Phi^+_V$ and $\Phi^-_V$ are submodules.

\item %%b
  The $osp (M|N)^{\hat{}}$-modules $\Phi^+_V$ and $\Phi^-_V$ are
  irreducible highest weight modules isomorphic to
  $L(-\tfrac{1}{2} \Lambda_0)$ and $L(-\tfrac{1}{2} \Lambda_0 -
  \tfrac{1}{2} \alpha_0)$ respectively, provided that $(M,N) \neq 
  (1,0)$ or $(2,0)$. 
  \end{enumerate}
\end{theorem}

\begin{proof}
  The proof that $\sigma$ is a representation is, as usual, a
  straightforward use of Wick's formula.  The proof of
  integrability of $\sigma$ is the same as in the proof of
  Theorem~\ref{th:3.1}.  This establishes~(a).

Note that, as before, $L_0$ commutes with $osp (M|N)$, and the
spectrum of $L_0$ on $\Phi^+_V$ (resp. $\Phi^-_V$) is $\ZZ_+$
(resp. $\tfrac{1}{2} +\ZZ_+$), the lowest eigenvalue eigenspace
being $S^{\pm}=\CC \vac$ (resp. $S^-=\{ \gamma_{-\frac{1}{2}} \vac 
| \gamma \in V \}$), which is the trivial $1$-dimensional
(resp. the standard) representation of $osp (M|N)$.  Provided
that $\Phi^{\pm}_V$ are irreducible $osp (M|N)^{\hat{}}$-modules, 
(b)~follows.

In order to prove irreducibility of $\Phi^{\pm}_V$, pick elements
$\psi^+,\psi^- \in V_{\bar{1}}$ as in Lemma~\ref{lem:7.1} and define the field $\ell (z)$
as in that lemma.  Let $\psi \in V_{\bar{1}}$ be an 
element orthogonal to both $\psi^+$ and $\psi^-$, and consider
the field $\displaystyle{\beta (z) =:\psi^+(z) \psi (z): \equiv
  \sum_{n \in \ZZ} \beta_n z^{-n-1}}$, so that
$\displaystyle{\beta_n=\sum_{j \in
    \frac{1}{2}+\ZZ}:\psi^+_j\psi_{n-j}:}$.  Since $\ell_0$
commutes with $\psi(z)$, we have by (\ref{eq:7.2}):
\begin{equation}
  \label{eq:7.4}
  [\ell_0 ,\beta_n] = \sum_{j \in \frac{1}{2}+\ZZ}
  j:\psi^+_j\psi_{n-j}: \, .
\end{equation}
Let $U \subset \Phi^{\pm}_V$ be an invariant with respect to $osp 
(M|N)^{\hat{}}$ subspace.  It follows from Lemma~\ref{lem:7.1}
and (\ref{eq:7.4}) that $v \in U$ implies that $((\ad
\ell_0)^s\beta_n) v \in U$, $s \in \ZZ_+$.  Hence $U$ is
invariant with respect to all operators $\psi^+_j\psi_k$, where $
\psi^+,\psi \in V$ are such that
$(\psi^+|\psi^+)=0=(\psi^+|\psi)$ and $j,k \in \tfrac{1}{2}
+\ZZ$.  

Hence, provided that $M \geq 3$, $U$ contains a non-zero purely
bosonic element, i.e.,~an element obtained by applying a
polynomial in the $\gamma_j \, \, ( \gamma \in V_{\bar{0}})$ to $\vac$.
Thus we reduced the problem to the purely bosonic case, i.e.,~the 
case when $M=0$.  In this case the irreducibility was proved in
\cite{L} using the character formula for modular invariant
representations of $\hat{C}_n$ from \cite{KW1} and formula~(12.13)
from \cite{K3} (the reference to (13.13) in \cite{L} is a
misprint).

The remaining cases, when $M=1$ or $2$ and $N=2n$ is even $\geq
2$ can be reduced again to the purely bosonic case by a direct
calculation.  We give below details in the $M=1$ case, the $M=2$ case
being similar.

The simple root vectors of $osp (1,N)^{\hat{}}=B(0,n)^{\hat{}}$ 
are as follows:
\begin{displaymath}
  \begin{array}{lclcl}
  e_0 &=& (\varphi^{1*}(z) \varphi^{1*}(z))_1
      &=& \sum_{s \in \ZZ} \varphi^{(1)*}_{-s+1/2}
          \varphi^{(1)*}_{s+1/2} \, , \\
  e_i &=& (\varphi^i (z) \varphi^{i+1*}(z))_0
      &=& \sum_{s \in \ZZ} \varphi^{(i)}_{-s-1/2}
         \varphi^{(i+1)*}_{s+1/2} \qquad (i=1, \ldots ,n-1) \, ,
         \\
   e_n &=& (\varphi (z) \psi (z))_0
       &=& \sum_{s \in \ZZ} \varphi^{(n)}_{-s-1/2}
       \psi_{s+1/2} \, .
  \end{array}
\end{displaymath}
Then the simple root vectors of $sp (N)^{\hat{}}=\hat{C}_n$ are $
e_0,e_1,\ldots e_{n-1}$ and $e'_n=[e_n,e_n]=$\break$\sum_{s \in \ZZ}
\varphi^{(n)}_{-s-1/2} \varphi^{(n)}_{s+1/2}$.

Any vector $v$ of $\Phi_V$ can be uniquely written in the form:
\begin{displaymath}
  v=\sum_{i_1 < \cdots <i_k} \psi_{i_1} \ldots \psi_{i_k}
  u_{i_1 , \ldots , i_k} \, ,
\end{displaymath}
where $u_{i_1,\ldots ,i_k}$ are purely bosonic elements
(i.e.,~obtained by applying polynomials in the $\varphi$'s to
$\vac$).  Now, if $v$ is a singular vector, i.e.,~$e_iv=0$ for
all $i=0,\ldots,n$, then, in particular, $e'_nv=0$, and since all 
$e_1 , \ldots , e_{n-1},e'_n$ commute with the $\psi$'s, we get:
\begin{displaymath}
  \sum_{i_1<\cdots <i_k} \psi_{i_1} \ldots \psi_{i_k}
     (e_iu_{i_1, \ldots ,i_k}) =0, \,\,
     \sum_{i_1<\cdots <i_k} \psi_{i_1} \ldots \psi_{i_k}
     (e'_n u_{i_1, \ldots, i_n}) =0 \, .
\end{displaymath}
It follows that all $u_{i_1 ,\ldots ,i_n}$ are purely bosonic
singular with respect to $\hat{C}_n$ vectors, hence, due to
irreducibility of $\Phi^{\pm}_V$ for $M=0$ mentioned above, we
obtain that all $u_{i_1 \ldots i_k}$ are linear combinations of
elements $\vac$ and $\varphi^{(1)}_{-1/2} \vac$.  Hence
\begin{displaymath}
  v=\sum_{i_1 <\cdots <i_k} a_{i_1 ,\ldots, i_k}
    \psi_{i_1} \ldots \psi_{i_k} \vac +
    \sum_{i_1 < \cdots < i_k} 
    b_{i_1,\ldots ,i_k} \psi_{i_1} \ldots \psi_{i_k}
    \varphi^{(1)}_{-1/2} \vac \, .
\end{displaymath}
Using that $e_nv=0$, we obtain:
\begin{eqnarray*}
  \sum^k_{r=1} \sum_{i_1 < \cdots < i_k} (-1)^{r-1}
     a_{i_1, \ldots ,i_k} \psi_{i_1} \ldots \hat{\psi}_{i_r}
     \ldots \psi_{i_k} \varphi^{(n)}_{i_r} \vac\\
+ \sum^k_{r=1} \sum_{i_1 < \ldots < i_k} (-1)^{r-1}
      b_{i_1 ,\ldots ,i_k} \psi_{i_1} \ldots \hat{\psi}_{i_r}
     \ldots \psi_{i_k} \varphi^{(n)}_{i_r} 
     \varphi^{(1)}_{-1/2} \vac =0\, ,
\end{eqnarray*}
which implies that $a_{i_1 ,\ldots, i_k}$ (resp. $b_{i_1, \ldots,
  i_k})=0$ if $k>0$.  Thus, the only singular vectors in
$\Phi^+_V$ (resp. $\Phi^-_V$) are scalar multiples of $\vac$
(resp. $\varphi^{(1)}_{-1/2} \vac$).

To conclude that the $B(0,n)^{\hat{}}$-modules $\Phi^{\pm}_V$ are 
irreducible, note that $\Phi_V$ carries a unique non-degenerate
Hermitian form $H(.,.)$ such that the square length of $\vac$ is
$1$ and the adjoint operators of $\varphi^{(j)}_k$ and $\psi_k$
are $\varphi^{(j)*}_{-k}$ and $\psi_{-k}$, respectively.  The
absence of non-trivial singular vectors in $\Phi^+_V$
(resp. $\Phi^-_V$) implies that the $B(0,n)^{\hat{}}$-submodules
$\Phi^{+\prime}_V$ (resp. $\Phi^{-\prime}_V$) generated by $\vac$ 
(resp. $\varphi^{(1)}_{-1/2} \vac$) is irreducible, hence the
restriction of $H$ to it is non-degenerate.  Hence the orthogonal 
complement to $\Phi^{+\prime}_V$ (resp. $\Phi^{-\prime}_V$) is a
complementary submodule which has no non-zero singular vectors,
hence it is zero, and $\Phi^{\pm}_V$ are irreducible.
\end{proof}

\begin{remark}
  \label{rem:7.2}

The irreducibility in the purely fermionic case was established
in \cite{KP1} by making use of the Weyl-Kac character formula.  An 
argument, using Virasoro operators, was given in \cite{F}.  
The method of using Virasoro operators to prove irreducibility
apparently works only in the presence of fermions
(cf.~Remark~\ref{rem:3.3}).  It is shown in \cite{L} that the
irreducibility claims of \cite{FF}, based on the use of Virasoro
operators, are false for the constructions of $A^{(2)}_{2 \ell
  -1}$ and $A^{(2)}_{2 \ell}$-modules.
\end{remark}

Using Theorem~\ref{th:7.1}, it is straightforward to write down the characters and
supercharacters for the integrable level~$1$ $osp
(M|N)^{\hat{}}$-modules.  We have:
\begin{equation}
  \label{eq:7.5}
  ch \Phi^+_V \pm ch \Phi^-_V = e^{-\frac{1}{2}\Lambda_0}
  \Pi^{\infty}_{k=1}
  \frac{(1\pm q^{k-1/2})^{p(M)}\Pi^m_{i=1}
    (1\pm e^{\epsilon_i}q^{k-1/2})(1\pm e^{-\epsilon_i}q^{k-1/2})}
  {\Pi^n_{j=1}(1\mp e^{\epsilon_{j+m}}q^{k-1/2})
    (1\mp e^{-\epsilon_{j+m}}q^{k-1/2})} \, ,
\end{equation}
where $p(M)=0$ (resp.~$1$) if $M$ is even (resp.~odd).  
A similar formula for supercharacters is obtained by reversing
signs in the numerator of the right-hand side of (\ref{eq:7.5}).

Letting all $\epsilon_i$ and $\Lambda_0$ equal $0$ in
(\ref{eq:7.5}), we obtain:
\begin{equation}
  \label{eq:7.6}
  tr_{\Phi^+_V} q^{L_0} \pm tr_{\Phi^-_V} q^{L_0} =
  \Pi^{\infty}_{k=1} \frac{(1\pm q^{k-\frac{1}{2}})^M}
  {(1\mp q^{k-\frac{1}{2}})^N} \, .
\end{equation}
Noticing that
\begin{equation}
  \label{eq:7.7}
  \Pi^{\infty}_{k=1} (1-q^{k-\frac{1}{2}}) =
  \frac{\varphi (q^{\frac{1}{2}})}{\varphi (q)}
  \hbox{ and }\Pi^{\infty}_{k=1} (1+q^{k-\frac{1}{2}})=
  \frac{\varphi (q)^2}{\varphi (q^{\frac{1}{2}}) \varphi (q^2)}
\end{equation}
and using the asymptotics (\ref{eq:4.12}) of $\eta (\tau )$, we
obtain the following asymptotics as $\tau \downarrow 0$:
\begin{equation}
  \label{eq:7.8}
  tr_{\Phi^{\pm}_V} q^{L_0} \sim \frac{1}{2^{n+1}}
  e^{\frac{\pi i}{12 \tau}(\frac{1}{2}M+N)} \, .
\end{equation}

\begin{remark}
  \label{rem:7.3}
The right-hand side of (\ref{eq:7.6}) multiplied by
$q^{(N-M)/48}$ is a modular function equal to a product of powers 
of functions $\eta (\tfrac{1}{2} \tau)/\eta (\tau)$ and $\eta
(\tau)^2/\eta (\tfrac{1}{2} \tau) \eta (2 \tau)$, and the same
holds if we replace $tr$ by $str$.  It is well known (and easy to 
see) that the above two modular functions along with the modular
function $\eta (2 \tau)/\eta (\tau)$ are transitively permuted
(with some constant factors) under the action of $SL (2,\ZZ)$.
Thus, the normalized by $q^{(N-M)/48}$ characters and
supercharacters of integrable level one $osp
(M|N)^{\hat{}}$-modules are modular functions, but
their $\CC$-span is not $SL (2,\ZZ)$-invariant.
\end{remark}

%Using their 
%transformation properties under $\tau \mapsto -1/\tau$ (we let,
%as usual, $q=e^{2 \pi i \tau}$) we obtain the following
%asymptotics as $\tau \downarrow 0$:
%%
%\begin{displaymath}
%  tr_{L(-\frac{1}{2} \Lambda_0)}q^{L_0} \sim
%  \left\{ 
%    \begin{array}{ll}
%      \frac{1}{2^{n+1}} e^{\frac{\pi i}{1 \, 2 \tau}(m+2n)}
%         & \hbox{for } D(m,n)^{\hat{}}\\
%%
%         \frac{1}{2^{n+1}} e^{\frac{\pi i}{1 \, 2 \tau}
%         (m+2n+\tfrac{1}{2})} & \hbox{for } B(m,n)^{\hat{}}.
%    \end{array}\right.
%\end{displaymath}

\section{On classification of modules over the associated vertex algebras}
\label{sec:8}

Define numbers $u$ and $h^{\vee}$ (the dual Coxeter number) by
\begin{equation}
  \label{eq:8.1}
  \hbox{level } (k u \Lambda_0)=k \, , 
  \hbox{level } (\rho)=h^{\vee} \, .
\end{equation}
Their values for all affine superalgebras are given in Table~\ref{tab:5}.

\begin{table}[h]
  \begin{center}
    \caption{ } %% {Table 5.}
    \vspace{1ex}
    \label{tab:5}
\begin{tabular}{c| c ccc  c c c cc}
  \hline\hline\\[1ex]
$\hat{\fg}$ & $A (m,n)^{\hat{}}$ & $B (m,n)^{\hat{}}$
& $C(n)^{\hat{}}$ & $D(m,n)^{\hat{}}$
& $D (2,1;a)^{\hat{}}$ & $F(4)^{\hat{}}$ & $G(3)^{\hat{}}$\\
\hline\\
$u$ & $1$ & $-1/2$ & $-1/2$ & $-1/2$ & $-(a+1)^{-1}$ & $-2/3$ &
$-3/4$\\
\hline
$h^{\vee}$ & $m-n$ & $2(m-n)-1$ & $n-1$ & $2(m-n-1)$ & $0$ & $3$ & $2$
\end{tabular}
  \end{center}
\end{table}
The following proposition is an immediate corollary of
Theorems~\ref{th:6.1} and \ref{th:6.2}.
\begin{proposition}
  \label{prop:8.1}
\alphaparenlist
  \begin{enumerate}
  \item %%a 
    If $\hat{\fg} =A(m,n)^{\hat{}}$ with $m \geq 1$,
    $B(m,n)^{\hat{}}$ with $m \geq 2$, $D(m,n)^{\hat{}}$,
    $F(4)^{\hat{}}$ or $G(3)^{\hat{}}$, then the
    $\hat{\fg}$-module $L(ku \Lambda_0)$ is principal integrable
    iff $k \in \ZZ_+$.

    The $B(1,n)^{\hat{}}$-module $L(ku\Lambda_0)$ is principal
    integrable iff $k \in \ZZ_+ \cup$ $\{ 2n-\tfrac{1}{2}+\ZZ_+ \}$.  The $D(2,1;a)^{\hat{}}$-module 
    $L(ku\Lambda_0)$   is principal
    integrable iff $k \in \ZZ_+ \cap \, a\ZZ_+$.

\item %%b
  If $\hat{\fg}=B(m,n)^{\hat{}}$, $C(n)^{\hat{}}$,
  $D(m,n)^{\hat{}}$, $D(2,1;a)^{\hat{}}$, $F(4)^{\hat{}}$ or
  $G(3)^{\hat{}}$, then the $\hat{\fg}$-module $L(k_0\Lambda_0)$
  is subprincipal integrable iff $k_0 \in \ZZ_+$.
  
  \end{enumerate}
\end{proposition}

Recall that the $\hat{\fg}$-module $V_k:=L(ku\Lambda_0)$ has a
canonical structure of a vertex algebra for any $k \in \CC$ (see
e.g.~\cite{K4}).  It is well known that any irreducible
$V_k$-module is one of the (irreducible) $\hat{\fg}$-modules
$L(\Lambda)$ of level~$k$, and it is an important problem of
vertex algebra theory to find out which of these $L(\Lambda)$ are 
actually $V_k$-modules.  A necessary condition is given by

\begin{proposition}
  \label{prop:8.2}
Suppose that $k$ is such that $L(ku\Lambda_0)$ is a principal
(resp. subprincipal) integrable $\hat{\fg}$-module.  If a
$\hat{\fg}$-module $L(\Lambda)$ of level~$k$ is a $V_k$-module,
then it must be a principal (resp. subprincipal) integrable.
\end{proposition}

\begin{proof}
  
Denote by $\hat{\fg}^0$ the subalgebra $\hat{\fg}'_{\bar{0}}$
(resp. $\hat{\fg}''_{\bar{0}}$) of $\hat{\fg}$ (see
\S~\ref{sec:6}).  This is an affine Lie algebra.  Denote by $V^0$ 
the vertex subalgebra $U(\hat{\fg}^0) v_{ku\Lambda_0}$ of $V_k$.
Since, by definition, $V^0$ is an integrable
$\hat{\fg}^0$-module, it follows that it is
$\hat{\fg}^0$-irreducible \cite{K3}, hence $V^0$ is a simple
affine vertex algebra of non-negative integral level.  But one
knows \cite{Z} that all irreducible modules over such a vertex
algebra are integrable $\hat{\fg}^0$-modules.  Using the complete 
reducibility of $\hat{\fg}^0$-modules \cite{K3}, we deduce that
any $V$-module, viewed as a $V^0$-module, is a direct sum of
irreducible integrable $\hat{\fg}^0$-modules, which proves the proposition.

\end{proof}

Let $\hat{\fg}_+=\CC [t] \otimes_{\CC} \fg + \CC d$ and consider
a $1$-dimensional module $\CC_k$ $(k \in \CC)$ over $\hat{\fg}_+
+ \CC K$ on which $\hat{\fg}_+$ acts trivially and $K=k$.  Then $
L(ku\Lambda_0)$ is a quotient of the induced $\hat{\fg}$-module
$\tilde{V}_k=U(\hat{\fg}) \otimes_{U(\hat{\fg}_++\CC K)} \CC_k$
by a left ideal $I_k$ of $U (\hat{\fg})$ applied to $1 \otimes
1$.  Suppose that $k$ is such that $L(ku \Lambda_0)$ is a
principal integrable $\hat{\fg}$-module.  As we have seen in the
proof of Proposition~\ref{prop:8.2}, viewed as a
$\hat{\fg}'_0$-module, $L(ku\Lambda_0)$ is a direct sum of
irreducible integrable highest weight modules.  All these modules 
have the same level~$\ell$ (resp. $\ell_+,\ell_-$ when
$\fg'_{\bar{0}}$ has two simple components) given in terms of $k$ 
as follows:
\begin{eqnarray*}
  \ell &=& \ell_+ =k \hbox{ if } \fg \neq B(1,n),\quad
  \ell = 2k \hbox{ if } \fg = B(1,n) \, ,\\
  \ell_- &=& k \hbox{ if } \fg = D(2,n),\quad
  \ell_- = a^{-1}k \hbox{ if } \fg =D(2,1;a) \, .
\end{eqnarray*}
In particular, $I_k$ contains the element
\begin{equation}
  \label{eq:8.2}
  e_{-\theta'}(1)^{\ell +1} \hbox{( resp. elements }
  e_{-\theta'_+}(1)^{\ell'_++1} \hbox{ and }
e_{-\theta'_-} (1)^{\ell'_- +1}) \, .
\end{equation}
If elements (\ref{eq:8.2}) generate the left ideal $I_k$, it
follows that a $\hat{\fg}$-module $L(\Lambda)$ of level~$k$ is a
$V_k$-module iff the field $e_{-\theta'}(z)^{\ell +1}$
(resp. fields $e^{\ell'_++1}_{-\theta'_+} (z)$ and
$e^{\ell'_-+1}_{-\theta'_-} (z)$) annihilate $L(\Lambda)$.  The
latter property implies that, viewed as a
$\hat{\fg}'_{\bar{0}}$-module, $L(\Lambda)$ is a direct sum of
irreducible integrable modules and therefore $L(\Lambda)$ is a principal
integrable $\hat{\fg}$-module.  We thus established a sufficient
condition for a $\hat{\fg}$-module $L(\Lambda)$ to be a
$V_k$-module:

\begin{proposition}
  \label{prop:8.3}
Let $k$ be such that $L(ku\Lambda_0)$ is a principal integrable $
\hat{\fg}$-module and suppose that the left ideal $I_k$ is
generated by (\ref{eq:8.2}).  Let $L(\Lambda)$ be a principal
integrable $\hat{\fg}$-module of level~$k$.  Then $L(\Lambda)$ is 
a $V_k$-module.
\end{proposition}

\begin{proposition}
  \label{prop:8.4}
Let $k$ be such that $L(ku\Lambda_0)$ is a principal integrable $
\hat{\fg}$-module.

\alphaparenlist
\begin{enumerate}
\item %%a
  Suppose that the highest weight $ku\Lambda_0$ is the only
  singular weight of the $\hat{\fg}$-module $\tilde{V}_k$ which
  is principal integrable.  Then elements~(\ref{eq:8.2})
    generate the left ideal~$I_k$.

\item %%b
  The assumption of~(a) holds if
  \begin{equation}
    \label{eq:8.3}
    k+ h^{\vee} \neq 0 \, ,
  \end{equation}
and for any principal integrable weight $\Lambda$ of level $k$
one has:
\begin{equation}
  \label{eq:8.4}
  \Lambda - ku \Lambda_0 \not\in \hat{Q}
  \backslash \ZZ \delta \, ,
\end{equation}
where $\hat{Q}= \sum_{i \in \hat{I}} \ZZ \alpha_i$ is the root
lattice of 
$\hat{\fg}$.
\end{enumerate}

\end{proposition}

\begin{proof}
  Let $I'_k$ $(\subset I_k)$ denote the left ideal of
  $U(\hat{\fg})$ generated by elements~(\ref{eq:8.2}).  Then the
  $\hat{\fg}$-module $V'_k=\tilde{V}_k / (I'_k (1 \otimes 1))$ is 
  principal integrable, hence each of its singular weights
  $\Lambda$ is integrable.  Hence, if the condition of~(a) holds, 
  the $\hat{\fg}$-module $V'_k$ is irreducible, and therefore
  $I'_k=I_k$.

Furthermore, obviously, $\Lambda - ku \Lambda_0 \in \hat{Q}$, hence (\ref{eq:8.4}) implies that
$\Lambda =ku \Lambda_0 + j\delta$ for some $j \in \ZZ$.  Using
the Casimir operator \cite{K3}, we obtain:
\begin{displaymath}
  (ku \Lambda_0 + \rho | ku \Lambda_0 + \rho)=
  (ku \Lambda_0+\rho -j\delta | ku \Lambda_0 +
  \rho -j \delta) \, ,
\end{displaymath}
which is equivalent to $j (k+ h^{\vee})=0$.  But then
(\ref{eq:8.3}) implies that $j=0$, proving~(b).
\end{proof}

\begin{theorem}
  \label{th:8.2}
  Let $\hat{\fg}$ be one of the affine superalgebras
  $A(m,n)^{\hat{}}$ with $m \geq 1$, $B(m,n)^{\hat{}}$ with $m
  \geq 1$, $D(m,n)^{\hat{}}$, $D(2,1;a)^{\hat{}}$ with $a^{-1}
  \in \NN$, $F(4)^{\hat{}}$ or $G(3)^{\hat{}}$.  Then all
  integrable $\hat{\fg}$-modules $L(\Lambda)$ of level~$1$ are
  $V_1$-modules (the complete list of these $\Lambda$'s is given by Remark~\ref{rem:6.3}).
\end{theorem}

\begin{proof}
  Note that in the $A(m,n)^{\hat{}}$ case $V_1$ is a subalgebra
  of the vertex subalgebra $F_0$ of $F$ (constructed in
  \S~\ref{sec:3}), while the highest component of the
  $F_0$-module $F_s$ restricted to $V_1$ is $L(\Lambda_{(s)})$.
  Since the $\Lambda_{(s)}$ exhaust all integrable highest
  weights of level~$1$, by Proposition~\ref{prop:8.1}, they give
  a complete list of irreducible $V_1$-modules.

In the $B(m,n)^{\hat{}}$ and $D(m,n)^{\hat{}}$ cases we note that 
$V_1$ is isomorphic to the vertex algebra $\Phi^+_V$ (see
Theorem~\ref{th:7.1}), $\Phi^-_V$ is its irreducible module, and
these two modules  produce all integrable highest weights of
level~$1$.

The cases $F(4)^{\hat{}}$ and $G(3)^{\hat{}}$ are obvious since $
V_1$ is the only irreducible integrable module of level~$1$ (see
Remark~\ref{rem:6.3}).

It remains to show that $L\left( -\frac{a+2}{a+1} \Lambda_0
  -\Lambda_1 + \frac{1-a}{a} \Lambda_3 \right)$ is a $V_1$-module
in the $D(2,1;a)^{\hat{}}$ case.  But $-\frac{a+2}{a+1} \Lambda_0
-\Lambda_1 + \frac{1-a}{a} \Lambda_3 =-\frac{1}{a+1} \Lambda_0 -
\left( \tfrac{1}{2} \alpha_0 + \frac{a-1}{2a} \alpha_3\right)$,
hence the difference of this weight and $u\Lambda_0$ does not
lie in the root lattice; we also have: $k=1$ and
level~$h^{\vee}=0$.  Hence we may apply
Propositions~\ref{prop:8.4} and \ref{prop:8.3}.

\end{proof}

\begin{remark}
  \label{rem:8.1}
The lowest energy $D (2,1;a)$-submodule of the module $L \left(
  -\frac{a+2}{a+1} \Lambda_0 - \Lambda_1 + \frac{1-a}{a}
  \Lambda_3\right)$ is the module $\bar{L}
(-\bar{\Lambda}_1+(a^{-1}-1)\bar{\Lambda}_3)$.  It has dimension
$4a^{-1}+2$.  For $a=1$ this is the defining module of $D(2,1)$;
for $a=\tfrac{1}{2}$ (resp.~$\tfrac{1}{3}$) this is the $10$-
(resp.~$14$-) dimensional module mentioned in
Remark~\ref{rem:6.8}.  As a $D(2,1;a)$-module, the even
(resp.~odd) part of this module is isomorphic to the irreducible
$s \ell(2)+s\ell(2)+s\ell(2)$-module
\begin{displaymath}
  \CC\boxtimes \CC^{a^{-1}} \boxtimes \CC^2
(\hbox{resp. } \CC^2 \boxtimes \CC^{a^{-1}+1}\boxtimes \CC )\, .
\end{displaymath}

\end{remark}

\begin{remark}
  \label{rem:8.2}
Let $V$ be a vertex algebra with a conformal vector such that 
$L_0$ is diagonizable with finite-dimensional eigenspaces and
rational eigenvalues.  It is a general belief that if $V$ has
finitely many irreducible modules, then the character $\tr_M
q^{L_0}$ of each of these modules $M$ becomes a modular function
when normalized, i.e.,~multiplied by a suitable power of $q$.
The example of the vertex algebra $V_1$ for $B(m,n)^{\hat{}}$ and 
$D(m,n)^{\hat{}}$ confirms this conjecture and leads to believe
that the same is true for $D(2,1;a)^{\hat{}}$ (with $a^{-1} \in \NN$),
$F(4)^{\hat{}}$ and $G(3)^{\hat{}}$.

The vertex algebra $V$ is called rational if $L_0$ has integral
eigenvalues, the number of irreducible $V$-modules is finite and
any $V$-module is completely reducible.  It follows from the
above discussion that the vertex algebra $V_1$ for
$B(m,n)^{\hat{}}$, $D(m,n)^{\hat{}}$, $D(2,1;a))$, $F(4))^{\hat{}}$ and
$G(3)^{\hat{}}$ is a rational vertex algebra, and that, moreover, 
the corresponding Zhu algebra \cite{Z} is finite-dimensional
semisimple (and even $1$-dimensional in the $F(4)^{\hat{}}$ and $
G(3)^{\hat{}}$ cases).

It was proved by Zhu \cite{Z} under certain technical assumptions 
that the $\CC$-span of normalized characters of irreducible
modules over a rational vertex algebra is $SL (2,\ZZ)$-invariant, 
and it was believed by many that the technical assumptions may be 
removed.  However, the above mentioned rational
vertex algebra $V_1$ shows that this is not the case.

\end{remark}

\begin{remark}
  \label{rem:8.3}
There are only two cases where there exists only a finite number
of essentially inequivalent subprincipal integrable
$\hat{\fg}$-modules of a given non-zero level~$k$:
\begin{displaymath}
  \hat{\fg} =F(4)^{\hat{}}, k=-\tfrac{3}{2} \hbox{ and }
  \hat{\fg} =G(3)^{\hat{}}, k=-\tfrac{4}{3} \, .
\end{displaymath}
In both cases the only subprincipal integrable $\hat{\fg}$-module 
is $L(\Lambda_0)$.  In both cases the associated vertex algebra
is rational with a unique irreducible module and the Zhu algebra
is $1$-dimensional.
\end{remark}

\section{Some remarks and open problems}
\label{sec:9}

\subsection{}\label{subsec:9.1}The calculation of characters of integrable highest weight
modules of arbitrary level~$k$ over affine superalgebras seems to
be a very difficult problem.  One may expect that the case of the 
``critical'' level $k=-h^{\vee}$ should be rather different from
other levels (as for the affine Lie algebras).  However, the
construction of level~$1$ integrable modules over $osp
(m|n)^{\hat{}}$ given in \S~\ref{sec:7} is the same for all
values $m$ and $n$ though $1$ is the critical level iff $m-n=1$.

Formula~(\ref{eq:5.12}) leads us to believe in the following conjecture:

Consider a principal integrable highest weight module
$L(\Lambda)$ over an affine superalgebra $\hat{\fg}$ and suppose
that one can choose a set of simple roots $\hat{\Pi}$ such that
it contains a maximal $\Lambda + \rho$-isotropic subset
$S_{\Lambda}$ of roots (i.e.,~all roots from $S_{\Lambda}$ are
pairwise orthogonal and orthogonal to $\Lambda +\rho$
\cite{KW}).  Let $\hat{W}^{\#}$ be the Weyl group 
of the integrable part $\hat{\fg}'_{\bar{0}}$ of $\hat{\fg}_{\bar{0}}$.  We conjecture
that  the following character formula holds:
\begin{equation}
\label{eq:9.1}
  e^{\rho} R ch L (\Lambda)
    = \sum_{w \in \hat{W}^{\#}}
    \epsilon (w)w 
    \frac{e^{\Lambda + \rho}}
    {\Pi_{\beta \in S_{\Lambda}} (1+e^{-\beta})}  \, .
\end{equation}
Note that the assumptions of this conjecture exclude the critical 
level, and include the level~$1$ integrable modules over
exceptional affine superalgebras.

\subsection{}\label{subsec:9.2}
In the papers \cite{KW1} and \cite{KW2} we proved character
formulas for a class of modules $L(\Lambda)$ over affine Lie
algebras, called admissible modules, which includes integrable
modules.  These character formulas imply that the normalized
specialized characters of admissible modules are modular
functions (and we conjecture that this property characterizes
admissible modules).  Of course, these character formulas break
down in the Lie superalgebra case.  However, in certain
exceptional situations, when the character admits a simple
product expansion in the Lie algebra case (see \cite{KW2},
Theorem~3.2), it seems that a similar product formula holds in
the Lie superalgebra case as well.

Concretely, let $u$ be a positive integer and let
\begin{equation}
  \label{eq:9.2}
  k=h^{\vee} (u^{-1}-1)
\end{equation}
(recall that in general the level~$k$ of an admissible module is
$\geq h^{\vee} (u^{-1}-1)$).  Let $y$ be an automorphism of the root lattice
$\hat{Q}$ such that all roots $\gamma_i=y((u-1)\delta_{i0}
K+\alpha^{\vee}_i)$ are positive $(i \in \hat{I})$.  The weights
of the form $y . k \Lambda_0$, where, as usual, $y . \lambda
=y(\lambda +\rho)-\rho$, are called \emph{admissible}.  We conjecture
that the following analog of formula~(3.3) from
\cite{KW2} holds:
\begin{eqnarray}
  \label{eq:9.3}
  ch L (y . (k \Lambda_0))=
  e^{y. (k \Lambda_0)} \left( \frac{\varphi (q^{u})}{\varphi (q)}
  \right)^{\ell }
  \Pi_{\substack{\alpha \in \bar{\Delta}_{\bar{0}}\\n \in   \NN}}
   \frac{1-q^{un}e^{y . \alpha}}{1-q^ne^{\alpha}}/
   \Pi_{\substack{\alpha \in \bar{\Delta}_{\bar{1}}\\n \in \NN}}
   \frac{1+q^{un}e^{y . \alpha}}{1+q^ne^{\alpha}} \, ,
\end{eqnarray}
where $\ell$ is the rank of $\fg$,  $\bar{\Delta}_{\bar{0}},
\bar{\Delta}_{\bar{1}}$ are the sets of even and odd roots of $\fg$, 
and $q=e^{-\delta}$.    

This conjecture agrees with
formula~(\ref{eq:7.5}) in the case $\hat{\fg}=osp (2|2)^{\hat{}}
= s \ell (2|1)^{\hat{}}$.  In this case $k=-1/2$ and
$h^{\vee}=1$, so that (\ref{eq:9.2}) holds for $u=2$.  All the admissible weights of
level~$-\tfrac{1}{2}$ are as follows:
\begin{displaymath}
  -\tfrac{1}{2} \Lambda_i (i=0,1,2), 
  -\tfrac{1}{2} \Lambda_0 - \tfrac{1}{2} \alpha_0 \, ,
\end{displaymath}
where the Dynkin diagram is chosen such that $\alpha_1$ and
$\alpha_2$ are odd roots (and $\alpha_0$ is even).  Character
formula~(\ref{eq:7.5}) gives:
\begin{eqnarray}
  \label{eq:9.4}
  \begin{array}{c}
    ch (-\frac{1}{2}\Lambda_0)\\
    ch(-\frac{1}{2} \Lambda_0-\frac{1}{2} \alpha_0)
  \end{array}
  = \tfrac{1}{2} e^{-\frac{1}{2} \Lambda_0}
   \left( \frac{\Psi (u^{-1}vq^{\frac{1}{2}};q)}
      {\Psi (-uvq^{\frac{1}{2}};q)} \pm
      \frac{\Psi (-u^{-1}vq^{\frac{1}{2}};q)}
         {\Psi (uvq^{\frac{1}{2}};q)}
  \right) \, ,
\end{eqnarray}
where
     $u = e^{-\frac{1}{2} \alpha_1}, 
     v=e^{-\frac{1}{2}\alpha_2}$ and
\begin{eqnarray}
\label{eq:9.5}
    \Psi (z;q) = \Pi^{\infty}_{k=1} (1+zq^{k-1})
    (1+z^{-1}q^k)\, ,
\end{eqnarray}
whereas formula~(\ref{eq:9.3}) gives:
\begin{subequations} \label{eq:9.6}
  \begin{eqnarray}
    \label{eq:9.6a}
    ch (-\tfrac{1}{2} \Lambda_0) &=& e^{-\frac{1}{2} \Lambda_0}
    \frac{\Psi (u^2 q;q^2) \Psi (v^2q;q^2) \varphi (q^2)^2}
      {\Psi (-uvq;q^2) \varphi (q)^2}\\[2ex]
\label{eq:9.6b}
    ch (-\tfrac{1}{2} \Lambda_0 - \tfrac{1}{2} \alpha_0) &=&
    e^{-\frac{1}{2} \Lambda_0 -\frac{1}{2}\alpha_0}
    \frac{\Psi (u^2;q^2) \Psi (v^2;q^2) \varphi (q^2)^2}
    {\Psi (-uvq;q^2) \varphi (q)^2} \, .
 \end{eqnarray}
\end{subequations}
However, the seemingly different expressions in the right-hand
sides of  (\ref{eq:9.4}) and (\ref{eq:9.6}) actually coincide due 
to one of the addition theta function formulas (cf.~\cite{M},
formula~(6.6) and notation on p.~17):
\begin{equation}
    \label{eq:9.7}
  \theta_{00} (\tau, z_1) \theta_{00} (\tau, z_2)
  = \theta_{00} (2\tau, z_1+z_2) \theta_{00} (2\tau,z_1-z_2)
 + \theta_{10} (2\tau,z_1+z_2) \theta_{10}(2\tau,z_1-z_2) \, ,
\end{equation}
if we let $u=e^{2\pi iz_1}, v=e^{2\pi i z_2}$.

Using \cite{KW}, formula~(6.1), it is immediate to show that the
span of supercharacters of the four admissible $s\ell
(2|1)^{\hat{}}$-modules of level $-1/2$ is $SL (2,
\ZZ)$-invariant.  Thus, it is natural to conjecture that this
modular invariance property of admissible characters holds for
any affine superalgebra $\hat{\fg}$ and any $k$ given by (\ref{eq:9.2}).

%%
%\begin{eqnarray}
%  \label{eq:9.7}
%  \frac{\Psi (u^{-1}vq^{\frac{1}{2}};q)}
%    {\Psi (-uvq^{\frac{1}{2}};q)} 
%  + \frac{\Psi (-u^{-1}vq^{\frac{1}{2}};q)}
%    {\Psi (uvq^{\frac{1}{2}};q)}
%  &=&  2 \frac{\Psi (u^2q;q^2) \Psi (v^2q:q^2)\varphi (q^2)^2}
%      {\Psi (-uvq;q^2) \varphi (q)^2} \, , \\[2ex]
%%
%\label{eq:9.8}
%     \frac{\Psi (u^{-1}vq^{\frac{1}{2}};q)}
%     {\Psi (-uvq^{\frac{1}{2}};q)}-
%%
%     \frac{\Psi (-u^{-1}vq^{\frac{1}{2}};q)}
%       {\Psi (uvq^{\frac{1}{2}};q)}
%%
%   &=& 2u^{-1}v^{-1}
%      \frac{\Psi (u^2;q^2) \Psi (v^2;q^2) \varphi (q^2)^2}
%      {\Psi (-uvq;q^2) \varphi (q)^2} \, .
%%
%\end{eqnarray}

Two other very interesting examples are provided by
Remark~\ref{rem:8.3}:  $\hat{\fg}=F(4)^{\hat{}}$ with $u=2$,
$y=1$ and $\hat{\fg}=G(3)^{\hat{}}$ with $u=3,y=1$.

\subsection{}
\label{subsec:9.3}
The first case not covered in \S~\ref{sec:5}, that when $m=n-1$,
is very interesting.  It connects the level~$1$ modules over $\fg
\ell (n-1|n)^{\hat{}}$ (or, equivalently the ``critical'' level~$-1$ 
modules over $g \ell (n|n-1)\hat{}$) to the denominator identity 
for $s \ell (n|n)^{\hat{}}$, which is unknown. Analyzing this
connection, we arrived at the following $s
\ell (2|2)^{\hat{}}$ denominator identity:
\begin{equation}
  \label{eq:9.9}
  e^{\rho}R = \sum_{w \in \hat{W}^{\#}} \epsilon (w)w
  \frac{e^{\rho}}{(1+e^{\alpha_0})\Pi^{\infty}_{j=1}(1+q^je^{\alpha_2})
     (1+q^{j-1} e^{-\alpha_2}) }
\end{equation}
where, as before, $q=e^{-\delta}$.  Here we use the Dynkin
diagram with the grey nodes $\alpha_0$ and $\alpha_2$.  If all four nodes are grey we get
the same identity with $\rho$ replaced by $0$; in yet another
form (\ref{eq:9.9}) can be written as follows:
\begin{eqnarray}
  \label{eq:9.10}
  \frac{R}{\varphi (q)} \Pi^{\infty}_{n=1} (1-q^{2n})
  (1+q^{2n-1} e^{\alpha_1+\alpha_3})
  (1+q^{2n-1}e^{-\alpha_1-\alpha_3})\\
\nonumber
= \sum_{w \in \hat{W}^{\#}} \epsilon (w)w
(\Pi^{\infty}_{n=1} (1+q^ne^{\alpha_1})
(1+q^ne^{\alpha_3})(1+q^{n-1}e^{-\alpha_1})(1+q^{n-1}e^{-\alpha_3})) 
\, .
\end{eqnarray}
The latter identity is equivalent to the following identity in
$u=e^{-\alpha_1}, x=e^{-\alpha_2}, v=e^{-\alpha_3}$ and $q$
(where $\Psi$ is defined by (\ref{eq:9.5})):
\begin{eqnarray*}
  \label{eq:9.11}
  \lefteqn{\hspace{-5ex}\Psi (uvq;q^2) \Psi (-ux;q) \Psi (-vx;q)}\\[1ex]
  &=& \Psi (uv^{-1}q;q^2) \Psi (x;q) \Psi (uvx;q)
  -x\Psi (uvx^2q;q^2) \Psi (u;q) \Psi (v;q)\, .
\end{eqnarray*}
In the notation of \cite{M} this identity can be rewritten in
terms of theta functions as
follows (if we let $u=e^{2\pi iz_1}$, $v=e^{2\pi i z_2}$,
$x=e^{2\pi iz_3}$):
\begin{eqnarray}
  \label{eq:9.12}
  \theta_{00} (2\tau ,z_1+z_2) \theta_{11}(\tau,z_1+z_3)
  \theta_{11} (\tau ,z_2+z_3)\\
\nonumber
  + \theta_{00} (2\tau ,z_1-z_2) \theta_{10} (\tau ,z_3)
   \theta_{11}(\tau , z_1 + z_2 + z_3)\\
\nonumber
  =\theta_{00} (2 \tau +z_1+z_2+2z_3) 
   \theta_{10} (\tau ,z_1) \theta_{10}(\tau,z_2) \, .
\end{eqnarray}
Identity (\ref{eq:9.11}) can be derived from (9.7) as follows.
Replacing $z_i$ by $z_i + \tfrac{1}{2} \tau$ (resp.~$z_i +
\tfrac{1}{2} (1+\tau)$) in (9.7), we obtain:

\begin{subequations}
  \begin{equation}
    \label{eq:9.11a}
    \theta_{10} (\tau ,z_1) \theta_{10}(\tau ,z_2)
    = \theta_{00} (2 \tau ,z_1+z_2) \theta_{10}(2\tau ,z_1-z_2)
    +\theta_{10}(2\tau,z_1+z_2) \theta_{00}(2\tau, z_1-z_2)
  \end{equation}
  \begin{equation}
    \label{eq:9.11b}
    \theta_{11} (\tau ,z_1) \theta_{11}(\tau ,z_2)
    = \theta_{00} (2 \tau ,z_1+z_2) \theta_{10}(2\tau ,z_1-z_2)
    -\theta_{10}(2\tau,z_1+z_2) \theta_{00}(2\tau, z_1-z_2)\, .
  \end{equation}

\end{subequations}

Substituting (\ref{eq:9.11b}) (resp. (\ref{eq:9.11a})) in the
first (resp. second) summand of the left-hand side of
(\ref{eq:9.11}), we obtain the product of $\theta_{00} (2\tau,
z_1+z_2+2z_3)$ and the right-hand side of (\ref{eq:9.11a}), and,
substituting its left-hand side, we obtain the left-hand side of
(\ref{eq:9.11}).

We also have a conjectural formula for an $s\ell (3|3)^{\hat{}}$
denominator identity, but it is too cumbersome to be reproduced
here.  We have no conjectures as how the $s\ell(n|n)^{\hat{}}$
denominator identity should look for $n>3$.

 Using the connection of the $s \ell (2|2)^{\hat{}}$ denominator
 identity to level~$1$ modules over $s \ell (2|1)^{\hat{}}$ we
 deduce from (\ref{eq:9.9}) $(k \in \ZZ_+)$:
 \begin{eqnarray}
   \label{eq:9.12x}
  \hspace*{8ex} ch L (k \Lambda_0 -(k+1) \Lambda_1)
   = \frac{\varphi (q)}{e^{\rho}R} \sum_{w \in \langle r_0 \rangle}
   \epsilon (w) w\sum_{j \in \ZZ_+} t_{j\alpha_0}
   \frac{e^{\Lambda +\rho}}
   {\Pi^{\infty}_{n=1} (1+q^{n-1} e^{-\alpha_2})
     (1+q^ne^{\alpha_2})} \, .
 \end{eqnarray}
Here the Dynkin diagram is chosen in such a way that $\alpha_0$
is even and $\alpha_1, \alpha_2$ are odd simple roots.

\subsection{}
\label{subsec:9.4}
Let $k$ be such that $V_k=L(uk \Lambda_0)$ is a (principal or
subprincipal) integrable $\hat{\fg}$-module of level~$k$.  Is it
always true that any integrable $\hat{\fg}$-module of level~$k$
can be extended to a module over the vertex algebra $V_k$?  Of
course, this question is closely related to the description of
generators of the left ideal $I_k$.  In the principal integrable
case $I_k$ contains elements (\ref{eq:8.2}), and the answer to
the above question in this case would be positive if $I_k$ were
generated by these elements.

Is it true that the normalized (by a power of $q$) characters $tr 
q^{L_0}$ (where $L_0$ is given by the Sugawara construction
\cite{K4}) of irreducible $V_k$-modules are modular functions,
provided that there are finitely many of them and $k+h^{\vee{}}
\neq 0$.  Is it true that for $k$ of the form (\ref{eq:9.2}), all
$V_k$-modules are admissible modules?

\subsection{}
\label{sec:9.5}

A few examples that we have worked out in the paper indicate that 
the theory of integrable highest weight modules over affine Lie
superalgebras is dramatically different from that in the Lie
algebra case.  The only exception is the case of
$\hat{\fg}=B(0,n)^{\hat{}}$.  The integrability conditions
are (see Table~\ref{tab:1} for its Dynkin diagram).
\begin{displaymath}
  k_i \in \ZZ_+ \hbox{ for all } i, \quad k_n \in 2\ZZ_+ \, , 
\end{displaymath}
hence the level $k=2k_0+\ldots +2k_{n-1}+k_n$ is a non-negative
even integer and the number of integrable highest weight
$B(0,n)^{\hat{}}$-modules is finite for each $k$.  Moreover, each 
of these modules extends to an irreducible $V_k$-module since
$I_k$ is generated by $e_{-\theta}(1)^{k+1}$ for each $k$, and these
are all irreducible modules over the vertex algebra $V_k\,\, (k \in
2\ZZ_+)$.

Furthermore, the character formula for all integrable
$B(0,n)^{\hat{}}$-modules $L(\Lambda)$ is known (see \cite{K2}),
and it is given by the same expression as that for the twisted
affine algebra $A^{(2)}_{2n}$ (replacing the black node by a
white one).  In order to derive the transformation formula of
$B(0,n)^{\hat{}}$ supercharacters from that of $A^{(2)}_{2n}$
characters, we need to go from the $A^{(2)}_{2n}$ coordinates,
which we call $A$-coordinates, to the $B(0,n)^{\hat{}}$
coordinates, which we will call $B$-coordinates.  This
calculation is explained below.

The $B$-coordinates $(\tau ,z_B,u_B)$ of $h \in \hat{\fh}$ are
defined by
\begin{displaymath}
  h=2\pi i (-\tau \tfrac{1}{2}\Lambda_0 + z_B+u_B\delta) \, ,
  \hbox{ where } z_B \in \fh \, .
\end{displaymath}
Let $\beta = \tfrac{1}{2} \sum^{n-1}_{j=0} (n-j)\alpha^{\vee}_{j}$.  Then $t_{\beta} (\Lambda_i)=\Lambda_i$ for
$i=1,\ldots,n-1$ and $\langle
t_{-\beta}(\tfrac{1}{2}\Lambda_0),\alpha^{\vee}_i \rangle
=\delta_{0n}$, hence we may take
$\Lambda_n=t_{-\beta}(\tfrac{1}{2}\Lambda_0)$ for the $0$\st{th}
fundamental weight of $A^{(2)}_{2n}$.  Hence the $A$-coordinates
are expressed via $B$-coordinates by
\begin{equation}
  \label{eq:9.12xx}
  \tilde{h}=t_{-\beta}(h)=2\pi i 
  (-\tau \Lambda_n +z_A + u_A \delta) \, .
\end{equation}

Recall that $SL (2,\ZZ)$ acts on functions in $\tau,z,u$ by the
formula [K3, Chapter~13]:
\begin{displaymath}
  F(\tau,z,u)|_{\left( 
      \begin{smallmatrix}
a&b\\c&d
      \end{smallmatrix}
\right)} = j (\tau)^{-n} F \left( \frac{a\tau +b}{c\tau +d}, \, 
\frac{z}{c\tau +d},
u-\frac{c(z|z)}{2(c\tau +d)}\right) \, .
\end{displaymath}
Furthermore, defining a new function $F^{\alpha ,\beta}$ by
\begin{displaymath}
  F^{\alpha ,\beta}(h)=F(t_{\beta}(h)+2\pi i \alpha
     - \pi i (\alpha | \beta)\delta) \, ,
\end{displaymath}
we have \cite {KP2}:
\begin{equation}
  \label{eq:9.13}
  F^{\alpha ,\beta}|_{\left(
      \begin{smallmatrix}
        a&b\\c&d
      \end{smallmatrix}\right)}= (F|_{\left(
      \begin{smallmatrix}
        a&b\\c&d
      \end{smallmatrix}\right)})^{d\alpha -b\beta,a\beta-c\alpha} 
  \, .
\end{equation}
We shall use the following connection between supercharacters of
$B(0,n)^{\hat{}}$ and characters of $A^{(2)}_{2n}$, which follows 
from (9.13) and definitions:
\begin{equation}
  \label{eq:9.14}
  sch L(\Lambda)(h)=chL(\Lambda)(\tilde{h})^{-\beta ,\beta}
  =(-1)^{nk/2} ch L (\Lambda)(\tilde{h})^{\beta,\beta} \, .
\end{equation}
Recall that the normalized $A^{(2)}_{2n}$ character
$\tilde{\chi}_{\Lambda}$ and the normalized $B(0,n)^{\hat{}}$
supercharacter $\chi_{\Lambda}$ are defined by:
\begin{displaymath}
  \tilde{\chi}_{\Lambda}=q^{\tilde{m}_{\Lambda}} chL (\Lambda),
  \, \chi_{\Lambda}=q^{m_{\Lambda}} sch L (\Lambda) \, ,
\end{displaymath}
where in $B$-coordinates:
\begin{displaymath}
  m_{\Lambda} =\frac{|\Lambda +\rho|^2}{2(k+h^{\vee})}
  - \frac{|\rho|^2}{2h^{\vee}}
  =\frac{(\Lambda +2\rho |\Lambda)}{2(k+h^{\vee})}
  -\frac{c_k}{24}, \, \quad 
  c_k=\frac{k \, \sdim B(0,n)}{k+h^{\vee}} \, ,
\end{displaymath}
and $\tilde{m}_{\Lambda}$ is defined by a similar formula in $A$-coordinates.
(Hence $\chi_{\Lambda} (\tau ,0,0)=str_{L(\Lambda)}
q^{L_0-c_k/24}$, as it should be.)

Let $S=\left( 
  \begin{smallmatrix}
0&-1\\1&0
  \end{smallmatrix}
\right) \in SL (2,\ZZ)$.  We shall denote by $S_A$ (resp. $S_B$)
the action of $S$ in $A$- (resp.~$B$-coordinates).  We have by
(\ref{eq:9.14}):
\begin{equation}
  \label{eq:9.15}
  \chi_{\Lambda} |_{S_B} =\tilde{\chi}^{-\beta,\beta}_{\Lambda}
  = (\tilde{\chi}_{\Lambda}|_{S_A})^{\beta,\beta} \, ,
\end{equation}
where the last equality holds due to (\ref{eq:9.13}).

But one has (see \cite {KP2},\cite{K3}, Theorem~13.8a):
\begin{displaymath}
  \tilde{\chi}_{\Lambda }|_{S_A}
  =\sum_{M\in P^k_+\!\!\!\! \mod \CC  \delta}
  S_{\Lambda M} \tilde{\chi}_M \, ,
 \end{displaymath}
where $(S_{\Lambda M})$ is an explicitly known matrix.  Hence,
continuing the calculation (\ref{eq:9.15}), we get, using
(\ref{eq:9.14}):
\begin{displaymath}
  \chi_{\Lambda} |_{S_B}
 =\sum_{M\in P^k_+\!\!\!\! \mod \CC  \delta}
  S_{\Lambda M}\tilde{\chi}^{\beta ,\beta}_M  =(-1)^{nk/2}\sum_M S_{\Lambda M} \tilde{\chi}^{-\beta ,\beta}_M \, .
\end{displaymath}
Using again (\ref{eq:9.14}), we obtain the final transformation
formula:
\begin{equation}
  \label{eq:9.16}
  \chi_{\Lambda}|_{S_B} =(-1)^{nk/2} 
  \sum_{M\in P^k_+\!\!\!\!\mod \CC \delta} S_{\Lambda M}\chi_M \, .
\end{equation}

It is clear from the above calculation that the
$SL(2,\ZZ)$-invariance of normalized $B(0,n)^{\hat{}}$ characters
$\tilde{\chi}_{\Lambda}$ does not hold, but the span of $\{
\tilde{\chi}_{\Lambda},\,\, \tilde{\chi}^{\beta ,0}_{\Lambda},\,\,
\tilde{\chi}^{0,\beta}_{\Lambda}\}_{\Lambda \in P^k_+ \!\!\mod \CC
  \delta}$ is $SL (2,\ZZ)$-invariant.

\subsection{}
\label{sec:9.6}

We use this opportunity to make some corrections to \cite{KW}.

Due to computer error the following lines disappeared from the paper:

\arabicparenlist

\begin{enumerate}

\item %%1
  Bottom of page 418:

  and  $4(n+1)^2$, respectively (given by Theorem~4.2; see also
  Examples~5.3 and %)

\item %%2
  Bottom of page 421:

  (this is independent of the choice of $B$), and let $W^{\#}$
  denote the subgroup of $W$ generated by reflections
  $r_{\alpha}$ with respect to all $\alpha \in \Delta^{\#}_0$.
  Denote by $(.|.)$ the even

\end{enumerate}

Also, the diagrams $B(0,n)$, $D(m,n)$ and $D(2,1;\alpha)$ on p.~429
should be as follows:

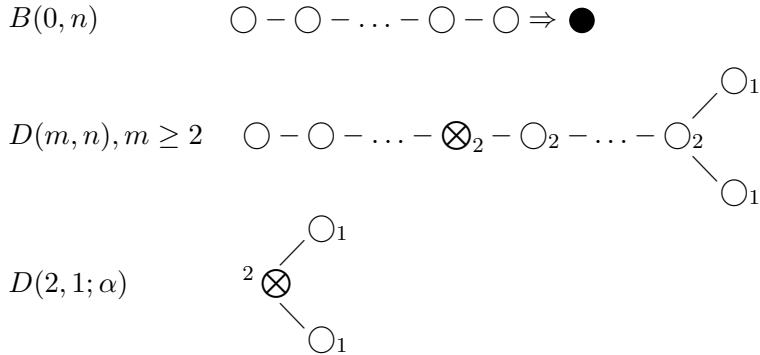
\begin{eqnarray*}
  \begin{array}{ll}
    B(0,n) & \bigcirc -\bigcirc - \ldots - \bigcirc - \bigcirc
    \Rightarrow  {\raisebox{-.75ex}{\Huge{\textbullet}}}\\[2ex]
    D(m,n),m \geq 2 &  
 \begin{array}{rll}
    & \hbox{\raisebox{-1.95ex}{\!\!\!\!\!\!\!\!$\diagup$}} 
    &\hspace{-2ex}\bigcirc_1\\
   \bigcirc - \bigcirc - \ldots - \bigotimes_2 
    - \bigcirc_2 - \ldots - \bigcirc_2 \\
    &  \hbox{\raisebox{1.95ex}{\!\!\!\!\!\!\!\!$\diagdown$}}
     & \hspace{-2ex}\bigcirc_1 
  \end{array}\\[3ex]
    D(2,1;\alpha) &  
 \begin{array}{rll}
    & \hbox{\raisebox{-1.95ex}{\!\!\!\!\!\!\!\!$\diagup$}} 
    &\hspace{-2ex}\bigcirc_1\\
    {}^2\bigotimes\\
    &  \hbox{\raisebox{1.95ex}{\!\!\!\!\!\!\!\!$\diagdown$}}
     & \hspace{-2ex}\bigcirc_1 
  \end{array}
  \end{array}
\end{eqnarray*}

Furthermore, the following corrections should be made:

\begin{list}{}{}
\item page 417, line 13$\uparrow$:  $\square (-q)$

\item page 432, line 12$\downarrow$: $M_{\Lambda}:=\{\alpha \in
  \bar{\Delta}_0|\alpha \perp S_{\Lambda}\}$,

\item page 434, lines 5,7$\downarrow$: $\alpha_2$ should be
  replaced by $\alpha_1$,

\item page 435, line 6$\downarrow$: $(\bigoplus_{j>0}t^j \otimes
  \fg)$,

\item page 438, line 1$\uparrow$: $:(\alpha_1 | \alpha_1)=0$,

\item page 449, line 3$\uparrow$: $(\alpha_2 | \alpha_2)=2$,

\item page 450, line 4$\uparrow$:    $\hat{R}=R_m
  \Pi^{\infty}_{n=1} \ldots$, where $R_m$ is the denominator
of $A_m$, not the one defined by (7.1),

\item page 453:  Theorem 8.1(a) as stated holds for the
  subprincipal integrable modules (cf.~Theorem~6.2 of the present 
  paper).

It is appropriate to mention here that the specialization (7.2)
of Conjecture 7.2 has been proved recently independently by
S.~Milne (by combinatorial methods) and by D.~Zagier (using
cusp forms). D.~Zagier also proved Conjecture 7.2 in the first
unknown case m=2.

\end{list}

In a slightly different form than in \cite{KW}, Conjecture~7.2
reads:
\begin{eqnarray*}
\Pi_{n=1}^{\infty} \left( \left( 
      \frac{1-q^{2n}}{1-q^{2n-1}} \right)^{2s}
    \Pi_{\alpha \in \Delta} 
    \frac{1-q^{2n}e^{\alpha}}{1-q^{2n-1} e^{\alpha}} \right)
  =\sum_{\substack{n_1,\ldots ,n_s \geq 0\\ k_1 \geq \ldots \geq
      k_s \geq 0}}
  ch L \left( \sum\limits^s_{i=1} k_i \gamma_i, A_m \right) \\[2ex]
\times \, q^{\sum\limits^s_{i=1} k_i (2n_i +1) + (m-2i+2) n_i} \, .
\end{eqnarray*}
Here $\Delta $ is the set of roots of $A_m, s=\left[
  \tfrac{m+1}{2}\right]$ and $\{ \gamma_1 , \ldots , \gamma_s \}$ 
is the set of positive pairwise orthogonal roots, $\gamma_1$
being the highest root.

\end{document}